\documentclass[%
    reprint,
    nofootinbib,
    amsmath,
    amssymb,
    aip,
    floatfix
]{revtex4-2}

\usepackage[utf8]{inputenc}
\usepackage[T1]{fontenc}
\usepackage{tikz}
\usepackage{amsmath}
\usepackage{mathtools}
\usepackage{url}
\usepackage{siunitx}
\usepackage{xcolor}
\usepackage{centernot}
\usepackage{placeins}
\usepackage{xparse}
\usepackage{float}

\DeclareSIUnit{\rad}{rad}

\newcommand{\I}{\mathrm{i}}
\newcommand{\E}{\mathrm{e}}
\newcommand{\dd}{\mathrm{d}}

\newcommand{\tsp}[1]{%
    \mathchoice
        {{\kern#1\dimexpr0.16667em\relax}}%
        {}%
        {}%
        {}%
}
\DeclarePairedDelimiter{\abs}{{\tsp{0.25}|}}{{|\tsp{0.25}}}

\DeclareMathOperator{\atanII}{arctan2}
\DeclareMathOperator{\sinc}{sinc}
\DeclareMathOperator{\cov}{cov}
\newcommand{\sgn}{\operatorname{sgn}}
\renewcommand{\Re}{\operatorname{Re}}
\newcommand{\hc}{\mathrm{h.c.}}

\newcommand{\comm}[2]{[{#1},{#2}]}
\DeclarePairedDelimiter{\ceil}{\lceil}{\rceil}
\DeclarePairedDelimiter{\floor}{\lfloor}{\rfloor}

\renewcommand{\vec}[1]{\boldsymbol{#1}}
\newcommand{\nullvec}{\boldsymbol{0}}

\newcommand{\T}[1]{{#1}^{\mathrm{T}}}

\newcommand{\inrowbvec}[2]{%
    {%
        \begin{bmatrix}%
            {#1}&{#2}%
        \end{bmatrix}%
    }%
}
\newcommand{\inrowbvecd}[3]{%
    {%
        \begin{bmatrix}%
            {#1}&{#2}&{#3}%
        \end{bmatrix}%
    }%
}

\newcommand{\leqdef}{\overset{\mathrm{def.}}{=}}
\newcommand{\reqdef}{\overset{\mathrm{def.}}{=}}

\newcommand{\of}[1]{%
    \mathchoice
        {\!\left({#1}\right)}
        {\!\left({#1}\right)}
        {\left({#1}\right)}
        {\left({#1}\right)}
}
\newcommand{\ofb}[1]{%
    \mathchoice
        {\!\left[{#1}\right]}
        {\!\left[{#1}\right]}
        {\left[{#1}\right]}
        {\left[{#1}\right]}
}

\counterwithin*{equation}{section}
\renewcommand{\theequation}{\arabic{section}.\arabic{equation}}
\let\appx\appendix
\renewcommand{\appendix}{\appx\renewcommand{\theequation}{\Alph{section}.\arabic{equation}}}

\usepackage{hyperref}
\hypersetup{%
    colorlinks,%
    linkcolor={blue},%
    citecolor={blue},%
    urlcolor={blue}%
}%

\usepackage{cleveref}
\Crefname{equation}{Eq.}{Eqs.}
\Crefname{figure}{Fig.}{Figs.}
\Crefname{tabular}{Tab.}{Tabs.}
\Crefname{section}{Sec.}{Secs.}
\newcommand{\crefrangeconjunction}{--}
\Crefmultiformat{equation}{Eqs.~(#2#1#3)}{ and~(#2#1#3)}{,~(#2#1#3)}{ and~(#2#1#3)}

\usepackage[caption=false]{subfig}
\newcommand{\pagecite}[2]{\cite{#2}\textsuperscript{, p.~{#1}}}
\newcommand{\citetwolastwithpage}[3]{%
    \cite{#1}\textsuperscript{;}%
    \cite{#2}\textsuperscript{, p.~{#3}}%
}
\newcommand{\Creffirstrangelast}[3]{%
    Eqs.~(\ref{#1}) and~(\ref{#2})\crefrangeconjunction(\ref{#3})%
}%

\ExplSyntaxOn
\NewDocumentCommand{\refcite}{m}{\refcite_aux:n{#1}}

\cs_new_protected:Npn \refcite_aux:n #1%
{%
    \seq_set_split:Nnn \l_tmpa_seq { , } { #1 }%
    \int_compare:nNnTF { \seq_count:N \l_tmpa_seq } > { 1 }%
        { Refs.\nobreakspace\onlinecite{#1} }%
        { Ref.\nobreakspace\onlinecite{#1} }%
}
\ExplSyntaxOff

\newcommand{\textsuoo}{{SU(1,1)}}

\newcommand{\eqset}{\overset{!}{=}}

\usepackage{tabularx}
\newcolumntype{Y}{>{\centering\arraybackslash}X}
\usepackage{multirow}

\begin{document}

    \title{Schmidt modes carrying orbital angular momentum generated by 
    cascaded systems pumped with Laguerre-Gaussian beams}

    \author{D.~Scharwald}
    \author{L.~Gehse}
    \author{P.~R.~Sharapova}
    \affiliation{%
        Department of Physics, Paderborn University,
        Warburger Straße 100, 33098 Paderborn, Germany
    }%

    \begin{abstract}
        Orbital Angular Momentum (OAM) modes are an important resource
        used in various branches of quantum science and technology due
        to their unique helical structure and countably infinite basis. 
        Generating light that simultaneously carries high-order orbital 
        angular momenta and exhibits quantum correlations is a
        challenging task.
        In this work, we present a theoretical approach to the
        generation of correlated Schmidt modes carrying OAM
        via parametric down-conversion (PDC) in
        cascaded nonlinear systems (nonlinear interferometers)
        pumped by Laguerre-Gaussian beams. We demonstrate how the number of
        generated modes and their population can be controlled
        by varying the pump parameters, the gain of the PDC process
        and the distance between the crystals. 
        We investigate the angular displacement sensitivity
        of these interferometers and demonstrate that
        it can overcome the classical shot noise limit.
    \end{abstract}

    \maketitle

    \section{INTRODUCTION}\label{sec:introduction}
        Structured light has been studied
        extensively over the past decade and
        currently plays an important role in
        various areas of
        physics~\cite{Bliokh2023}. One important
        type of structured light
        is light that carries orbital angular momentum
        (OAM)~\cite{Forbes2022,PhysRevA.45.8185}.
        These beams have a helical azimuthal
        phase dependency $\E^{-\I l \varphi}$ with a 
        central singularity, where $\varphi$ is the
        azimuthal angle and $l\in\mathbb{Z}$ the OAM quantum
        number (topological charge
        number) of the mode~\cite{Bliokh2023,PhysRevA.45.8185,Wang2021}.
        Due to their unique helical structure
        and countably infinite basis, OAM modes are an
        important resource for quantum communication,
        quantum computing, quantum information
        processing~\cite{Mair2001,Shen2019} and
        quantum cryptography~\cite{Sit:17}. 
        They can be generated using spatial light
        modulators~\cite{Fickler2012}, plasmonic
        nanostructures~\cite{Karimi2014} or
        in parametric
        down-conversion~\cite{Miatto2012,PRA91,Beltran2017,Shen2019}~(PDC)
        and four-wave
        mixing~\cite{Offer2018}~(FWM)
        processes and can be
        transmitted using
        multimode fibers~\cite{PhysRevApplied.11.064058,Wang2021}.
        Recently, the transmission of OAM
        modes and their superpositions over
        long distances has been demonstrated~\cite{Krenn2016},
        which is an important element for the creation of
        free-space optical communication links~\cite{Xie:15}.
        In addition to the applications mentioned above, OAM modes
        play an important role in metrology:
        It has been
        shown that entangled OAM states
        can be used to improve the
        \textit{angular displacement 
        sensitivity}~\cite{Fickler2012,PhysRevLett.127.263601,PhysRevA.83.053829}
        by a factor of~$1/l$.
        \par Thus, the generation and
        manipulation of correlated light
        carrying high-order orbital angular
        momenta is a vital problem. 
        Cascaded systems of PDC processes,
        in particular nonlinear
        interferometers, constitute an important
        tool that can be used for
        these purposes since they make it
        possible to manipulate
        the set of populated OAM modes~\cite{PRA91,Perez:14}. 
        Additionally, nonlinear interferometers have
        taken an important role in quantum metrology as they
        can be used to overcome the shot noise level
        for the phase sensitivity,
        which is the limit of
        any classical
        interferometer~\cite{Manceau2017,Chekhova:16,Frascella:19}.
        \par In this work, we present a theoretical
        framework to describe PDC and nonlinear interferometers pumped by
        Laguerre-Gaussian beams. Our framework is based
        on the Schmidt-mode formalism developed in
        \refcite{PRA91} and provides valuable analytical insights
        that would otherwise not be possible with rigorous methods,
        such as the integro-differential equations
        approach~\cite{PRR2}, that include
        time-ordering effects and can in general
        only be evaluated numerically.
        \par \citet{PhysRevA.106.063711}
        have also investigated
        PDC with Laguerre-Gaussian pump beams to
        analyze the role of the Gouy phase in SPDC. For this,
        the SPDC state is expanded over Laguerre-Gaussian modes
        for the signal and idler beams.
        However, outside of the
        double-Gaussian approximation,
        the Schmidt modes of PDC are generally not exactly given
        by Laguerre-Gaussian
        modes~\cite{Miatto2012,PRA91} and only look comparatively
        similar. Therefore, in \Cref{sec:model}, 
        we start by explicitly evaluating
        the spatial integral
        of the PDC-Hamiltonian and arrive
        at the two-photon amplitude~(TPA) and the Schmidt
        decomposition for PDC with a
        Laguerre-Gaussian pump beam. Along the way,
        we will highlight similarities and differences that arise 
        compared to the case of Gaussian pump~\cite{PRA91}
        and illustrate the conservation of angular
        momentum~\cite{Frascella:19}.
        Furthermore, since we present a general
        derivation allowing
        for arbitrary OAM and radial indices of
        the pump beam, our work
        can serve as a starting point for a 
        theoretical framework
        with arbitrarily shaped
        pump beams (pump engineering), since a
        given pump field can always be
        expanded in terms of the OAM modes~\cite{PhysRevA.106.063711}.
        This work may also
        provide the basis for future work on the description of
        PDC with Laguerre-Gaussian pump beams using the
        integro-differential equations approach~\cite{PRR2}.
        \par Next, in \Cref{sec:application_smode_structure}, we
        investigate the modal structure of
        the PDC radiation generated by pumping nonlinear
        interferometers with Laguerre-Gaussian beams with different
        OAM and radial numbers. To this end,
        we analyze the distribution of the weights
        (Schmidt eigenvalues)
        of the generated Schmidt modes and the
        output intensity profiles.
        In \Cref{sec:angular_displacement}, we investigate
        the angular displacement sensitivity
        of these interferometers.
        Finally, we draw our conclusion in \Cref{sec:conclusion}.
        
    \section{THEORETICAL MODEL}\label{sec:model}

        \subsection{The PDC Hamiltonian for Laguerre-Gaussian pump beams}
            In its general form,
            the Hamiltonian describing
            the parametric down-conversion~(PDC)
            process is given by~\cite{PRA91,Karan2020,klyshko1988photons}:
            \begin{align}\label{eq:hamiltonian_general}
                \hat{H}&\!\propto\!\int\!\dd^3r\,\chi^{\left(2\right)}\of{\vec{r}}
                    \hat{E}_p^{\left(+\right)}\of{\vec{r},t} \hat{E}_s^{\left(-\right)}\of{\vec{r},t}
                    \hat{E}_i^{\left(-\right)}\of{\vec{r},t} + \hc,
            \end{align}
            where $\chi^{\left(2\right)}$ is the second order (nonlinear)
            susceptibility of the PDC section and
            $\hat{E}_j\of{\vec{r},t}$ are the electric field operators, while
            the labels $j=p,s,i$ refer to the pump, signal and idler
            fields, respectively. In the following, we will assume
            $\chi^{\left(2\right)}$ to be independent of the
            position space coordinate vector
            ${\vec{r}=\T{\inrowbvecd{x}{y}{z}}}$.
            Furthermore, in this work, we restrict ourselves to
            monochromatic signal and
            idler fields, so that their field operators
            can be written as~\cite{Karan2020}:
            \begin{align}\label{eq:signal_idler_field_operators}
                \hat{E}_j^{\left(-\right)}\of{\vec{r},t} &= \epsilon_{j} \int\!\dd^2 q_{j}\,
                \E^{-\I\left[\vec{q}_j\cdot\vec{r}_{\perp}+k_{j,z}
                \of{\vec{q}_j,\omega_j} z - \omega_j t\right]} 
                \hat{a}^{\dagger}_j\of{\vec{k}_{j}} ,
            \end{align}
            where $\epsilon_{j}$ is a normalization constant, $j=s,i$,
            and the transverse wave vector and the transverse coordinate
            vector
            are given by $\vec{q}_j=\T{\inrowbvec{k_{j,x}}{k_{j,y}}}$ and
            $\vec{r}_{\perp}=\T{\inrowbvec{x}{y}}$, respectively.
            Importantly, note that there are only two integrals over the
            transverse wave vectors and no integral over the frequencies
            $\omega_j$. As a consequence, $k_{j,z}$ is fixed and can be
            written as a function of $\vec{q}_j$ and $\omega_j$, as indicated
            above.
            \par The pump is described by a
            strong classical field with a
            transverse Laguerre-Gaussian profile, so
            that~\cite{PhysRevA.45.8185,Offer2018,Pampaloni2004}
            \begin{align}\label{eq:laguerre_gaussian_pump}
                \begin{split}
                    E_p^{\left(+\right)}\of{\vec{r},t} &= E_0
                        \frac{r_\perp^{\abs{l_p}}}{w_0^{\abs{l_p}+1}}
                        L^{\abs{l_p}}_{m_p}\of{\frac{2r_\perp^2}{w_0^2}}
                        \E^{-\left(\frac{r_\perp}{w_0}\right)^2} \\
                    &\qquad\times \E^{-\I l_p \varphi} 
                        \E^{\I\left(\vec{k}_p\cdot\vec{r}-\omega_p t\right)},
                \end{split}
            \end{align}
            where $r_\perp=\abs{\vec{r}_\perp}=\sqrt{x^2+y^2}$ and
            $\varphi=\atanII\of{y,x}$ is\footnote{Throughout
            this work, we will use
            the two-argument arctangent function $\atanII$ for
            clarity, when appropriate.} the azimuthal 
            angle in coordinate space.
            $L^{\abs{l_p}}_{m_p}$ denotes the Laguerre polynomials,
            so that $l_p\in\mathbb{Z}$ and $m_p\in\mathbb{N}_0$
            are the OAM and radial quantum number of the
            Laguerre-Gaussian pump profile,
            respectively. $w_0$ is the width of
            the Gaussian term so that the full width at half maximum (FWHM)
            of the intensity distribution is given
            by $w_0\sqrt{2\ln{2}}$. More specifically,
            $w_0$ can be understood as the $1/\E^2$-radius of the intensity distribution
            of the Gaussian term.
            We have therefore also assumed
            that the beam waist diameter is constant over
            the setup under consideration. This is usually
            satisfied if the PDC sections are sufficiently thin.
            \par After plugging \Cref{eq:signal_idler_field_operators,eq:laguerre_gaussian_pump}
            into \Cref{eq:hamiltonian_general} and assuming perfect frequency
            matching $\omega_p=\omega_s+\omega_i$ (neglecting time-ordering effects),
            the Hamiltonian can be written in the form
            \begin{subequations}
                \begin{align}
                    \hat{H} &= \I\hbar\Gamma \iint\!\dd^2q_s\dd^2q_i\,F\of{\vec{q}_s,\vec{q}_i}
                        \hat{a}^{\dagger}_s\of{\vec{q}_s} \hat{a}^{\dagger}_i\of{\vec{q}_i} + \hc
                    \label{eq:hamiltonian_with_tpa}
                \end{align}
                where
                \begin{align}\label{eq:small_f_gxy_gz}
                    F\of{\vec{q}_s,\vec{q}_i} &= C g_{xy}\of{\vec{q}_s, \vec{q}_i} 
                        g_z\of{\vec{q}_s, \vec{q}_i}
                \end{align}
            \end{subequations}
            is the two-photon amplitude~(TPA)
            of the PDC process and $C>0$~is an
            $l_p$- and $m_p$-dependent
            normalization constant to ensure that
            the TPA is normalized to unity:
            ${\iint\!\dd^2q_s\dd^2q_i\,\abs{F\of{\vec{q}_s,\vec{q}_i}}^2=1}$.
            The newly introduced constant
            $\Gamma=\varepsilon_0\frac{\chi^{\left(2\right)}}{3}\frac{1}{C}
            \epsilon_s\epsilon_i E_0$ is the theoretical gain parameter.
            We have factorized the TPA so that the function $g_{xy}$
            contains the spatial integrals over
            the (transverse) $xy$-plane and $g_z$ contains the
            integral along the $z$-axis,
            both of which originate from the PDC Hamiltonian in 
            \Cref{eq:hamiltonian_general}.

        \subsection{Evaluation of the transverse spatial integral}
            The integral over the $xy$-plane of the nonlinear medium
            corresponds to the Fourier
            transform of the Laguerre-Gaussian envelope. In polar coordinates,
            $\vec{q}_j\equiv\left(q_j,\phi_j \right)$ (with~$j=s,i$) and
            $\vec{r}_{\perp}\equiv\left(r,\varphi \right)$, this integral
            takes the form:
            \begin{widetext}
                \begin{align}
                    g_{xy}\of{q_s, q_i, \phi_s, \phi_i} &= \int\!\dd r\int\!\dd\varphi\,
                        \left(\frac{r}{w_0}\right)^{\abs{l_p}+1}
                        L^{\abs{l_p}}_{m_p}\of{\frac{2r^2}{w_0^2}}
                        \E^{-\left(\frac{r}{w_0}\right)^2} 
                         \E^{-\I l_p \varphi}
                        \E^{-\I r \left[q_s\cos\left(\varphi-\phi_s\right)
                            +q_i\cos\left(\varphi-\phi_i\right)\right]}.
                \end{align}
                Since the last exponent is a superposition of
                weighted cosine terms, it can be rewritten in terms of a single
                sine function~\pagecite{84}{Bronshtein2015}:
                \begin{subequations}
                    \begin{align}
                            q_s\cos\of{\varphi-\phi_s}+q_i\cos\of{\varphi-\phi_i}
                                = \xi\of{q_s, q_i, \phi_s,\phi_i}\sin\ofb{\varphi+\psi\of{q_s, q_i, \phi_s, \phi_i}},
                    \end{align}
                    where
                    \begin{align}
                        &\xi\of{q_s, q_i, \phi_s, \phi_i} = \sqrt{q_s^2+q_i^2+2q_s q_i
                            \cos\of{\phi_s-\phi_i}}, \label{eq:def_beta} \\
                        &\psi\of{q_s, q_i, \phi_s, \phi_i} = \atanII\of{q_s\cos\phi_s+q_i\cos\phi_i,
                            q_s\sin\phi_s+q_i\sin\phi_i}. \label{eq:def_psi}
                    \end{align}
                \end{subequations}
                Note that $\xi=\abs{\vec{q}_p}$~\cite{PhysRevA.106.063711}.
                The transverse integral can then be rewritten in a more simple form:
                \begin{equation}
                    \begin{aligned}
                        g_{xy}\of{q_s, q_i, \phi_s, \phi_i} &= \int\!\dd r\, \left(\frac{r}{w_0}\right)^{\abs{l_p}+1}
                            L^{\abs{l_p}}_{m_p}\of{\frac{2r^2}{w_0^2}} \E^{-\left(\frac{r}{w_0}\right)^2}
                            \int_0^{2\pi}\!\dd\varphi\,\E^{\I\left(-r\xi \sin\of{\varphi+\psi}-l_p\varphi\right)}.
                    \end{aligned}\label{eq:gxy_phiintsimplified}%
                \end{equation}
                The integration over the polar angle $\varphi$ can be performed
                by using the definition of
                the Bessel functions of the first kind 
                $J_n$~\citetwolastwithpage{PhysRevA.106.063711}{watson}{20}
                and can be evaluated as:
                \begin{align}
                    \begin{split}
                        &\int_0^{2\pi}\!\dd\varphi\,\E^{\I\left(-r\xi \sin\of{\varphi+\psi}-l_p\varphi\right)}
                        = 2\pi \left[-\sgn\of{l_p}\right]^{l_p} \E^{\I l_p\psi} J_{\abs{l_p}}\of{r\xi},
                    \end{split}
                \end{align}
                where $\sgn$ is the sign function.
                The remaining integral over the radial coordinate $r$ then becomes 
                a Hankel transformation integral of order~$\abs{l_p}$~\pagecite{768}{Bronshtein2015}:
                \begin{align}
                    g_{xy}\of{q_s, q_i, \phi_s, \phi_i}
                    & = 2\pi \left[-\sgn\of{l_p}\right]^{l_p} \E^{\I l_p\psi}
                        \int_0^\infty\!\dd r\, \left(\frac{r}{w_0}\right)^{\abs{l_p}+1}
                        L^{\abs{l_p}}_{m_p}\of{\frac{2r^2}{w_0^2}} \E^{-\left(\frac{r}{w_0}\right)^2}
                        J_{\abs{l_p}}\of{r\xi}.
                         \label{eq:g_intermediate}
                \end{align}
                Substituting $x=\sqrt{2}r/w_0$, this integral can be directly evaluated
                using the following Hankel transform integral formula~\pagecite{42}{bateman}:
                \begin{align}
                    \begin{split}
                        \int_0^\infty\!\dd x\, x^{\nu+\frac{1}{2}}
                            \E^{-\frac{1}{2}x^2} L^{\nu}_n\of{x^2} J_{\nu}\of{xy}
                            \left(xy\right)^{\frac{1}{2}}
                        = \left(-1\right)^n \E^{-\frac{1}{2}y^2}
                            y^{\nu+\frac{1}{2}} L^{\nu}_n\of{y^2},
                    \end{split}
                \end{align}
                which is valid for $\Re\of{\nu}>-1$ and $y>0$.
                Finally, the function $g_{xy}$ is given by:
                \begin{align}
                    \begin{split}\label{eq:g_form_4}
                        g_{xy}\of{q_s, q_i, \phi_s, \phi_i} &= 
                            2\pi \left[\sgn\of{l_p}\right]^{\abs{l_p}} \left(-1\right)^{l_p+m_p}
                            \left(\frac{w_0}{2}\right)^{\abs{l_p}+1}
                        \E^{-\left(\frac{w_0\xi}{2}\right)^2} \E^{\I l_p\psi}\xi^{\abs{l_p}}
                        L^{\abs{l_p}}_{m_p}\of{\frac{\left(w_0\xi\right)^2}{2}}.
                    \end{split}
                \end{align}
            \end{widetext}

        \subsection{Separation of the signal and idler variables}
            It can be shown that $\psi\of{q_s, q_i, \phi_s, \phi_i}$
            defined in \Cref{eq:def_psi}
            and appearing in the exponential function in \Cref{eq:g_form_4} and,
            therefore, the entire TPA, cannot simply be written in terms
            of the signal-idler angle difference ${\phi_s-\phi_i}$. This means that the
            results derived in \refcite{PRA91} for the Gaussian pump cannot
            be directly applied for the Laguerre-Gaussian
            beam because the one-dimensional Fourier decomposition
            as given in Eq.~(3) therein
            cannot be applied to the TPA derived above. However,
            as will be shown below,
            it is possible to find a similar
            decomposition for the TPA derived
            in this work.
            \par To see this, first, note that
            \begin{subequations}
                \begin{align}
                    \sin\of{\psi} &= \frac{q_s \cos\of{\phi_s}+q_i\cos\of{\phi_i}}{\xi},
                        \label{eq:sin_cos_psi_sin}\\
                    \cos\of{\psi} &= \frac{q_s \sin\of{\phi_s}+q_i\sin\of{\phi_i}}{\xi}.
                        \label{eq:sin_cos_psi_cos}
                \end{align}
            \end{subequations}
            Using Euler's formula for the exponential function and taking into account
            \Cref{eq:sin_cos_psi_sin,eq:sin_cos_psi_cos}, the function
            $\E^{\I l_p \psi}$ can be rewritten
            as\footnote{To see this, write $l_p=\abs{l_p}\sgn\of{l_p}$,
            apply Euler's formula to the exponential term in $\left[\exp\of{\I\sgn\of{l_p}\psi}\right]^{\abs{l_p}}$,
            then \Cref{eq:sin_cos_psi_sin,eq:sin_cos_psi_cos},
            use the fact that $\cos$ and $\sin$ are even and odd functions in their arguments, respectively,
            to multiply the argument with $\sgn\of{l_p}$ when necessary
            and finally apply Euler's formula backwards.}
            \begin{subequations}
                \begin{align}
                    \E^{\I l_p \psi}
                        &=\left(\frac{\I \sgn\of{l_p}}{\xi}\right)^{\abs{l_p}}
                            W\of{q_s,q_i,\phi_s-\phi_i} \E^{-\I l_p \phi_i},
                \end{align}
                where
                \begin{align}\label{eq:def_W_term}
                    W\of{q_s,q_i,\phi_s-\phi_i} &=
                    \left[q_s \E^{-\I \sgn\of{l_p} \left(\phi_s-\phi_i\right)}+ q_i \right]^{\abs{l_p}}.
                \end{align}
            \end{subequations}
            Here, we have chosen to factor out the 
            exponential term containing the idler angular variable
            $\phi_i$ from $W$. However, it is also possible to
            instead factor out the exponential term containing
            $\phi_s$ or both.
            The TPA can then be combined as:
            \begin{align}
                F\of{q_s, q_i, \phi_s, \phi_i} &= R\of{q_s,q_i,\phi_s-\phi_i}
                    \E^{-\I l_p \phi_i}, \label{eq:F_R_exp}
            \end{align}
            with the newly defined function
            \begin{align}
                \begin{split}
                    R\of{q_s,q_i,\phi_s-\phi_i} &= 2\pi C \left(-1\right)^{l_p+m_p} \I^{\abs{l_p}}
                            \left(\frac{w_0}{2}\right)^{\abs{l_p}+1} \\
                        &\qquad\times L^{\abs{l_p}}_{m_p}\of{\frac{\left(w_0\xi\right)^2}{2}}
                            \E^{-\left(\frac{w_0\xi}{2}\right)^2} \\
                        &\qquad\times \left[q_s \E^{-\I \sgn\of{l_p} \left(\phi_s-\phi_i\right)}
                            + q_i \right]^{\abs{l_p}} \\
                        &\qquad\times g_z\of{q_s,q_i,\phi_s-\phi_i}, \label{eq:def_R}
                \end{split}
            \end{align}
            which depends only on the signal-idler
            angle difference ${\phi_s-\phi_i}$. Intuitively, this result for $R$ is plausible, since
            Laguerre-Gaussian functions are fixed points/eigenfunctions
            of the Fourier transform~\cite{Fontaine2019}.
             The normalization constant $C$ must be chosen so that
            $\iiint\!\dd q_s\dd q_i \dd \phi\, q_s q_i \abs{R\of{q_s,q_i,\phi}}^2=1/\!\left(2\pi\right)$,
            which is a consequence of the fact that we write $R$ in
            terms of the signal-idler angle difference $\phi$.
            \par As mentioned above, the full TPA $F$ cannot be written purely in
            terms of ${\phi_s-\phi_i}$. 
            Therefore, we apply the one-dimensional Fourier expansion only to $R$:
            \begin{align}\label{eq:def_R_decomp}
                R\of{q_s,q_i,\phi_s-\phi_i} &= \frac{1}{2\pi} \sum_{n=-\infty}^{\infty} \chi_n\of{q_s, q_i}
                    \E^{-\I n\left(\phi_s-\phi_i\right)}.
            \end{align} 
            The Fourier coefficients
            $\chi_n\of{q_s, q_i}$ can be
            obtained by computing the Fourier coefficient integral:
            \begin{align}\label{eq:chi_calc_integral}
                \chi_n\of{q_s,q_i} = \int_0^{2\pi}\!\dd\!\left(\phi_s-\phi_i\right)
                        R\of{q_s,q_i,\phi_s-\phi_i} \E^{\I n\left(\phi_s-\phi_i\right)}.
            \end{align}
            \par Plugging the Fourier
            decomposition [\Cref{eq:def_R_decomp}]
            into \Cref{eq:F_R_exp}, one can obtain
            the expansion of the
            full TPA in terms of the
            coefficients $\chi_n\of{q_s, q_i}$ as
            \begin{align}
                F\of{q_s, q_i, \phi_s, \phi_i} = \frac{1}{2\pi} \sum_{n} \chi_n\of{q_s, q_i}
                    \E^{-\I n \phi_s} \E^{-\I \left(l_p-n\right)\phi_i}.
                    \label{eq:decomposition}
            \end{align}
            For $l_p=0$, the regular Fourier
            decomposition over the angle
            difference is recovered~\cite{PRA91},
            resulting in perfectly anti-correlated
            OAM for the
            signal and idler photons.
            The normalization
            condition for the TPA now implies
            $\sum_n\iint\!\dd q_s \dd q_i\,q_s q_i \abs{\chi_n\of{q_s,q_i}}^2=1$
            (corresponding to
            \textit{Parseval's theorem}~\pagecite{637}{Bronshtein2015}).
            \par The Schmidt decomposition for
            the expansion coefficients reads~\cite{PRA91}:
            \begin{align}
                \chi_n\of{q_s, q_i} &= \sum_{m=0}^{\infty} \sqrt{\lambda_{mn}} \frac{u_{mn}\of{q_s}}{\sqrt{q_s}} \frac{v_{mn}\of{q_i}}{\sqrt{q_i}}, \label{eq:chi_n_sdecomp}
            \end{align}
            where $\lambda_{mn}$ are the Schmidt
            eigenvalues and $u_{mn}$ and $v_{mn}$
            are the eigenfunctions.
            The properties of the Schmidt decomposition
            ensure that the radial mode functions $u_{mn}$
            and $v_{mn}$ fulfill the following
            orthonormality conditions:
            \begin{subequations}
                \begin{align}
                    \int\!\dd q_s\, u_{mn}\of{q_s} u_{kn}^*\of{q_s} &= \delta_{mk}, \label{eq:orthonorm_u} \\
                    \int\!\dd q_i\, v_{mn}\of{q_i} v_{kn}^*\of{q_i} &= \delta_{mk}.
                \end{align}
            \end{subequations}
            Furthermore, the normalization condition of the TPA
            now requires ${\sum_{m,n}\lambda_{mn}=1}$.
            \par Finally, the TPA can be written in the following form:
            \begin{widetext}
                \begin{align}\label{eq:full_decomp_tpa}
                    F\of{q_s, q_i, \phi_s, \phi_i} &= \frac{1}{2\pi} \sum_{m,n} \sqrt{\lambda_{mn}} \frac{u_{mn}\of{q_s}}{\sqrt{q_s}} \frac{v_{mn}\of{q_i}}{\sqrt{q_i}} \E^{-\I n \phi_s} \E^{-\I \left(l_p-n\right)\phi_i}.
                \end{align}
            \end{widetext}
            From this expression, it can be seen that for the fixed summation
            indices $m$ and $n$, the signal and idler photons have different
            OAM, namely,
            $n$ and $l_p-n$, respectively.
            This form of the TPA allows us to introduce
            the spatially broadband Schmidt
            operators $\hat{A}_{mn}$, $\hat{B}_{mn}$ and diagonalize
            the full Hamiltonian:
            \begin{align}\label{eq:Hamiltonian_diagonalized}
                \hat{H} &=
                    \I\hbar\Gamma\sum_{m,n}\sqrt{\lambda_{mn}}
                    \left(\hat{A}_{mn}^{\dagger} \hat{B}_{mn}^{\dagger}-
                    \hat{A}_{mn}\hat{B}_{mn}\right),
            \end{align}
            where the Schmidt operators are defined as
            \begin{subequations}
                \begin{align}
                    \hat{A}_{mn}^{\dagger} &= \frac{1}{\sqrt{2\pi}}
                        \int\!\dd^2q_s\,\frac{u_{mn}\of{q_s}}{\sqrt{q_s}}\E^{-\I n \phi_s} \hat{a}_s^{\dagger}\of{\vec{q}_s},
                        \label{eq:defschmidtopa} \\
                    \hat{B}_{mn}^{\dagger} &= \frac{1}{\sqrt{2\pi}} \int\!\dd^2q_i\,\frac{v_{mn}\of{q_i}}{\sqrt{q_i}}\E^{-\I \left(l_p-n\right) \phi_i} \hat{a}_i^{\dagger}\of{\vec{q}_i}. \label{eq:defschmidtopb}
                \end{align}
            \end{subequations}
            Note that since we work in polar coordinates,
            the area element
            is given by $\dd^2q_j=q_j\dd q_j\dd\phi_j$ for $j=s,i$.
            \par From the expressions of the Schmidt operators one
            can see that the creation operator
            $\hat{A}_{mn}^{\dagger}$ creates a
            spatially multimode signal photon over the plane-wave basis with
            OAM $n$, while the creation
            operator $\hat{B}_{mn}^{\dagger}$ creates an idler photon with
            the complementary OAM of
            $l_p-n$. Together, these two operators always create
            two photons in their corresponding Schmidt modes
            while conserving the OAM of the pump photons:
            $n+\left(l_p-n\right)=l_p$.

        \subsection{Symmetries and commutation relations}\label{sec:symms_comm_rels}
            For several important systems, the function $g_z$ containing the spatial
            $z$-axis integral
            have the symmetry property:
            \begin{subequations}
                \label{eq:phi_symm_props_gz}
                \begin{align}
                    g_z\of{q_s,q_i,\phi_s-\phi_i} &= g_z\of{q_s,q_i,\phi_i-\phi_s},
                    \label{eq:phi_symm_gz}
                \end{align}
                which also always holds for the function
                $\xi$ defined in \Cref{eq:def_beta}:
                \begin{align}
                    \xi\of{q_s,q_i,\phi_s-\phi_i} &= \xi\of{q_s,q_i,\phi_i-\phi_s}.
                    \label{eq:phi_symm_beta}
                \end{align}
            \end{subequations}
            In other words, these two functions are even in the
            angle difference ${\phi_s-\phi_i}$. Additionally,
            these functions are also ${2\pi}$-periodic in the angle
            difference.
            \par In this work, we will restrict ourselves to analyzing
            PDC from single crystals and from two-crystal setups [\textsuoo{}
            interferometers].
            Furthermore, we will only consider
            degenerate type-I PDC, where
            the signal and idler photons have
            the same frequency and polarization.
            From the concrete expressions of $g_z$ for these systems,
            see \Cref{eq:gz_one_crystal,eq:gz_twocystals}
            below\footnote{Note that the symmetry properties of $g_z$
            shown in \Cref{eq:phi_symm_gz,eq:qsi_symm_gz}            
            also hold for the exact
            expressions without the Fresnel approximation, see
            \Cref{eq:phase_mismatch_def,eq:fresnell} and the
            surrounding text.
            Additionally, note that
            $k_{p,z}\of{\vec{q}_s,\vec{q}_i}=
            \sqrt{\abs{\vec{k}_p}^2-\xi^2\of{q_s,q_i,\phi_s-\phi_i}}$,
            where $\abs{\vec{k}_p}$ is a constant depending on the pump
            frequency and refractive index.},
            it can be directly seen that the symmetry property in
            \Cref{eq:phi_symm_gz} holds. Additionally,
            this property also holds for the TPA derived
            in \Cref{sec:angular_displacement} since
            the dependencies on $\vec{q}_s$ and $\vec{q}_i$
            can be reduced to those of a single crystal.
            \par Since we are working in the frequency-degenerate regime,
            the refractive indices for the signal and idler beam
            are the same. As a consequence,  for the fixed phase difference $\phi$, the function $g_z$ is invariant under the exchange of the signal and idler
            variables, which also (always) holds for $\xi$:
            \begin{subequations}
                \label{eq:qsi_symm_props}
                \begin{align}
                    g_z\of{q_s,q_i,\phi} &= g_z\of{q_i,q_s,\phi},
                    \label{eq:qsi_symm_gz}\\
                    \xi\of{q_s,q_i,\phi} &= \xi\of{q_i,q_s,\phi}.
                    \label{eq:qsi_symm_beta}
                \end{align}
            \end{subequations}
            Note that these properties do not apply to $R$
            defined in \Cref{eq:def_R} due to the presence of
            the factor $W$, as defined in \Cref{eq:def_W_term}.
            However, for this term,
            \begin{align}\label{eq:W_lp_symmetry}
                \E^{\I l_p \left(\phi_s-\phi_i\right)} W\of{q_s,q_i,\phi_s-\phi_i}
                &= W\of{q_i,q_s,\phi_i-\phi_s}.
            \end{align}
            Combined,
            \Cref{eq:phi_symm_props_gz,eq:qsi_symm_props,eq:def_W_term,eq:W_lp_symmetry}
            imply
            \begin{align}\label{eq:R_lp_property}
                \E^{\I l_p \left(\phi_s-\phi_i\right)} R\of{q_s,q_i,\phi_s-\phi_i}
                &= R\of{q_i,q_s,\phi_i-\phi_s}.
            \end{align}
            Applying this
            to the inverse transform of the Fourier decomposition given
            in \Cref{eq:def_R_decomp} reveals:
            \begin{align}\label{eq:chi_symm_qsi}
                \chi_{l_p-n}\of{q_s,q_i} &= \chi_n\of{q_i,q_s}.
            \end{align}
            Due to this relation, the Schmidt decomposition components have the following
            helpful properties:
            \begin{subequations}
                \begin{align}
                    \lambda_{m,l_p-n} &= \lambda_{mn}, \label{eq:lp_rel_lambda} \\
                    u_{m,l_p-n} &= v_{mn}, \label{eq:lp_rel_uv}\\
                    v_{m,l_p-n} &= u_{mn}, \label{eq:lp_rel_vu}%
                \end{align}%
            \end{subequations}%
            which can also be applied to the Schmidt operators.
            Due to
            the fact that we are working in the degenerate
            regime, we obtain:
            \begin{subequations}
                \label{eq:schmidt_operators_symmetry_relation}
                \begin{align}
                    \hat{A}_{m,l_p-n}^{\dagger} &= \hat{B}_{mn}^{\dagger}, 
                        \label{eq:schmidt_operators_symmetry_relation_A} \\
                    \hat{B}_{m,l_p-n} &= \hat{A}_{mn}.
                        \label{eq:schmidt_operators_symmetry_relation_B}
                \end{align}
            \end{subequations}
            \par Furthermore, note that the full TPA is invariant under the
            $\vec{q}_s\leftrightarrow \vec{q}_i$ exchange:
            \begin{align}
                F\of{q_s,q_i,\phi_s,\phi_i} = F\of{q_i,q_s,\phi_i,\phi_s},
                \label{eq:tpa_si_symmetry}
            \end{align}
            which can be seen directly from \Cref{eq:F_R_exp,eq:R_lp_property}
            and reflects the degeneracy of the PDC process.
            
            \par Using the definition of the Schmidt operators
            [\Cref{eq:defschmidtopa,eq:defschmidtopb}]
            it can be seen that the commutation relation of the Schmidt operators 
            $\hat{A}_{mn}$ and $\hat{B}_{mn}$ with 
            their respective hermitian conjugates are:
            \begin{subequations}%
                \begin{align}%
                    \comm{\hat{A}_{mn}}{\hat{A}_{kl}^{\dagger}} &= 
                        \delta_{mk} \delta_{nl}, \label{eq:comm_AmnAkl} \\
                     \comm{\hat{B}_{mn}}{\hat{B}_{kl}^{\dagger}} &= \delta_{mk} \delta_{nl}.
                        \label{eq:comm_BmnBkl}
                \end{align}
                These coincide with the relations already
                found for the case $l_p=0$ in \refcite{PRA91}.
                From the connections between the Schmidt operators
                given in \Cref{eq:schmidt_operators_symmetry_relation_A,%
                eq:schmidt_operators_symmetry_relation_B} it is immediately
                clear that the commutation relation between the signal and idler
                Schmidt operators is given by
                \begin{align}
                    \comm{\hat{A}_{mn}}{\hat{B}_{kl}^{\dagger}} &= \delta_{mk} \delta_{n,l_p-l},
                    \label{eq:comm_ab}
                \end{align}
            \end{subequations}
            which reduces to the result found in \refcite{PRA91}
            for $l_p=0$, as expected.
            \par It should be emphasized that the
            labeling of ``signal'' and ``idler''
            photons is now, due to the degeneracy of the process
            under consideration, only meaningful in the sense that
            the photons of each signal-idler pair
            carry complementary orbital angular momenta
            with respect to the pump.
            This can also be seen from the
            symmetry properties given in \Cref{eq:chi_symm_qsi,eq:lp_rel_lambda,eq:lp_rel_uv,eq:lp_rel_vu,eq:schmidt_operators_symmetry_relation}:
            Swapping the OAM index $n$ with its
            complementary value $l_p-n$ is equivalent to switching between
            the signal and idler subsystems.
            \par If we had assumed that the signal
            and idler plane-wave
            operators describe distinguishable
            photons, for example due to different polarizations
            or frequencies, $A_{mn}$ and $B_{kl}^{\dagger}$ would commute:
            $\comm{A_{mn}}{B_{kl}^{\dagger}}=0$.
            However, since this is not the case in our consideration, the two Schmidt-mode
            operators commute only when they differ in at least one of the two indices.

        \subsection{Equations of motion and intensity spectrum}
            Using the commutation relations for the Schmidt operators
            [\Cref{eq:comm_AmnAkl,eq:comm_BmnBkl}], one can
            obtain their equations of motion
            in the Heisenberg picture:
            \begin{subequations}
                \begin{alignat}{3}
                    \frac{\dd \hat{A}_{mn}}{\dd t} &= 
                        \Gamma\sqrt{\lambda_{mn}}\left(\hat{A}_{m,l_p-n}^{\dagger}+\hat{B}_{mn}^{\dagger}\right) &&=2\Gamma\sqrt{\lambda_{mn}}\hat{B}_{mn}^{\dagger},
                        \label{eq:schmidt_op_diffeq_AB} \\
                    \frac{\dd \hat{B}_{mn}^{\dagger}}{\dd t} &= 
                        \Gamma\sqrt{\lambda_{mn}}\left(\hat{A}_{mn}+\hat{B}_{m,l_p-n}\right) &&=2\Gamma\sqrt{\lambda_{mn}}\hat{A}_{mn},
                \end{alignat}
            \end{subequations}
            where the relationships between the Schmidt operators given in
            \Cref{eq:schmidt_operators_symmetry_relation_A,eq:schmidt_operators_symmetry_relation_B}
            have been used in the second step.
            \par The solutions to this set of differential equations
            are the well-known Bogoliubov transforms~\cite{PRA91}:
            \begin{subequations}
                \begin{align}
                    \hat{A}_{mn}^{\mathrm{out}} &= \hat{A}_{mn}^{\mathrm{in}}
                        \cosh\of{G\sqrt{\lambda_{mn}}}
                        + \left(\hat{B}_{mn}^{\mathrm{in}}\right)^{\dagger}
                        \sinh\of{G\sqrt{\lambda_{mn}}}, \label{eq:bogoliubov_Aout} \\
                    \hat{B}_{mn}^{\mathrm{out}} &= \hat{B}_{mn}^{\mathrm{in}}   
                        \cosh\of{G\sqrt{\lambda_{mn}}}
                        + \left(\hat{A}_{mn}^{\mathrm{in}}\right)^{\dagger} 
                        \sinh\of{G\sqrt{\lambda_{mn}}},
                \end{align}
            \end{subequations}
            where $G=\int\!\dd t\, 2\Gamma$ is the theoretical
            parametric gain and the integral is taken over
            the interaction time. The operators denoted with
            the~\textsuperscript{in} superscript are the input
            Schmidt operators at the beginning of the interaction
            (at the input of the PDC section),
            while operators on the right hand side, denoted
            with~\textsuperscript{out}, are the output operators
            at the end of the interaction (at the output of the PDC
            section).
            \par The commutators of the plane-wave
            annihilation operator and
            the Schmidt operators are given by
            \begin{subequations}
                \begin{align}
                    \comm{\hat{a}\of{\vec{q}}}{\hat{A}_{mn}^{\dagger}} &= 
                        \frac{1}{\sqrt{2\pi}}\frac{u_{mn}\of{q}}{\sqrt{q}}\E^{-\I n \phi},
                        \label{eq:comm_a_A}\\
                    \comm{\hat{a}\of{\vec{q}}}{\hat{B}_{mn}^{\dagger}} &= 
                        \frac{1}{\sqrt{2\pi}}\frac{v_{mn}\of{q}}{\sqrt{q}}\E^{-\I \left(l_p-n\right) \phi}.
                        \label{eq:comm_a_B}
                \end{align}
            \end{subequations}
            Note that these hold regardless of whether
            $\hat{a}\of{\vec{q}}$ belongs to the signal or idler
            subsystem due to the degeneracy
            condition
            mentioned in \Cref{sec:symms_comm_rels},
            which is also why the index $j$ referring to either
            the signal or idler subsystem was dropped
            [here and in the following, $\vec{q}\equiv\left(q,\phi\right)$].
            If the signal and idler
            plane-wave operators were
            describing distinguishable photons, the
            commutators given in
            \Cref{eq:comm_a_A,eq:comm_a_B}
            would vanish if the plane-wave annihilation operator
            corresponds to the opposite subsystem as the
            Schmidt operator.
            \par Using these commutation relations, one can obtain
            the equations of motion
            for the plane-wave annihilation operator using the
            Hamiltonian given in \Cref{eq:Hamiltonian_diagonalized}:
            \begin{align}\label{eq:diffeq_pw}
                \frac{\dd \hat{a}\of{\vec{q}}}{\dd t} &= \frac{2\Gamma}{\sqrt{2\pi}}
                \sum_{m,n} 
                    \sqrt{\lambda_{mn}}\frac{u_{mn}\of{q}}{\sqrt{q}}\E^{-\I n \phi} \hat{B}_{mn}^{\dagger},
            \end{align}
            where we have shifted a
            summation index and used 
            \Cref{eq:lp_rel_lambda,eq:lp_rel_uv,eq:lp_rel_vu,%
            eq:schmidt_operators_symmetry_relation_A}
            so that only $B_{mn}^{\dagger}$ remains in the result.
            \par By using \Cref{eq:schmidt_op_diffeq_AB,eq:bogoliubov_Aout},
            the solution to the differential equation (\ref{eq:diffeq_pw})
            can be found:
            \begin{widetext}
                \begin{align}
                    \begin{split}
                        \hat{a}^{\mathrm{out}}\of{\vec{q}} &= \hat{a}^{\mathrm{in}}\of{\vec{q}} +
                        \frac{1}{\sqrt{2\pi}}\sum_{m,n} \frac{u_{mn}\of{q}}{\sqrt{q}}
                        \E^{-\I n \phi} \left\lbrace\hat{A}_{mn}^{\mathrm{in}}\left[\cosh\of{G\sqrt{\lambda_{mn}}}-1\right]+\left(\hat{B}_{mn}^{\mathrm{in}}\right)^{\dagger}\sinh\of{G\sqrt{\lambda_{mn}}}\right\rbrace,
                    \end{split}
                    \label{eq:in_out_plane_wave}
                \end{align}
            \end{widetext}
            where the plane-wave operators with
            the~\textsuperscript{in} and~\textsuperscript{out}
            superscripts
            are the input and output operators of the PDC section,
            respectively, analogously to the Schmidt operators with
            the same superscripts. Note that the input Schmidt
            mode operators depend on the input plane-wave operators.
            \par Using the definition of the plane-wave
            photon number operator,
            $\hat{N}\of{\vec{q}}={\left[\hat{a}^{\mathrm{out}}\of{\vec{q}}\right]^{\dagger}
            \hat{a}^{\mathrm{out}}\of{\vec{q}}}$,
            it is easy to see that 
            the intensity distribution
            (mean photon number distribution) is given by:
            \begin{align}
                \langle \hat{N}\of{\vec{q}}\rangle &=\frac{1}{2\pi}\sum_{m,n} \frac{\abs{u_{mn}\of{q}}^2}{q}\Lambda_{mn},
                \label{eq:mean_photon_number}
            \end{align}
            with the \textit{high-gain eigenvalues}
            \begin{align}
                \Lambda_{mn} &= \sinh^2\of{G\sqrt{\lambda_{mn}}}.
                \label{eq:hgeigs}
            \end{align}
            This expression for the intensity distribution
            does not depend on the azimuthal angle~$\phi$
            (cylindrical symmetry) and is identical to the one found in 
            \refcite{PRA91} for the Gaussian pump. As was
            also discussed therein, it is possible to define
            weights $\Lambda_{mn}'$ which describe the contribution
            to the intensity spectrum
            of each Schmidt mode depending on the gain:
            \begin{align}
                \Lambda_{mn}' = 
                    \frac{\Lambda_{mn}}{\sum_{m,n}\Lambda_{mn}}.
                    \label{eq:renormal_eigenvalues}%
            \end{align}
            The $\Lambda_{mn}'$ can be understood as the new Schmidt
            eigenvalues, re-normalized for the high-gain
            regime~\cite{PRA91}. At low-gain,
            $G\ll1$, $\Lambda'_{mn} \approx \lambda_{mn}$.

    \section{SCHMIDT MODE STRUCTURE}\label{sec:application_smode_structure}
        First, we start the analysis by taking a look at the
        modal structure of the PDC radiation generated
        from a Laguerre-Gaussian pump beam.
        To this end, we consider systems of a single BBO
        crystal and two BBO crystals with an air gap
        in-between [\textsuoo{} interferometer].
        Each crystal has a length of $L=\SI{2}{\milli\meter}$
        and the system is pumped by the Laguerre-Gaussian beam with
        a width of $w_0=\SI{50}{\micro\meter}$ and a
        wavelength of $\SI{354.7}{\nano\meter}$.
        For the presented results, we will restrict ourselves to
        $l_p\geq 0$. However, as is shown in \Cref{sec:angular_displacement},
        the results for $l_p<0$ would not differ qualitatively
        from the ones presented here.
        In the following, the intensity profiles are plotted over the
        external angle $\Theta_s\approx q_s/k_s^{\mathrm{air}}$,
        where $k_s^{\mathrm{air}}$ is the 
        modulus of the wave vector of the signal photons in air.
        This approximation is valid for small angles withing
        the paraxial approximation.
        \par In order to determine the connection between the
        \textit{experimental} gain $G_{\mathrm{exp}}$ and the theoretical gain
        parameter $G$, we fit~\cite{PRR5,PRR2,Spasibko:12}
        the collinear output
        intensity $\langle \hat{N}\of{0}\rangle\dd q \dd \phi$
        of a single crystal with a function of the form
        $y\of{\gamma} = B\sinh^2\of{\tilde{A}\gamma}$, where
        $\tilde{A}$ and $B$ are the fitting parameters and 
        $G=2\mathcal{D}\gamma$. The factorization of $G$
        is performed to simplify the 
        numerical fitting procedure. Here,
        $\mathcal{D}$ is a suitable scaling constant;
        possible choices are
        for example $\mathcal{D}=1$ or $\mathcal{D}=1/C$.
        The experimental gain
        $G_{\mathrm{exp}}$ is then given by
        $G_{\mathrm{exp}} = \tilde{A}\gamma = ACG/2$, 
        where the additionally defined constant
        $A=\tilde{A}/\!\left(C\mathcal{D}\right)$
         does not depend on the choice of $\mathcal{D}$.
        Instead of $\tilde{A}$, we will only refer to $A$
        as the ``fit constant'' in the
        following since,
        once $A$ is known, the theoretical
        gain parameter $G$ can be calculated given the
        experimental gain $G_{\mathrm{exp}}$
        \par In the context of this
        work, we mainly use the experimental low-gain value of
        $G_{\mathrm{exp}} = 0.01$ and the high-gain value of 
        $G_{\mathrm{exp}} = 4$, unless specified otherwise.
        The values for $A$ are listed in
        \Cref{tab:G_and_A_of_Gamma}. Due to the multimode
        structure of the PDC radiation, this constant may
        differ when the fit is performed for the low- or
        the high-gain regime~\cite{PRR5},
        which is why the table lists differing
        values for $A$ around $G_{\mathrm{exp}}=0.01$
        and $G_{\mathrm{exp}}=4$.
        \begin{table}
            \renewcommand{\arraystretch}{1.3}
            \begin{center}
                \caption{%
                    Fitting parameter $A$ for the selected orbital $l_p$ and radial $m_p$ numbers of the pump. \label{tab:G_and_A_of_Gamma}%
                }%
                \begin{tabularx}{\linewidth}{YYYY}
                    \hline\hline
                    \multirow{2}{*}{$l_p$} & \multirow{2}{*}{$m_p$} & \multicolumn{2}{c}{Fitting parameter $A$} \\
                    \cline{3-4}
                    \phantom{} & \phantom{} & $G_{\mathrm{exp}}=0.01$ & $G_{\mathrm{exp}}=4$ \\
                    \hline
                    0 & 0 & $14.2$ & $17.2$ \\
                    0 & 7 & $4.40$ & $9.27$ \\
                    0 & 18 & $2.58$ & $5.75$ \\
                    7 & 0 & $6.64\times 10^{-9}$ & $7.22\times 10^{-9}$ \\
                    7 & 7 & $1.99\times 10^{-7}$ & $2.56\times 10^{-7}$ \\
                    7 & 18 & $1.56\times 10^{-6}$ & $2.08\times 10^{-6}$ \\
                    18 & 0 & $5.44\times 10^{-20}$ & $5.87\times 10^{-20}$ \\
                    18 & 7 & $2.03\times 10^{-17}$ & $ 2.35\times 10^{-17}$ \\
                    18 & 18 & $1.95\times 10^{-15}$ & $2.35\times 10^{-15}$ \\
                    \hline\hline
                \end{tabularx}
            \end{center}
        \end{table}

        \subsection{Single-crystal setup}\label{sec:modes_struct_single_cryst}
            For a single crystal, the function $g_z$ containing
            the spatial integral over the $z$-axis
            reads~\cite{PRA91,Miatto2012}:
            \begin{align}
                g_{z}\of{q_s, q_i, \phi_s, \phi_i} &=
                    L\sinc\of{L\frac{\left(\vec{q}_s-\vec{q}_i\right)^2}{4k_p}}
                    \E^{\I L\frac{\left(\vec{q}_s-\vec{q}_i\right)^2}{4k_p}},
                    \label{eq:gz_one_crystal}%
            \end{align}
            where $\sinc\of{x}=\sin\of{x}\!/x$, $L$ is the length of the PDC section,
            $k_p=\abs{\vec{k}_p}$ is the modulus of the pump wave vector
            and
            $\left(\vec{q}_s-\vec{q}_i\right)^2 = q_s^2+q_i^2-2q_sq_i\cos\of{\phi_s-\phi_i}$. Note that
            here, the Fresnel approximation (paraxial approximation)
            has been used to express
            the collinear phase mismatch
            \begin{align}\label{eq:phase_mismatch_def}
                \Delta k\of{\vec{q}_s,\vec{q}_i}=k_{p,z}\of{\vec{q}_s,\vec{q}_i}
                    -k_{s,z}\of{\abs{\vec{q}_s}}-k_{i,z}\of{\abs{\vec{q}_i}}
            \end{align}
            in an
            analytically more convenient form by expanding the
            $z$-components of the wave vectors to the second order in
            terms of the transverse
            wave vectors $\abs{\vec{q}_j}\ll\abs{\vec{k}_{j}}$~\cite{Karan2020,PhysRevA.106.063711}:
            \begin{align}\label{eq:fresnell}
                k_{j,z} = \sqrt{\abs{\vec{k}_{j}}^2-\abs{\vec{q}_j}^2}
                    \approx \abs{\vec{k}_{j}} \left(1-\frac{\abs{\vec{q}_j}^2}{2\abs{\vec{k}_{j}}^2}\right),
            \end{align}
            where $j=p,s,i$ and $\vec{q}_p=\vec{q}_s+\vec{q}_i$.
            \par
\begin{figure*}[!htbp]%
                \centering%
                \subfloat{\includegraphics[width=0.5\linewidth]{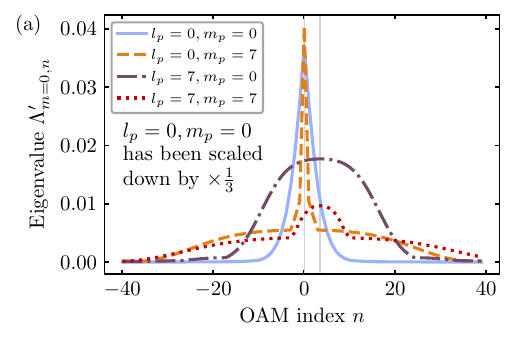}%
                    \label{fig:Intensity_mulitplot_cut_low_high_7_0.0a}}%
                \subfloat{\includegraphics[width=0.5\linewidth]{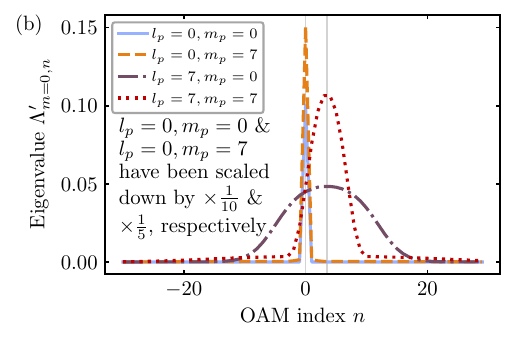}%
                    \label{fig:Intensity_mulitplot_cut_low_high_7_0.0b}}\\%
                \subfloat{\includegraphics[width=0.5\linewidth]{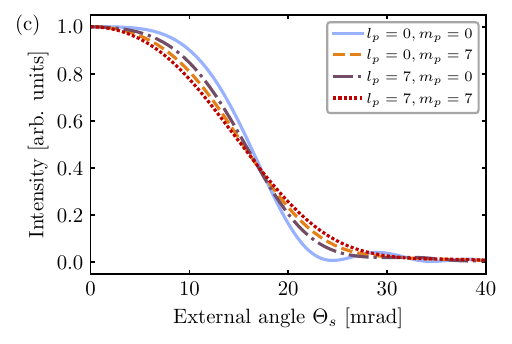}%
                    \label{fig:Intensity_mulitplot_cut_low_high_7_0.0c}}%
                \subfloat{\includegraphics[width=0.5\linewidth]{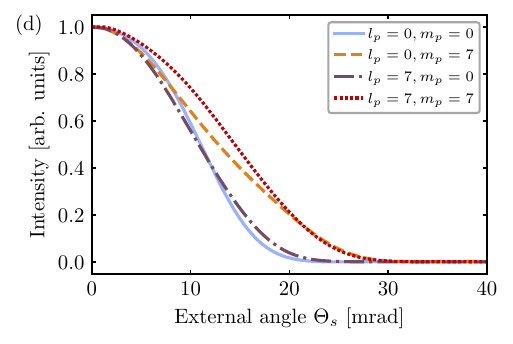}%
                    \label{fig:Intensity_mulitplot_cut_low_high_7_0.0d}}\\%
                \caption{%
                    Cuts of the normalized modal weights $\Lambda'_{m=0,n}$ (${\sum_{m,n} \Lambda'_{mn}}=1$) for the
                    (a)~low-gain and (b)~high-gain regime for a single
                    BBO crystal of length $L=\SI{2}{\milli\meter}$.
                    The thin gray lines indicate $n=0$ and $n=3.5$,
                    which are the symmetry lines for $l_p=0$ and $l_p=7$,
                    respectively.
                    The plots (a)~and (b)~clearly show an increase in the width 
                    in the orbital direction (along $n$) when increasing
                    the radial
                    $m_p$ and orbital $l_p$ numbers of the pump.
                    Normalized
                    intensity spectra
                    for the (c)~low-gain and (d)~high-gain
                    regime.  For $l_p = 0$ and $m_p = 0$ the intensity distribution
                    has a $\sinc^2$-shape. The intensity profile becomes more
                    Gaussian as $l_p$ and $m_p$ increases.
                }
                \label{fig:Intensity_mulitplot_cut_low_high_7_0.0}
            \end{figure*}

 Performing the Schmidt decomposition of the Fourier coefficients
            $\chi_n\of{q_s, q_i}$ [\Cref{eq:chi_n_sdecomp}] of the two-photon amplitude
            as defined in \Cref{eq:small_f_gxy_gz} with $g_{xy}$ and $g_z$ defined
            in \Cref{eq:g_form_4,eq:gz_one_crystal}, respectively, we obtain
            the normalized eigenvalues $\lambda_{mn}$
            (so that ${\sum_{m,n} \lambda_{mn}}=1$).
            \begin{figure*}[!htpb]
                \centering%
                \includegraphics[width=\linewidth]{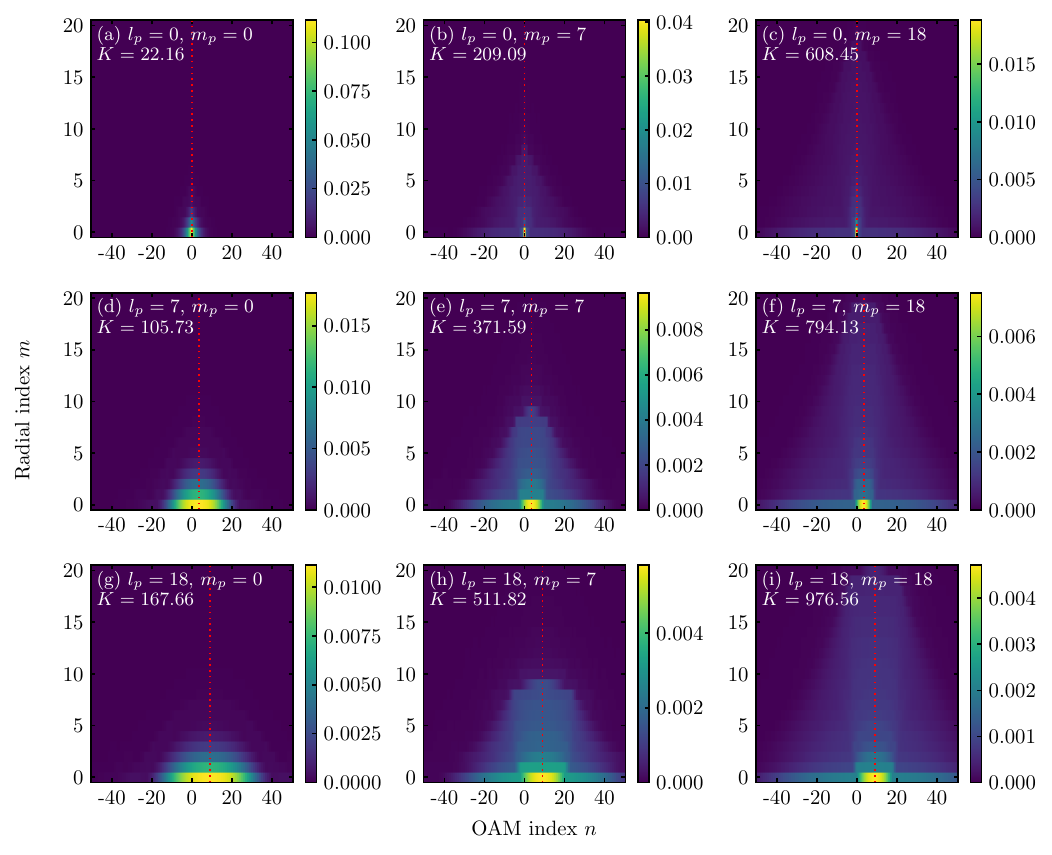}%
                \caption{%
                    Normalized modal weights $\Lambda'_{mn}$
                    (${\sum_{m,n} \Lambda'_{mn}}=1$) for the
                    low-gain regime ($G=0.01$), for a single
                    BBO crystal of length $L=\SI{2}{\milli\meter}$ 
                    pumped by a Laguerre-Gaussian pump with different combinations
                    of orbital $l_p$ and radial $m_p$ indices as mentioned in the plots.
                    The red dotted line indicates 
                    the symmetry axis of the eigenvalue distribution at
                    $n=l_p/2$ ($n=0$, $n=3.5$ and $n=9$,
                    for $l_p=0$, $l_p=7$ and $l_p=18$, respectively).
                    The graphs clearly show an increase in the Schmidt number $K$
                    and the width of the distribution
                    in the radial 
                    (along $m$)
                    and orbital (along $n$) directions 
                    with increasing radial
                    $m_p$ and orbital $l_p$ numbers of the pump,
                    respectively. The width $w_0=\SI{50}{\micro\meter}$ and
                    wavelength $\SI{354.7}{\nano\meter}$ of the pump are
                    fixed for all further plots.%
                }%
                \label{fig:2D_eigen_Multi_7_0.0}%
            \end{figure*}
            These eigenvalues determine the
            gain-dependent 
            weights $\Lambda_{mn}'$
            [see \Cref{eq:hgeigs,eq:renormal_eigenvalues}]
            of the corresponding Schmidt
            modes and are presented in \Cref{fig:2D_eigen_Multi_7_0.0} for
            different orbital $l_p$ and radial $m_p$ 
            numbers of the pump. It can be seen that the
            widths of the distribution in the orbital~$n$ and
            radial~$m$ direction become
            larger with increasing 
            orbital $l_p$ and radial $m_p$ numbers of the pump,
            respectively
            [see also
            \Cref{fig:Intensity_mulitplot_cut_low_high_7_0.0a}].
            Therefore, by varying the parameters $l_p$ and $m_p$
            of the pump beam, one can significantly
            increase the number of populated OAM and radial modes.
            To further quantify this, we also provide
            the Schmidt number (effective mode number)~\cite{PRA91,PRR5}
            \begin{align}\label{eq:schmidt_number}
                K &= \left[\sum_{m,n} \left(\Lambda_{mn}'\right)^2\right]^{-1}
            \end{align}
            in each 2D plot of eigenvalue distributions. As expected,
            the Schmidt number increases strongly
            as $l_p$ and $m_p$ are increased [see \Cref{fig:2D_eigen_Multi_7_0.0}].
            \par Furthermore, it is important
            to point out that the eigenvalue distribution is symmetric
            with respect to the axis $n=l_p/2$,
            see \Cref{fig:2D_eigen_Multi_7_0.0,fig:Intensity_mulitplot_cut_low_high_7_0.0a,fig:Intensity_mulitplot_cut_low_high_7_0.0b}.
            This means that for odd values of $l_p$,
            the distribution must be
            flat around the point $l_p/2$, since the eigenvalues
            to the left ($n=\floor{l_p/2}$) and to the right
            ($n=\ceil{l_p/2}$)
            of this point must be
            identical, see
            \Cref{fig:Intensity_mulitplot_cut_low_high_7_0.0a}.           
            \par For the low-gain regime, the shapes of the
            intensity spectra for different orbital and radial numbers are shown in
            \Cref{fig:Intensity_mulitplot_cut_low_high_7_0.0c}.
            They are formed by
            several Schmidt modes with the weights
            $\lambda_{mn}$ according to
            \Cref{eq:mean_photon_number,eq:renormal_eigenvalues}.
            For $l_p = 0$ and $m_p = 0$ the intensity distribution
            has a $\sinc^2$-shape. The intensity profile becomes more
            Gaussian with increasing $m_p$, since the
            contribution of the fundamental mode (which
            is Gaussian) becomes higher compared to other
            modes, see orange curve in
            \Cref{fig:Intensity_mulitplot_cut_low_high_7_0.0a},
            see also~\refcite{PRA91}.
            For nonzero $l_p$, even though
            the mode with the largest population
            is not Gaussian, the number of modes
            contributing to the total intensity is
            large enough, so the intensity profile still
            has a shape which is close to Gaussian.
            \par As we move to the high-gain
            regime, shown in \Cref{fig:Intensity_mulitplot_cut_low_high_7_0.0b},
            the eigenvalues are redistributed compared to the low-gain
            regime~\cite{PRA91}
            [see \Cref{fig:Intensity_mulitplot_cut_low_high_7_0.0a}]
            according to \Cref{eq:renormal_eigenvalues} and
            fewer Schmidt modes contribute to the shape of the intensity
            spectrum. By reducing the number of contributing
            modes the eigenvalue
            distributions and intensity profiles get narrower.
            Moreover, it can be seen that as $l_p$ and $m_p$ increase,
            the intensity distribution becomes wider due to
            larger mode widths for higher $l_p$ and $m_p$.

        \subsection{Two-crystal setup}\label{sec:two_crystal_setup}
            An \textsuoo{} interferometer consists of two PDC sections
            with a linear medium in-between. This linear medium
            imprints a (possibly spatially varying)
            phase to the pump, signal and idler
            radiation~\cite{Frascella:19,PRR2,PRR5,Chekhova:16}.
            In this section, we consider a phase which is simply given by
            the air gap of length $d$ between the two crystals.
            The function $g_z$ for the two-crystal setup can then
            be written in the form~\cite{PRA91}:
            \begin{widetext}
                \begin{align}
                    \begin{split}
                        g_{z}\of{q_s, q_i, \phi_s, \phi_i} &=
                        2L\sinc\of{L\frac{(\vec{q}_s-\vec{q}_i)^2}{4k_p}}
                        \cos\ofb{L\frac{(\vec{q}_s-\vec{q}_i)^2}{4k_p}+ \frac{\Delta n k_s d}{n_s}+ n_s d \frac{(\vec{q}_s-\vec{q}_i)^2}{4k_p}} \\
                        &\qquad\times \exp\ofb{\I\left(L\frac{(\vec{q_s}-\vec{q}_i)^2}{2k_p}+ \frac{\Delta n k_s d}{n_s}+ n_s d \frac{(\vec{q}_s-\vec{q}_i)^2}{4k_p}\right)},
                        \label{eq:gz_twocystals}%
                    \end{split}
                \end{align}
            \end{widetext}
            where
            $\Delta n = n^{\mathrm{air}}_p-\frac{1}{2}(n^{\mathrm{air}}_s+n^{\mathrm{air}}_i)=n^{\mathrm{air}}_p-n^{\mathrm{air}}_s$
            and $n^{\mathrm{air}}_j$ ($j=p,s,i$) are the
            refractive indices of the pump,
            signal and idler radiation in the air gap
            between the two crystals, while
            $n_s$ is the refractive index of
            the signal beam inside the crystal.
            Here, the Fresnel approximation has again been
            applied and perfect phase
            matching is assumed for the collinear direction
            (where $\vec{q}_s=\vec{q}_i=\nullvec$), just as in 
            \Cref{sec:modes_struct_single_cryst} for the single crystal.
            \begin{figure}%
                \centering%
                \includegraphics[width=\linewidth]{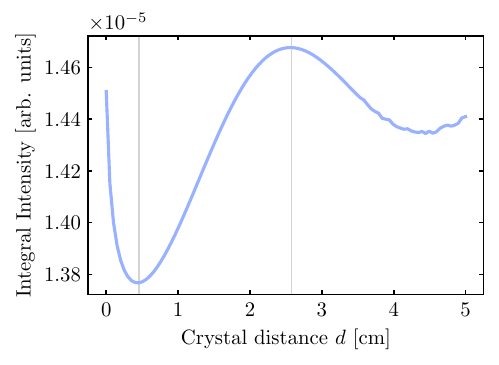}%
                \caption{%
                    Integral intensity of the \textsuoo{}
                    interferometer output over the distance
                    between the two crystals.
                    The thin gray lines indicate the distances
                    for which the first
                    minimum at $d=\SI{0.42}{\centi\meter}$ and the first
                    maximum at $d=\SI{2.52}{\centi\meter}$ are observed.
                    Note that the position of the bright and dark
                    fringe seem to be independent of $l_p$ and $m_p$.
                    To verify this, we compared all 64
                    combinations of $l_p$ and $m_p$ from 0 to 7
                    and found that for different
                    $l_p$ and $m_p$ the integral
                    intensities show no significant differences
                    compared to each other and only differ
                    by a scaling factor.%
                }%
                \label{fig:Integral_int}%
            \end{figure}%
            \par Figure~\ref{fig:Integral_int} presents the integral
            intensity over the distance $d$ between the crystals that is obtained  by
            integrating the intensity distribution
            \Cref{eq:mean_photon_number} over the radial and azimuthal
            variables. Here, starting near
            $d=\SI{3.2}{\centi\meter}$, numerical noise becomes visible.
            This is caused by the interference in the second crystal
            which results in peaks in the intensity spectrum that 
            become denser as $d$ increases and are at some point no 
            longer resolved with the chosen numerical grid.
            Below we will only pay attention to the two special points
            of this dependence, namely, the first minimum
            ($d=\SI{0.42}{\centi\metre}$,
            \textit{dark fringe}) and the first maximum
            ($d=\SI{2.52}{\centi\metre}$,
            \textit{bright fringe}), which are shown by the
            thin gray lines in \Cref{fig:Integral_int}. 

            \subsubsection{Dark fringe}
\begin{figure*}
                    \centering
                    \includegraphics[width=\linewidth]{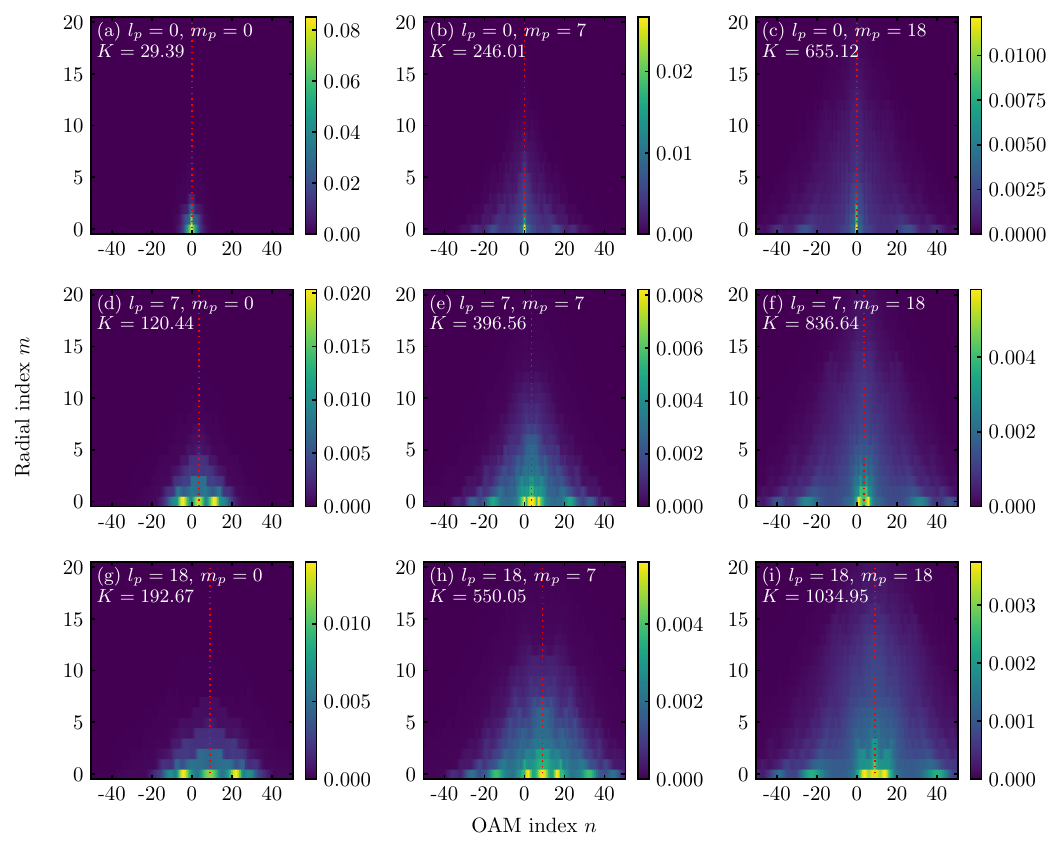}
                    \caption{%
                        Normalized modal weight spectra
                        $\Lambda'_{mn}$ (${\sum_{m,n} \Lambda'_{mn}}=1$)
                        for the low-gain regime ($G=0.01$),
                        for the two-crystal setup (two BBO crystals, each of length
                        $L=\SI{2}{\milli\metre}$) with an air gap of length
                        $d=\SI{0.42}{\centi\metre}$ (dark fringe),
                        pumped by a Laguerre-Gaussian
                        pump with different combinations of orbital $l_p$ and
                        radial $m_p$ indices as written in the plots.
                        The red dotted lines indicate the symmetry axis of the
                        eigenvalue distribution at $n=l_p/2$
                        ($n=0$, $n=3.5$ and $n=9$, for $l_p=0$, $l_p=7$
                        and $l_p=18$, respectively).
                        The graphs clearly show non-monotonic behavior
                        for the eigenvalue distributions
                        along the orbital direction $n$ and
                        an increase in the Schmidt number compared
                        to the single-crystal case (presented in
                        \Cref{fig:2D_eigen_Multi_7_0.0}).
                    }
                    \label{fig:2D_Eigenvaludis_mulitplot_18_0.5}
                \end{figure*}

                Figure~\ref{fig:2D_Eigenvaludis_mulitplot_18_0.5} shows
                the normalized eigenvalue distribution for the first
                minimum of the total intensity distribution
                ($d=\SI{0.42}{\centi\meter}$) for different orbital
                $l_p$ and radial $m_p$ numbers of the pump. 
                It can be seen that the eigenvalue distribution becomes
                broader and the Schmidt number $K$ is increased for
                an \textsuoo{} interferometer setup compared to the
                single-crystal setup for the same values of
                $l_p$ and $m_p$
                (compare \Cref{fig:2D_eigen_Multi_7_0.0,fig:2D_Eigenvaludis_mulitplot_18_0.5}).
                \par
\begin{figure*}%
                    \vspace*{-0.5em}%
                    \centering%
                    \subfloat{%
                        \includegraphics[width=0.5\linewidth]{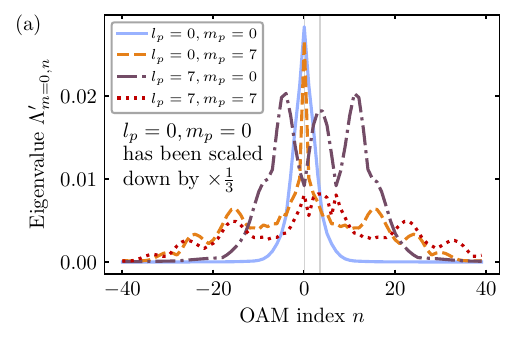}%
                        \label{fig:fig_Multiplot_cut_7_0.5_a}%
                    }%
                    \subfloat{%
                        \includegraphics[width=0.5\linewidth]{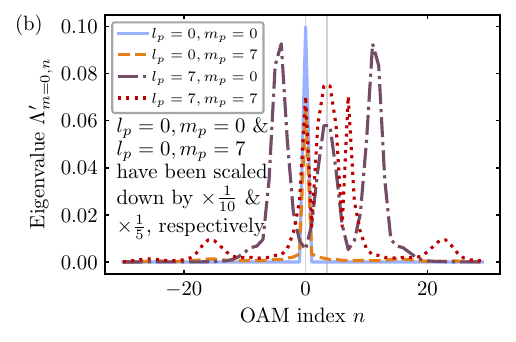}%
                        \label{fig:fig_Multiplot_cut_7_0.5_b}%
                    }\\[-1em]%
                    \subfloat{%
                        \includegraphics[width=0.5\linewidth]{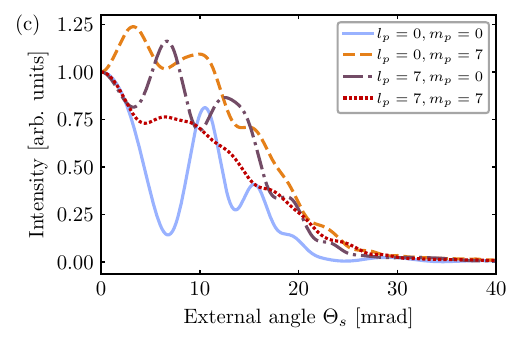}%
                        \label{fig:fig_Multiplot_cut_7_0.5_c}%
                    }%
                    \subfloat{%
                        \includegraphics[width=0.5\linewidth]{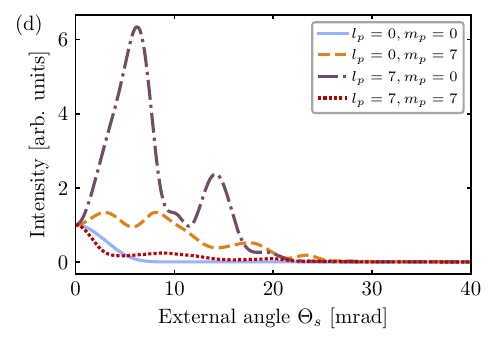}%
                        \label{fig:fig_Multiplot_cut_7_0.5_d}%
                    }%
                    \caption{%
                        Cuts of the normalized modal weights $\Lambda'_{m=0,n}$ 
                        (${\sum_{m,n} \Lambda'_{mn}}=1$) for the
                        (a)~low-gain and (b)~high-gain regime for two BBO crystals
                        of length $L=\SI{2}{\milli\meter}$,
                        an air gap between
                        the crystals of length
                        $d=\SI{0.42}{\centi\meter}$ (dark fringe)
                        and different orbital $l_p$ and radial $m_p$ numbers of 
                        the pump as written in the plots.
                        The thin gray lines indicate $n=0$ and $n=3.5$,
                        which are the symmetry lines for $l_p=0$ and $l_p=7$, 
                        respectively. For a non-Gaussian pump, a non-monotonic
                        behavior in the eigenvalue distribution 
                        can be observed. 
                        Normalized intensity spectra for the (c)~low-gain and
                        (d)~high-gain regime. Due to the non-trivial
                        behavior of the eigenvalue distribution,
                        the intensities have a complex non-Gaussian
                        shape at low and high gain.%
                    }
                    \label{fig:Low_High_different_distances_0.5_1.5}%
                \end{figure*}

 Furthermore, one can
                observe that, contrary to the single-crystal
                configuration, the normalized eigenvalue distributions
                in the dark-fringe case are characterized by
                non-monotonic behavior along the orbital direction $n$.
                This non-monotonic behavior is observed for nonzero
                orbital/radial pump numbers and is more clearly visible
                in the $\lambda_{m=0,n}$ cuts of the modal weight distribution
                presented in \Cref{fig:fig_Multiplot_cut_7_0.5_a}.   
                Furthermore, it can be seen that for nonzero orbital
                pump numbers, the orbital number
                of the PDC modes with the largest population
                is no longer $l_p/2$.
                For example, when $l_p=7$ and $m_p=0$, the 
                mode with OAM $n=11$ has the
                highest population. In the high-gain regime,
                a repopulation of the modes
                takes place~\cite{PRA91}, and the modes with
                the largest weights are
                strongly amplified, see \Cref{fig:fig_Multiplot_cut_7_0.5_b}.
                This means that in the dark fringe, 
                modes with an OAM higher than
                the pump OAM can be efficiently populated and
                filtered at high gain.
                \par These non-trivial eigenvalue profiles
                result in the non-Gaussian
                intensity distributions with a complex shape presented
                in \Cref{fig:fig_Multiplot_cut_7_0.5_c,fig:fig_Multiplot_cut_7_0.5_d}
                for the low- and high-gain regimes,
                respectively. Some Schmidt modes which contribute to
                the intensity distribution are presented
                in \Cref{fig:multiplot_modes} in \Cref{sec:modeprof2d}.
                A high population of modes with high-order OAM in the case of
                $l_p=7, m_p=0$ leads to a donut-shaped intensity profile
                with a local minimum in the center, which is most
                pronounced in the high-gain regime.
                \par As already mentioned above,
                the eigenvalues are symmetrically populated around the point 
                $l_p/2$, due to the fact that $\lambda_{m,l_p-n} = \lambda_{mn}$,
                see \Cref{eq:lp_rel_lambda}. 
                This means that if modes other than $l_p/2$ are most populated, then
                there is no single mode that uniquely determines the shape of the
                intensity distribution in the high-gain regime: There are at least
                two equally populated modes.
                \begin{figure*}%
                    \subfloat{%
                        \includegraphics[width=0.5\linewidth]{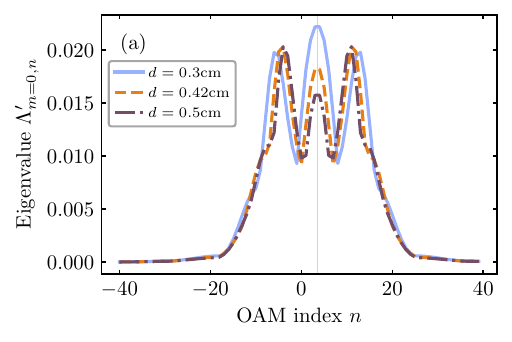}}%
                    \subfloat{%
                        \includegraphics[width=0.5\linewidth]{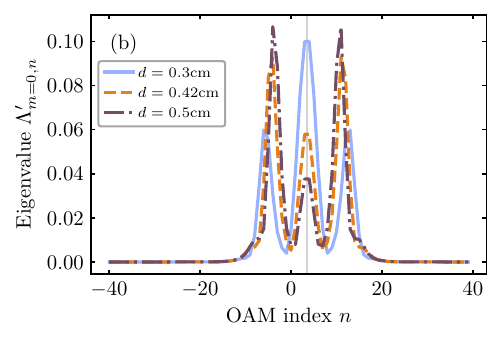}}%
                    \caption{%
                        Cuts of the normalized modal weights
                        $\Lambda'_{m=0,n}$ (${\sum_{m,n} \Lambda'_{mn}}=1$) for $l_p=7$, $m_p=0$ 
                        for the (a)~low-gain and (b)~high-gain regime
                        for different distances between the
                        two crystals of an \textsuoo{} interferometer
                        around the dark fringe ($d=\SI{0.42}{\centi\meter}$).
                        The thin gray lines indicate $n=l_p/2=3.5$.
                        For the distances
                        $d=\SI{0.42}{\centi\metre}$ and 
                        $d=\SI{0.5}{\centi\metre}$,
                        the most populated OAM mode appears at $n = 11$,
                        which is higher than the pump OAM $l_p = 7$.
                        According to \Cref{eq:renormal_eigenvalues},
                        the most populated OAM modes are strongly amplified in the high-gain
                        regime~\cite{PRA91}. See also
                        \Cref{fig:Low_High_cut_different_distances_0.5_1}
                        in \Cref{sec:modal_weights_app} which briefly
                        discusses the case $l_p=18$.
                    }%
                    \label{fig:Low_High_cut_different_distances_0.5}
                \end{figure*}%
                By changing the distance between the crystals one can strongly modify the
                eigenvalue distribution and the OAM of the most populated mode. For example, varying
                the distance around the dark fringe point, one can change the OAM of the most populated
                mode from $n=4$ for $d=\SI{0.3}{\centi\metre}$ to $n=11$ for $d=\SI{0.42}{\centi\metre}$
                and $d=\SI{0.5}{\centi\metre}$, see \Cref{fig:Low_High_cut_different_distances_0.5}.
                See also \Cref{fig:Low_High_cut_different_distances_0.5_1}
                in \Cref{sec:modal_weights_app} where the case $l_p=18$
                is discussed.

            \subsubsection{Bright fringe} 
\begin{figure*}
                    \centering
                    \includegraphics[width=\linewidth]{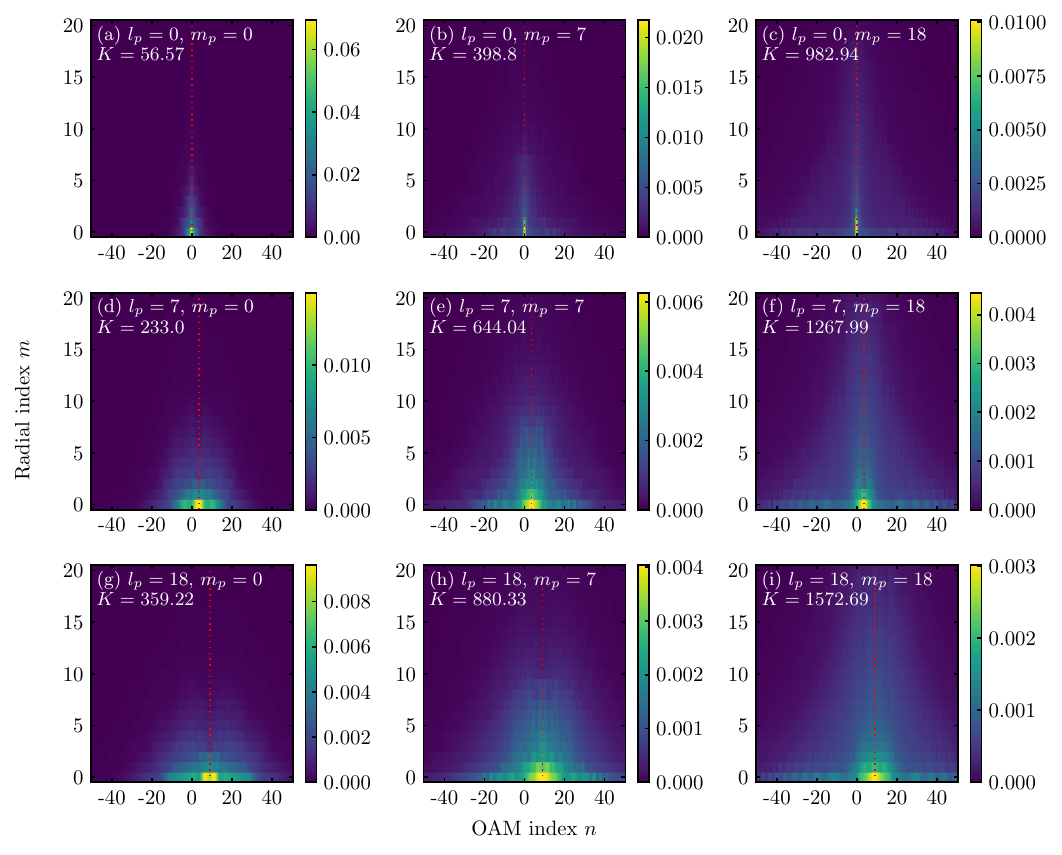}
                    \caption{%
                        Normalized modal weight spectra $\Lambda'_{mn}$
                        (${\sum_{m,n} \Lambda'_{mn}}=1$) for the
                        low-gain regime ($G=0.01$), for the
                        two-crystal setup (two BBO crystals, each of
                        length $L=\SI{2}{\milli\metre}$)
                        with an air gap of length 
                        $d=\SI{2.52}{\centi\metre}$ (bright fringe),
                        pumped by a Laguerre-Gaussian pump with different
                        combinations of orbital $l_p$ and radial $m_p$
                        indices as written in the plots. The red dotted
                        line indicates the symmetry axis of the eigenvalue
                        distribution at $n=l_p/2$ ($n=0$, $n=3.5$ and
                        $n=9$, for  $l_p=0$, $l_p=7$ and $l_p=18$, respectively). 
                    }
                    \label{fig:2D_Eigenvaludis_mulitplot_18_3.0}
                \end{figure*}

\begin{figure*}
                    \subfloat{%
                        \label{fig:Multiplot_cut_intensity_18_0.5a}%
                        \includegraphics[width=0.5\linewidth]{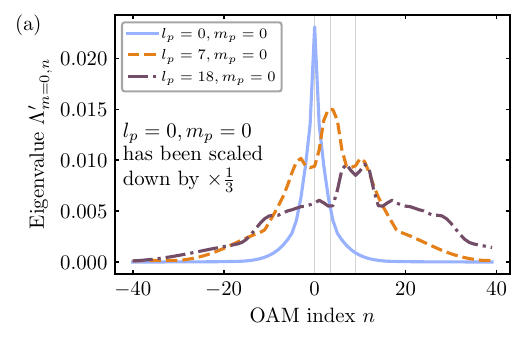}%
                    }%
                    \subfloat{%
                        \label{fig:Multiplot_cut_intensity_18_0.5b}
                        \includegraphics[width=0.5\linewidth]{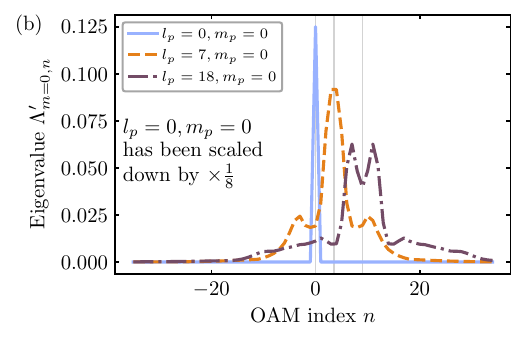}%
                    }\\[-1em]%
                    \caption{%
                        Cuts of the normalized modal weights $\Lambda'_{m=0,n}$ (${\sum_{m,n} \Lambda'_{mn}}=1$) for 
                        the (a)~low-gain and
                        (b)~high-gain regime for two BBO 
                        crystals of length $L=\SI{2}{\milli\meter}$, 
                        an air gap of length $d=\SI{2.52}{\centi\meter}$
                        (bright fringe)
                        between the crystals
                        and different orbital $l_p$ and radial $m_p$ numbers
                        of the pump as written in the plots. The thin gray
                        lines indicate $n=0$, $n=3.5$ and $n=9$,
                        which are the symmetry lines for $l_p=0$, $l_p=7$ and $l_p=18$, 
                        respectively. Increasing the pump
                        OAM $l_p$ changes the
                        order of the most populated OAM modes.
                    }
                    \label{fig:Multiplot_cut_intensity_18_0.5}
                \end{figure*}

                In the case of the bright fringe ($d=\SI{2.52}{\centi\meter}$),
                the non-monotonic behavior is less pronounced compared to the
                dark fringe, see \Cref{fig:2D_Eigenvaludis_mulitplot_18_3.0},
                which shows the normalized eigenvalue distributions 
                for the bright fringe for different
                orbital $l_p$ and radial $m_p$ numbers of the pump.
                However, the cuts of the modal weights $\Lambda_{m=0,n}$
                presented in \Cref{fig:Multiplot_cut_intensity_18_0.5}
                clearly demonstrate that this behavior also occurs
                here, while changing in the pump OAM
                modifies the OAM of the most populated mode
                [see \Cref{fig:Multiplot_cut_intensity_18_0.5a}].
                Similarly to the dark fringe, the high-gain regime allows us to
                filter these most populated modes, see
                \Cref{fig:Multiplot_cut_intensity_18_0.5b}.

        \subsection{Non-monotonic behavior}
            As was demonstrated in the previous section,
            configurations of two crystals with a nonzero distance
            between the crystals and non-Gaussian pumping can result
            in a non-monotonic structure of the eigenvalue distribution.
            This non-monotonic behavior is not observed for a single
            crystal pumped by a non-Gaussian pump and leads to configurations
            where the highest population of modes occurs
            at an OAM higher than the pump OAM.
            \par To investigate the non-monotonic behavior in more detail,
            we focus on the \textit{orbital eigenvalues}
            in the low-gain regime
            \begin{align}
                \varLambda_n \leqdef \sum_m \lambda_{mn},
            \end{align}
            which are the probabilities to
            find a Schmidt mode with OAM $n$ in the
            signal/idler field.
            The orbital eigenvalues can also be obtained by
            integrating the
            Fourier coefficients $\chi_n\of{q_s,q_i}$
            introduced in \Cref{eq:chi_n_sdecomp},
            namely, 
            \begin{align}
                \varLambda_n=\iint\!\dd q_s \dd q_i\, q_s q_i \abs{\chi_n\of{q_s,q_i}}^2.
                \label{eq:OAM_eigenvalues}
            \end{align}
            Since the non-monotonic behavior has been
            observed for any nonzero
            $l_p$ and any $m_p$, we set $m_p=0$ for simplicity, so that
            the Laguerre polynomials in
            the pump envelope
            [\Cref{eq:laguerre_gaussian_pump}]
            are equal to~$1$, regardless of the value of $l_p$.
            We also define the (complex) constant
            \begin{align}
                \bar{A} \leqdef 4CL\pi^2\left(-1\right)^{l_p}
                    \I^{|l_p|}\left(\frac{w_0}{2}\right)^{|l_p|+1},
            \end{align}
            which includes all prefactors of $R$ that do
            not depend on $q_s$, $q_i$, $\phi_s$ or $\phi_i$
            [compare \Cref{eq:def_R}].
            \par In order to allow for an analytical treatment,
            we use the
            Gaussian approximation
            $\sinc\of{b^2 x^2}\approx\exp\of{-\alpha^2 b^2 x^2}$
            for the
            $\sinc$-function appearing in $g_z$ in \Cref{eq:gz_twocystals},
            where ${\alpha=0.65}$ is an additional heuristic scaling factor
            to achieve a good approximation~\cite{Miatto2012}.
            \par
\begin{figure*}%
                \subfloat{%
                    \label{fig:Lambda_n_7_0_a}%
                    \includegraphics[width=0.5\linewidth]{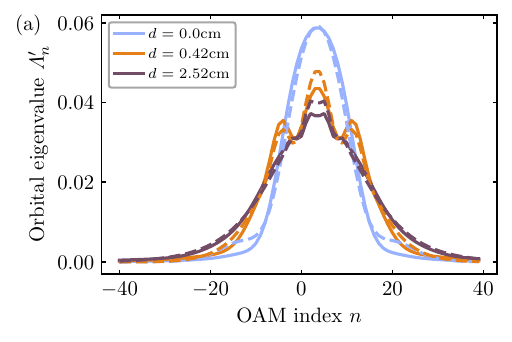}%
                }%
                \subfloat{%
                    \label{fig:Lambda_n_7_0_b}%
                    \includegraphics[width=0.5\linewidth]{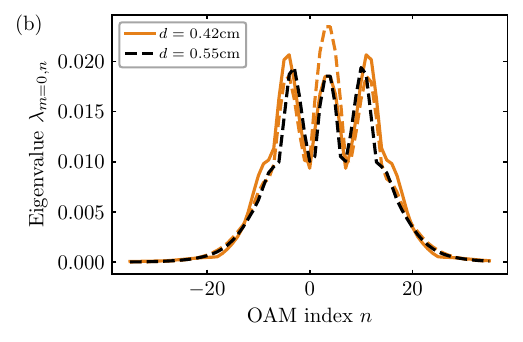}%
                }%
                \caption{%
                    (a)~Normalized orbital eigenvalues $\varLambda_n'=\varLambda_n/\sum_n\varLambda_n$
                    and (b)~cuts of the normalized low-gain modal weights $\lambda_{m=0,n}$ (${\sum_{m,n} \lambda_{mn}}=1$) 
                    for different distances~$d$ between the two crystals of
                    the \textsuoo{} interferometer. Pump parameters are  $l_p=7$ and $m_p=0$. The curves are calculated using $g_z$ in the form of
                      the $\sinc$-term as in
                    \Cref{eq:gz_twocystals} (solid lines)
                    and using the double Gaussian approximation
                    [\Cref{eq:Analytical_expression_Lambda_n}, dashed lines].
                   }%
                \label{fig:Lambda_n_7_0}
            \end{figure*}

 The function $g_z$ is then given by
            \begin{widetext}
                \begin{align}
                    \begin{split}
                        g_{z}\of{q_s, q_i, \phi_s, \phi_i} &=
                        2L\exp\of{-\alpha^2 L\frac{(\vec{q}_s-\vec{q}_i)^2}{4k_p}}
                        \cos\of{L\frac{(\vec{q}_s-\vec{q}_i)^2}{4k_p}+ \frac{\Delta n k_s d}{n_s}+ n_s d \frac{(\vec{q}_s-\vec{q}_i)^2}{4k_p}} \\
                        &\qquad\times \exp\ofb{\I\left(L\frac{(\vec{q_s}-\vec{q}_i)^2}{2k_p}+ \frac{\Delta n k_s d}{n_s}+ n_s d \frac{(\vec{q}_s-\vec{q}_i)^2}{4k_p}\right)}.
                        \label{eq:gz_twocystals_doublegauss}%
                    \end{split}
                \end{align}
                \par Next, we expand $W\of{q_s,q_i,\phi_s-\phi_i}$ 
                defined in \Cref{eq:def_W_term} using
                the binomial theorem and abbreviate the angle difference
                as $\phi_s-\phi_i\reqdef\phi$.
                The Fourier integral from \Cref{eq:chi_calc_integral} is then given by:
                \begin{align}
                    \begin{split}
                    \chi_n\of{q_s,q_i} &= \bar{A} \int_0^{2\pi}\!\dd\phi\,
                    \exp\ofb{-L\alpha^2\frac{(\vec{q}_s-\vec{q}_i)^2}{4k_p}-\left(\frac{w_0\xi}{2}\right)^2+\I\left(L\frac{(\vec{q}_s-\vec{q}_i)^2}{2k_p}+\frac{\Delta n k_s d}{n_s}+n_s d \frac{(\vec{q}_s-\vec{q}_i)^2}{4k_p}\right)}
                    \\ 
                    &\qquad\times 
                        \cos\of{L\frac{\left(\vec{q}_s-\vec{q}_i\right)^2}{4k_p}+ \frac{\Delta n k_s d}{n_s}+ n_s d 
                        \frac{\left(\vec{q}_s-\vec{q}_i\right)^2}{4k_p}}
                        \sum_{\nu=0}^{l_p}\binom{l_p}{\nu} q_s^{\nu} q^{l_p-\nu}_i \E^{\I\left(n-\nu\right)\phi}.
                    \end{split}
                \end{align}
            \end{widetext}
            By analytically evaluating the
            integral over $\phi$ in the expression above
            and integrating $\abs{\chi_n\of{q_s,q_i}}^2$ over
            $q_s$ and $q_i$, one can obtain  
            the orbital eigenvalues $\varLambda_{n}$ according
            to \Cref{eq:OAM_eigenvalues}:
            \begin{subequations}
                \begin{widetext}
                    \begin{align}
                        \begin{split}
                        \varLambda_{n} &= A' \int_0^{\infty}\!\int_0^{\infty}\!\dd q_s \dd q_i\,q_s q_i \E^{-a\left(q_s^2+q_i^2\right)} \left|\sum_{\nu=0}^{l_p}\binom{l_p}{\nu} q_s^{\nu}q^{l_p-\nu}_i
                        \left[\E^{\I\frac{E}{2}(q_s^2+q_i^2)} \E^{\I\mu}J_{n-\nu}\of{q_s q_i V}
                        +\E^{-\I\frac{E}{2}(q_s^2+q_i^2)}\E^{-\I\mu}J_{n-\nu}\of{q_s q_i U}\right]\right|^2, 
                        \end{split}
                        \label{eq:Analytical_expression_Lambda_n}
                    \end{align}
                \end{widetext}
                where $J_{n-\nu}$ denotes the Bessel functions of order $n-\nu$ and
                \begingroup
                    \allowdisplaybreaks
                    \begin{align}
                        A' &= \bar{A}\pi, \\
                        a &= \frac{w_0^2}{2}+\frac{L\alpha^2}{2k_p}, \\
                        E &= \frac{L}{2k_p}+\frac{n_sd}{2k_p}, \\
                        \mu &=\frac{\Delta n k_s d}{n_s}, \\
                        V &= \frac{\I w_0^2}{2}-\frac{\I L\alpha^2}{2k_p}-\frac{3L}{2k_p}-\frac{n_sd}{k_p}, \\
                        U &= \frac{\I w_0^2}{2}-\frac{\I L\alpha^2}{2k_p}-\frac{L}{2k_p}.
                    \end{align}
                \endgroup
            \end{subequations}
            From \Cref{eq:Analytical_expression_Lambda_n} it
            can be seen that for $d=0$, the profile of
            orbital eigenvalue
            distribution is defined by a Binomial distribution
            and the most populated mode has the OAM $n=l_p/2$,
            while the distribution has a monotonic behavior.
            However, for $d>0$, a
            complicated interference takes place
            (see the term in the brackets) due to the
            nonzero $\mu=\Delta n k_s d/n_s$ factor. This leads to 
            the suppression of low-order Bessel terms and
            enhanced contribution of high-order Bessel functions
            to the eigenvalue distribution,
            resulting in the non-monotonic behavior.
            To demonstrate this, we plot the 
            $\varLambda_n$
            obtained from the analytical expression 
            [\Cref{eq:Analytical_expression_Lambda_n}] with
            the Double-Gaussian approximation
            for different distances between the crystals,
            see \Cref{fig:Lambda_n_7_0_a} (dashed lines).
            \par It can be  seen that a non-monotonic behavior takes
            place for both constructive and destructive interference
            points, however, it is more pronounced for the destructive
            interference configuration. For $d=0$ the distribution is
            centred at $n=l_p/2$ and, as predicted, has a monotonic
            behavior. In the same plot, the numerically calculated
            orbital eigenvalues [obtained using $g_z$ with the
            $\sinc$-term as in
            \Cref{eq:gz_twocystals}] are presented by solid
            lines. The good agreement between the numerically
            calculated and analytically evaluated orbital
            eigenvalues demonstrates the validity of the
            approximated
            analytical expression \Cref{eq:Analytical_expression_Lambda_n}.
            \par The modal weight cuts also demonstrate a qualitative similarity
            between the two models, see \Cref{fig:Lambda_n_7_0_b}, where
            they are presented for the dark fringe
            of the interferometer. It is worth noting that
            in the Double-Gaussian approximation, the point
            of the dark fringe is shifted, and therefore, the side
            peaks for the crystal distance $d=\SI{0.42}{\centi\meter}$
            are not as pronounced as in the case including the
            $\sinc$-term. However, for the distance
            ${d=\SI{0.55}{\centi\meter}}$, we obtain the same picture
            with more pronounced side peaks compared to the symmetry
            point $l_p/2$ in the Double-Gaussian approximation.

    \section{ANGULAR DISPLACEMENT MEASUREMENT}\label{sec:angular_displacement}
\begin{figure*}
            \centering%
            \hspace*{-0.5em}%
            \def\svgwidth{\linewidth}%
             \input{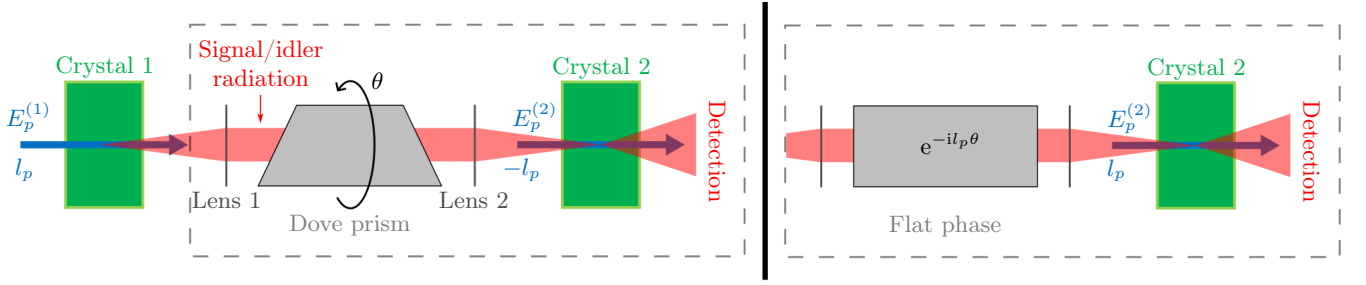}%
            \caption{%
                Left: Sketch of the \textsuoo{} interferometer in
                \textit{diffraction compensated} configuration
                (\textit{4f-lens system})~\cite{PRR5}
                with a Dove prism rotated by an angle $\theta$
                and located between the two lenses in the signal and idler
                field. The first crystal is
                pumped by a Laguerre-Gaussian beam
                $E_p^{\left(1\right)}$ with
                OAM $l_p$, while the second crystal is
                pumped by a Laugerre-Gaussian beam
                $E_p^{\left(2\right)}$ with OAM $-l_p$.
                In an experimental setup, the pump fields
                can usually be removed after the first
                crystal and before the detection stage using
                a dichroic mirror (not shown).
                Mathematically, the lenses
                are taken into account
                via \Cref{eq:tparelcomp}~\cite{PRR5}.
                Right: Interpretation of
                the dashed rectangle on the left.
                As described in the text, the
                setup shown on the left is
                mathematically equivalent
                to instead pumping the second crystal
                with a Laguerre-Gaussian
                beam $E_p^{\left(2\right)}$ with
                OAM $l_p$ (same sign as for the first crystal)
                and adding a flat
                (constant with respect to~$\vec{q}_p$,
                $\vec{q}_s$ and~$\vec{q}_i$) phase
                profile $\E^{-\I l_p \theta}$ to the
                signal and idler radiation
                (or, equivalently, to the pump beam),
                which results in an additional
                phase factor in the second crystal's
                TPA,
                see \Cref{eq:dove2d2connection}.
                The structure of the system is then
                equivalent to the compensated
                \textsuoo{} interferometer
                considered in \refcite{PRR5},
                except for the pump shape and the fact
                that therein, only one transverse
                dimension is considered. Note that in both
                the left and right part of the figure,
                the superscript~{$^{(2,D)}$} used in the text
                refers to the subsystem marked with the dashed
                rectangle.
            }%
            \label{fig:sketch_su11_dove}%
        \end{figure*}

        In this section, we finalize our analysis of PDC with
        Laguerre-Gaussian pump beams by analyzing the
        application of the \textsuoo{} interferometer for angular
        displacement detection. As it has already been mentioned in
        \Cref{sec:introduction}, it is possible to use entangled
        photon states to reach supersensitivity for angular displacement
        detection~\cite{Fickler2012,PhysRevLett.127.263601,PhysRevA.83.053829}.
        In this section, to beat the classical limit,
        we consider an \textsuoo{} interferometer which has
        a Dove prism placed between the two crystals.
        A sketch of this setup is presented in \Cref{fig:sketch_su11_dove}.
        \par A Dove prism modifies the electric signal and idler fields
        in two ways, provided that the fields are not too tightly
        focused~\cite{Gonzalez:06,PhysRevA.83.053829}:
        First, it swaps the helicity of the OAM components
        of the electric field.
        For example, if the signal field inside the
        interferometer (the radiation
        generated from the first crystal) has an azimuthal
        dependency
        $\E^{\I n \phi_s}$ for some $n\in\mathbb{Z}$, then
        after the prism it is given by
        $\E^{-\I n \phi_s}$ and, analogously,
        the helicity is swapped
        for the idler field. Second, the prism introduces
        a phase $\E^{-\I n {\theta}}$ to the
        field, where $\theta$ is the rotation angle of the prism.
        Note that in our system, this phase is added for both
        the signal and idler field.
        \par To understand the impact these
        actions have on the 
        \textsuoo{} interferometer, we first analyze the symmetries
        of the system under a sign
        change of the pump OAM index.
        Intuitively, if the sign of $l_p$
        is changed, $l_p\rightarrow-l_p$, the results
        presented in this work should
        not be affected qualitatively and only change by certain
        symmetry transformations. To analyze this more rigorously,
        we add a superscript $^{\left[l_p\right]}$ or
        $^{\left[-l_p\right]}$ to the quantities in question, 
        depending on the sign of the pump OAM number.
        \par With this notation, it is immediately clear that for the function $W$
        as defined in \Cref{eq:def_W_term},
        \begin{align}
            W^{\left[l_p\right]}\of{q_s,q_i,\phi_s-\phi_i} &=
                W^{\left[-l_p\right]}\of{q_s,q_i,\phi_i-\phi_s}.
        \end{align}
        As a consequence, for $R$ as defined in \Cref{eq:def_R}, we
        can see with the help of \Cref{eq:phi_symm_gz,eq:phi_symm_beta} that
        \begin{align}
            R^{\left[l_p\right]}\of{q_s,q_i,\phi_s-\phi_i} &=
                R^{\left[-l_p\right]}\of{q_s,q_i,\phi_i-\phi_s}.
        \end{align}
        \par Applying this to the Fourier coefficient
        integral for $\chi_n$ as given in
        \Cref{eq:chi_calc_integral} and using the fact
        that $R$ is ${2\pi}$-periodic in the angle difference,
        it can be seen that
        \begin{align}\label{eq:chi_lp_sign_swap}
            \chi_{n}^{\left[l_p\right]}\of{q_s,q_i} &=
                \chi_{-n}^{\left[-l_p\right]}\of{q_s,q_i},
        \end{align}
        which therefore also holds
        for the Schmidt decomposition
        quantities:
        \begin{subequations}
            \begin{align}
                \lambda_{mn}^{\left[l_p\right]} &=
                    \lambda_{m,-n}^{\left[-l_p\right]}, \\
                u_{mn}^{\left[l_p\right]} &=
                    u_{m,-n}^{\left[-l_p\right]}, \\
                v_{mn}^{\left[l_p\right]} &=
                    v_{m,-n}^{\left[-l_p\right]}.
            \end{align}
        \end{subequations}
        Therefore, the results for the eigenvalue distributions
        presented in this work change
        only in the sense that $n\rightarrow -n$ if the
        sign of $l_p$ is swapped.
        \par Next, we will derive the TPA describing the Dove prism
        and the second crystal of the interferometer. 
        In order to differentiate quantities related to each crystal,
        we will use superscripts~{$^{(1)}$}
        and~{$^{(2,D)}$}
        for quantities related to the first crystal, and the second crystal
        including the Dove prism, respectively.
        Note that the latter superscript refers to the subsystem
        marked with the dashed rectangle in \Cref{fig:sketch_su11_dove}. 
        \par Assuming that the second crystal is pumped with
        a Laguerre-Gaussian beam with OAM quantum number $-l_p$,
        the Fourier decomposition of its TPA
        $F^{\left(2,D\right),\left[-l_p\right]}$
        [compare \Cref{eq:decomposition}]
        can be written in the following form, taking
        into account the additional phase factors and
        sign changes due to the Dove prism as described above:
        \begin{widetext}
            \begin{align}
                F^{\left(2,D\right),\left[-l_p\right]}\of{q_s, q_i, \phi_s, \phi_i} &=
                    \frac{1}{2\pi} \sum_{n} \chi_n^{\left(2\right),\left[-l_p\right]}\of{q_s, q_i}
                    \E^{\I n \phi_s} \E^{\I n \theta} \E^{\I \left(-l_p-n\right)\phi_i} \E^{\I \left(-l_p-n\right)\theta}.
            \end{align}
            Note that $\chi_n$, 
            which for each $n$ only contains
            information regarding the radial parts of the Schmidt
            modes, is not affected by the Dove prism and therefore
            only has a superscript~{$^{(2)}$}.
            After using $n+\left(-l_p-n\right)=-l_p$ and flipping the sign
            of the summation index ${n\rightarrow -n}$, the equation above
            can be written as
            \begin{align}
                F^{\left(2,D\right),\left[-l_p\right]}\of{q_s, q_i, \phi_s, \phi_i} &=
                    \frac{\E^{-\I l_p\theta}}{2\pi} \sum_{n} \chi_{n}^{\left(2\right),\left[l_p\right]}\of{q_s, q_i}
                    \E^{-\I n \phi_s} \E^{-\I \left(l_p-n\right)\phi_i},
                    \label{eq:decomposition_prism_simplified}
            \end{align}
        \end{widetext}
        where \Cref{eq:chi_lp_sign_swap} has been used to rewrite
        $\chi_{-n}^{\left(2\right),\left[-l_p\right]}\of{q_s, q_i} = \chi_{n}^{\left(2\right),\left[l_p\right]}\of{q_s, q_i}$.
        Clearly,
        the right-hand side of \Cref{eq:decomposition_prism_simplified}
        represents the decomposition of the TPA that corresponds to the second crystal 
        without a prism, 
        pumped by a beam with an OAM  $l_p$ and multiplied with an additional
        phase factor~$\E^{-\I l_p\theta}$:
        \begin{align}
            F^{\left(2,D\right),\left[-l_p\right]}\of{\vec{q}_s, \vec{q}_i} &=
                \E^{-\I l_p\theta} F^{\left(2\right),\left[l_p\right]}\of{\vec{q}_s, \vec{q}_i}.
            \label{eq:dove2d2connection}
        \end{align}
        This expression provides an alternative way to understand how
        the Dove prism affects the interferometer: Placing the Dove prism in the
        signal and idler radiation arm is effectively equivalent to pumping
        the second crystal of an interferometer
        without the Dove prism with the OAM of opposite
        helicity and adding an extra
        phase factor $\E^{-\I l_p\theta}$ to the TPA of the second
        crystal.
        This interpretation of the setup is shown on the right
        hand side in \Cref{fig:sketch_su11_dove}.
        \par Furthermore, note that in the \textit{diffraction compensated}
        \textsuoo{} interferometer configuration~\cite{PRR5} (``wide-field \textsuoo{}
        interferometer''~\cite{Frascella:19})
        the TPAs of the first and second
        crystal are related via
        \begin{align}\label{eq:tparelcomp}
            F^{\left(2\right),\left[l_p\right]}\of{\vec{q}_s, \vec{q}_i} &=
                F^{\left(1\right),\left[l_p\right]}\of{\vec{q}_s, \vec{q}_i},
        \end{align}
        which can be seen by applying the results of
        Sec.~II of \refcite{PRR5} to the function
        $g_z$ for the second crystal: For the non-compensated
        case, this function is described by the integral 
        \begin{align}
            g_z^{\left(2\right)}\of{\vec{q}_s, \vec{q}_i}
                &= \int_L^{2L}\!\dd z\,\E^{\I\Delta k z},
        \end{align}
        while for the compensated setup, this function
        is instead given by\footnote{The apparent
        difference of the first
        integrand in \Cref{eq:integ_gz2_comp} to
        the function defined in Eq.~(2.9)
        of \refcite{PRR5} comes from
        the fact that therein, the $z$-axis integration for the second
        crystal is taken over the interval
        $\left[0,L\right]$ (reflection geometry)
        instead of $\left[L,2L\right]$
        (transmission geometry), as in the current paper.}
        \begin{align}
            g_z^{\left(2\right)}\of{\vec{q}_s, \vec{q}_i}
                &= \int_L^{2L}\!\dd z\,\E^{-\I\Delta k \left[z-2L\right]}
                = \int_0^{L}\!\dd z\,\E^{\I\Delta k z},
            \label{eq:integ_gz2_comp}
        \end{align}
        which is identical to the integral for the
        first crystal. Hence, in this
        configuration, the TPA of the second crystal
        can ultimately be written as:
        \begin{align}\label{eq:tpa_21_connection}
            F^{\left(2,D\right),\left[-l_p\right]}\of{\vec{q}_s, \vec{q}_i} &=
                F^{\left(1\right),\left[l_p\right]}\of{\vec{q}_s, \vec{q}_i} \E^{-\I l_p\theta}.
        \end{align}
        Using this relationship, in the following, we will express 
        all quantities via the quantities of the first
        crystal. Furthermore, for readability,
        we will omit the~{$^{\left[\pm l_p\right]}$}
        superscripts below.
        \par One way of connecting the TPAs of the first and
        the second crystal including the Dove prism is by using the
        fact that the output plane-wave operators of the first crystal
        must coincide with the input plane-wave operators of the
        subsystem consisting of the Dove prism and the second crystal.
        This approach has previously been used for the description of
        \textsuoo{} interferometers in \refcite{PRR5}.
        \par To implement this approach for the
        angular displacement measurements, we
        first bring \Cref{eq:in_out_plane_wave}
        into the form
        \begin{align}\label{eq:in_out_transfer_funcs}
            \begin{split}
                \hat{a}^{\mathrm{out}}\of{\vec{q}} &=
                    \int\!\dd^2 q'\,\tilde{\eta}\of{\vec{q},\vec{q}'}
                    \hat{a}^{\mathrm{in}}\of{\vec{q}'} \\
                    &\qquad+ \int\!\dd^2 q'\,\beta\of{\vec{q},\vec{q}'} \left[\hat{a}^{\mathrm{in}}\of{\vec{q}'}\right]^{\dagger}
            \end{split}
        \end{align}
        and find that the \textit{transfer functions}
        $\tilde{\eta}$ and $\beta$ connecting the input and
        output plane-wave operators are given by
        \begin{widetext}
            \begin{subequations}
                \begin{align}
                    \tilde{\eta}\of{\vec{q},\vec{q}'} &=
                        \frac{1}{2\pi} \sum_{m,n} \sqrt{\tilde{\Lambda}_{mn}} \frac{u_{mn}\of{q}}{\sqrt{q}}
                        \frac{u_{mn}^{*}\of{q'}}{\sqrt{q'}} \E^{-\I n \left(\phi-\phi'\right)},
                        \label{eq:tf_etat} \\
                    \beta\of{\vec{q},\vec{q}'} &=
                        \frac{1}{2\pi} \sum_{m,n} \sqrt{\Lambda_{mn}} \frac{u_{mn}\of{q}}{\sqrt{q}}
                        \frac{v_{mn}\of{q'}}{\sqrt{q'}}
                        \E^{-\I n \phi} \E^{-\I \left(l_p-n\right) \phi'}, \label{eq:tf_beta}
                \end{align}
            \end{subequations}
        \end{widetext}
        where $\Lambda_{mn}$ is defined as in \Cref{eq:hgeigs}
        and where
        \begin{align}
            \tilde{\Lambda}_{mn} &= \cosh^2\of{G\sqrt{\lambda_{mn}}}. \label{eq:Lambda_tilde_mn_cosh}
        \end{align}
        Note that the expression for $\beta$ is similar to the
        TPA $F$ as given in \Cref{eq:full_decomp_tpa}, except for the
        fact that the singular values $\sqrt{\lambda_{mn}}$
        are replaced with the square roots of the
        high-gain eigenvalues $\sqrt{\Lambda_{mn}}$.
        \par In this formalism, each subsystem (the first crystal
        and the second crystal with the Dove prism) has its own set of
        transfer functions which we will mark with the
        corresponding superscripts
        as introduced above. The transfer functions
        connecting the input
        and output operators of the entire
        interferometer are then given by~\cite{PRR5}:
        \begin{subequations}
            \begin{align}
                \begin{split}
                    \tilde{\eta}^{\left(\mathrm{SU}\right)}\of{\vec{q},\vec{q}'} &= \int\!\dd^2\bar{q}\,
                        \tilde{\eta}^{\left(2,D\right)}\of{\vec{q},\bar{\vec{q}}} 
                        \tilde{\eta}^{\left(1\right)}\of{\bar{\vec{q}},\vec{q}'} \\
                    &\qquad+ \int\!\dd^2\bar{q}\,\beta^{\left(2,D\right)}\of{\vec{q},\bar{\vec{q}}}
                        \left[\beta^{\left(1\right)}\of{\bar{\vec{q}},\vec{q}'}\right]^{*},
                    \label{eq:composite_tf_eta_nophase}
                \end{split}\\
                \begin{split}
                    \beta^{\left(\mathrm{SU}\right)}\of{\vec{q},\vec{q}'} &= \int\!\dd^2\bar{q}\,
                        \tilde{\eta}^{\left(2,D\right)}\of{\vec{q},\bar{\vec{q}}} 
                        \beta^{\left(1\right)}\of{\bar{\vec{q}},\vec{q}'} \\
                    &\qquad+ \int\!\dd^2\bar{q}\,\beta^{\left(2,D\right)}\of{\vec{q},\bar{\vec{q}}}
                        \left[\tilde{\eta}^{\left(1\right)}\of{\bar{\vec{q}},\vec{q}'}\right]^{*}.
                        \label{eq:composite_tf_beta_nophase}
                \end{split}
            \end{align}
        \end{subequations}
        For both subsystems, the transfer functions can be
        obtained by applying the decomposition of the corresponding
        TPA to their definitions \Cref{eq:tf_etat,eq:tf_beta}.
        \par Furthermore, note that since the additional
        phase factor $\E^{-\I l_p\theta}$ in \Cref{eq:tpa_21_connection} does
        not depend on $\vec{q}_s$ or $\vec{q}_i$, one can find a direct connection
        between the Schmidt modes and the eigenvalues of both subsystems:
        \begin{subequations}
            \begin{align}
                u_{mn}^{\left(1\right)}\of{q} \E^{-\I\frac{l_p}{2} \theta} &= u_{mn}^{\left(2,D\right)}\of{q}, \\
                v_{mn}^{\left(1\right)}\of{q} \E^{-\I\frac{l_p}{2} \theta} &= v_{mn}^{\left(2,D\right)}\of{q}, \\
                \lambda_{mn}^{\left(1\right)} &= \lambda_{mn}^{\left(2,D\right)}.
            \end{align}
        \end{subequations}
        Substituting this to \Cref{eq:tf_etat,eq:tf_beta}, it becomes immediately clear that the transfer
        functions of both subsystems are
        connected
        via
        \begin{subequations}
            \begin{align}
                \tilde{\eta}^{\left(2,D\right)}\of{\vec{q}, \vec{q}'} &=
                    \tilde{\eta}^{\left(1\right)}\of{\vec{q}, \vec{q}'}, \label{eq:eta_21_connection} \\
                \beta^{\left(2,D\right)}\of{\vec{q}, \vec{q}'} &=
                    \beta^{\left(1\right)}\of{\vec{q}, \vec{q}'} \E^{-\I l_p\theta}. 
                    \label{eq:beta_21_connection}
            \end{align}
        \end{subequations}
        These expressions are
        analogous\footnote{In \refcite{PRR5} these relationships were
        derived directly from the integro-differential equations describing
        the evolution of the plane-wave operators.
        Note that from \Cref{eq:tf_etat} it
        follows that
        $\tilde{\eta}\of{\vec{q}, \vec{q}'}
        =\left[\tilde{\eta}\of{\vec{q}',\vec{q}}\right]^{*}$.
        However,
        it should be emphasized that this does not hold at
        high-gain when time ordering effects
        are included~\cite{PRR2,PRR5}.}
        to Eqs.~(4.1a) and~(4.1b) of
        \refcite{PRR5} and greatly simplify the following steps.
        For example, using \Cref{eq:composite_tf_beta_nophase}
        and the properties of the eigenvalues given in
        \Cref{eq:lp_rel_lambda,eq:lp_rel_uv,eq:lp_rel_vu},
        it can be shown that the high-gain eigenvalues of the
        entire interferometer are given by
        \begin{align}\label{eq:eigs_hg_tf}
            \Lambda_{mn}^{\left(\mathrm{SU}\right)} =
                4\Lambda_{mn}^{\left(1\right)}
                \left(\Lambda_{mn}^{\left(1\right)} + 1\right)\cos^2\of{\frac{l_p}{2}\theta}.
        \end{align}
        \par The sensitivity for the measurement of the rotation angle
        $\theta$ of the Dove prism
        can be obtained from the error propagation relation in terms
        of the integral output intensity
        ${\langle \hat{N}_{\mathrm{tot}} \rangle 
        = \int\!\dd^2 q\,\langle \hat{N}_s\of{\vec{q}}
        +\hat{N}_i\of{\vec{q}}\rangle
        = 2\int\!\dd^2 q\,\langle \hat{N}_s\of{\vec{q}}\rangle}$
        and output 
        covariance function $\cov\of{\vec{q}_s,\vec{q}_i}$ of the
        interferometer~\cite{PhysRevA.83.053829,PRR5}:
        \begin{align}\label{eq:delta_theta_def}
            \Delta\theta
                = \frac{\langle\Delta\hat{N}_{\mathrm{tot}}\rangle}{\left|\frac{\dd \langle \hat{N}_{\mathrm{tot}} \rangle}{\dd\theta}\right|}
                = \frac{\sqrt{\iint\!\dd^2q_s\dd^2q_i \cov\of{\vec{q}_s,\vec{q}_i}}}{\left|\frac{\dd \langle \hat{N}_{\mathrm{tot}} \rangle}{\dd\theta}\right|}.
        \end{align}
        Note that this expression is identical to the
        definition of the phase sensitivity in \textsuoo{}
        interferometers. Therefore, the angular
        sensitivity for the interferometer can be rewritten by using
        Eqs.~(\ref{eq:hgeigs}),
        (\ref{eq:in_out_transfer_funcs})%
        \crefrangeconjunction%
        (\ref{eq:Lambda_tilde_mn_cosh}),
        (\ref{eq:eta_21_connection}) and
        (\ref{eq:beta_21_connection})
        and following the necessary steps
        as in Sec.~IV of \refcite{PRR5}:
        \begin{subequations}%
            \begin{align}%
                \Delta\theta_{\mathrm{TF}} = 
                    \frac{\sqrt{2}}{2}\,\frac{\sqrt{\mathcal{A}+4\mathcal{B}\cos^2\of{\frac{l_p}{2}\theta}}}{\mathcal{A}\left|l_p\right|\left|\sin\of{\frac{l_p}{2}\theta}\right|},
                \label{eq:delta_theta_tf}
            \end{align}%
            with
            \begin{align}%
                \mathcal{A} &=
                    \sum_{mn} \Lambda_{mn}^{\left(1\right)}
                    \left(\Lambda_{mn}^{\left(1\right)} + 1\right),
                    \label{eq:delta_theta_tf_AA} \\
                \mathcal{B} &= \sum_{mn} \left[ \Lambda_{mn}^{\left(1\right)}
                    \left(\Lambda_{mn}^{\left(1\right)} + 1\right) \right]^2,
                    \label{eq:delta_theta_tf_BB}
            \end{align}%
        \end{subequations}
        where we have added the subscript~\textsubscript{TF}
        to $\Delta\theta$ to indicate that it was obtained 
        by connecting the two subsystems using the
        transfer functions. The additional factor $\sqrt{2}$
        results from the fact that we consider indistinguishable
        signal and idler photons, see \Cref{sec:symms_comm_rels}.
        \par Clearly, the above expression for
        $\Delta\theta_{\mathrm{TF}}$ diverges for ${l_p=0}$. In this case,
        the pump beam is a Gaussian for ${m_p=0}$ and
        circularly symmetric for any $m_p$. Since there is no
        preferred transverse direction in the system, the
        combined signal-idler field must
        also be circularly symmetric. This can be seen from the fact
        that for $l_p=0$, the TPA only depends on the 
        signal-idler angle difference ${\phi_s-\phi_i}$, but
        not on the absolute values of the angles, see \Cref{eq:full_decomp_tpa}
        and \refcite{PRA91}. Due to this symmetry, the Dove prism
        does not modify the signal-idler field generated by the
        first crystal in a detectable manner
        and it is therefore not possible to detect the rotation angle
        of the Dove prism for $l_p=0$.
        \par Alternatively to the \textit{transfer-function approach}
        presented above,
        there is a purely Schmidt-mode theory
        based approach to obtain an expression for $\Delta\theta$.
        In this case, the TPA $F^{\left(\mathrm{SU}\right)}$ for
        the \textsuoo{} interferometer can  be written as
        \begin{align}\label{eq:TPA_interf_dove_c1c2}
            F^{\left(\mathrm{SU}\right)}\of{\vec{q}_s,\vec{q}_i} &= 
                F^{\left(1\right)}\of{\vec{q}_s,\vec{q}_i}
                + F^{\left(2,D\right)}\of{\vec{q}_s,\vec{q}_i}.
        \end{align}
        This is analogous to the way the TPA is calculated for the
        \textsuoo{} interferometers and results in $g_z$
        as given in
        \Cref{eq:gz_twocystals,eq:gz_twocystals_doublegauss}.
        \par An alternative way to obtain this result
        can be seen from
        \Creffirstrangelast{eq:hgeigs}{eq:tf_etat}{eq:Lambda_tilde_mn_cosh}
        as follows:
        For ${G\ll 1}$,
        ${\sinh\of{G\sqrt{\lambda_{mn}}}\approx G\sqrt{\lambda_{mn}}}$
        and ${\cosh\of{G\sqrt{\lambda_{mn}}}\approx 1}$, meaning
        \begin{subequations}
            \begin{align}
                \tilde{\eta}\of{\vec{q},\vec{q}'} &\approx \delta^{\left(2\right)}\of{\vec{q}-\vec{q}'}, \\
                \beta\of{\vec{q},\vec{q}'} &\approx G\,F\of{\vec{q},\vec{q}'},
            \end{align}
        \end{subequations}
        where $\delta^{\left(2\right)}$ is the two-dimensional
        Dirac delta function
        With these two expressions for the transfer functions,
        \Cref{eq:composite_tf_beta_nophase}
        simplifies to \Cref{eq:TPA_interf_dove_c1c2}.
        Note that there is no distinction between whether
        $\vec{q}$ and $\vec{q}'$ correspond to $\vec{q}_s$ or $\vec{q}_i$
        due to the degeneracy, see \Cref{sec:symms_comm_rels}.
        \par Next, applying the connection
        between the TPAs for the two subsystems given in
        \Cref{eq:tpa_21_connection} to \Cref{eq:TPA_interf_dove_c1c2}
        yields:
        \begin{align}\label{eq:TPA_angular_sens}
            F^{\left(\mathrm{SU}\right)}\of{\vec{q}_s,\vec{q}_i} &= 
                F^{\left(1\right)}\of{\vec{q}_s,\vec{q}_i}
                \left(1+\E^{-\I l_p\theta}\right).
        \end{align}
        This connection between the TPA
        of the interferometer
        and the first crystal allows
        us to define the eigenvalues of
        the Schmidt decomposition of the interferometer as
        \begin{align}
            \lambda_{mn}^{\left(\mathrm{SU}\right)} =
                4\lambda_{mn}^{\left(1\right)}\cos^2\of{\frac{l_p}{2}\theta},
        \end{align}
        since, as mentioned above,
        the rotation angle $\theta$ does not depend on
        the transverse wave-vector components
        $\vec{q}_s$ and $\vec{q}_i$.
        From there, the high-gain eigenvalues follow as
        \begin{align}
            \Lambda_{mn}^{\left(\mathrm{SU}\right)} =
                \sinh^2\ofb{2G\sqrt{\lambda_{mn}}
                \cos\of{\frac{l_p}{2}\theta}}.
                \label{eq:hg_lambda_cos_arg}
        \end{align}
        Note that this expression is not exactly
        equivalent to the \textit{transfer function} case considered above,
        compare \Cref{eq:eigs_hg_tf}. However, for $G\ll 1$,
        \Cref{eq:hg_lambda_cos_arg,eq:eigs_hg_tf} coincide.
        \par Using \Cref{eq:hg_lambda_cos_arg}, one can evaluate 
        the expression for
        $\Delta\theta$ given by \Cref{eq:delta_theta_def} as follows:
        \begin{widetext}
            \begin{align}
                \Delta\theta_{\mathrm{SMT}} &= 
                    \frac{\sqrt{2}}{2}
                    \frac{1}{G\left|l_p\right|\left|\sin\of{\frac{l_p}{2}\theta}\right|}
                    \frac{
                        \sqrt{
                            \sum_{mn}
                            \sinh^2\ofb{4G\sqrt{\lambda_{mn}}\cos\of{\frac{l_p}{2}\theta}}
                        }
                    }{
                        \sum_{mn}
                        \sqrt{\lambda_{mn}}
                        \sinh\ofb{4G\sqrt{\lambda_{mn}}\left|\cos\of{\frac{l_p}{2}\theta}\right|}},
                \label{eq:delta_theta_smt}
            \end{align}
        \end{widetext}
        where we have added the subscript~\textsubscript{SMT} to indicate
        that this expression was obtained by using the Schmidt-mode theory,
        instead of the transfer functions as in \Cref{eq:delta_theta_tf}.
        Note that this expression is, in general, not equal to $\Delta\theta_{\mathrm{TF}}$
        as defined in \Cref{eq:delta_theta_tf}. However, for low-gain $G\ll 1$,
        both of these expressions coincide:
        \begin{align}\label {eq:delta_theta_lowgain}
            \Delta\theta_{G\ll 1} &\approx
            \frac{\sqrt{2}}{2}
            \frac{1}{G\left|l_p\right| \left|\sin\of{\frac{l_p}{2}\theta}\right|}.
        \end{align}
        Clearly, the differences in $\Delta_{\mathrm{TF}}$ and
        $\Delta_{\mathrm{SMT}}$ must be related to the
        fact of how we introduce the displacement phase:
        In the transfer functions approach,
        both crystals are treated separately and
        the angular displacement phase is applied
        to the solution for the operators of the second crystal,
        thereby introducing ordering in
        time, which is not the case for the Schmidt-mode
        theory approach where both TPAs are directly
        added together [see \Cref{eq:TPA_interf_dove_c1c2}].
        \par Usually, the sensitivity $\Delta\theta$ is normalized
        with respect to the classical limit, that is, the shot noise level  (SNL, the standard quantum limit~\cite{PhysRevA.83.053829}),
        which is defined by
        \begin{align}
            \Delta\theta_{\mathrm{SNL}} &= \frac{1}{\sqrt{\langle\hat{N}_{\mathrm{tot}}^{\left(1\right)}\rangle}},
        \end{align}
        where $\langle\hat{N}_{\mathrm{tot}}^{\left(1\right)}\rangle$ is the
        total intensity inside the interferometer that
        interacts with the Dove prism, that is, the total intensity
        of the signal and idler radiation generated by the first
        crystal: $\langle\hat{N}_{\mathrm{tot}}^{\left(1\right)}\rangle
        =2\langle\hat{N}_{s,\mathrm{tot}}^{\left(1\right)}\rangle
        =2\int\!\dd^2 q\,\langle \hat{N}_s^{\left(1\right)}\of{\vec{q}}\rangle$.
        Then, the normalized angular displacement sensitivity reads
        \begin{align}\label{eq:f_def}
            f &= \frac{\Delta\theta}{\Delta\theta_{\mathrm{SNL}}}
        \end{align}
        and can be evaluated using
        either the transfer functions approach
        [\Cref{eq:delta_theta_tf,eq:delta_theta_tf_AA,eq:delta_theta_tf_BB}]
        or the Schmidt-mode theory
        [\Cref{eq:delta_theta_smt}].
        \par
\begin{figure*}
            \subfloat{%
                \label{fig:f_hg_1_0}%
                \includegraphics[width=0.5\linewidth]{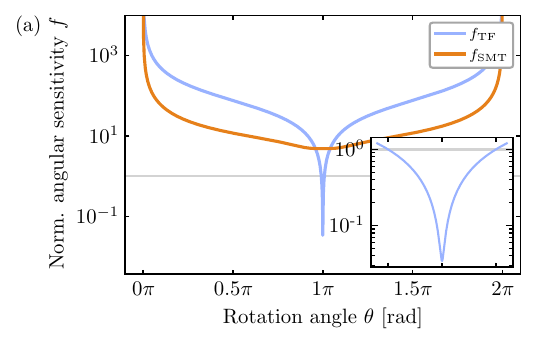}%
            }%
            \subfloat{%
                \label{fig:f_hg_7_0}%
                \includegraphics[width=0.5\linewidth]{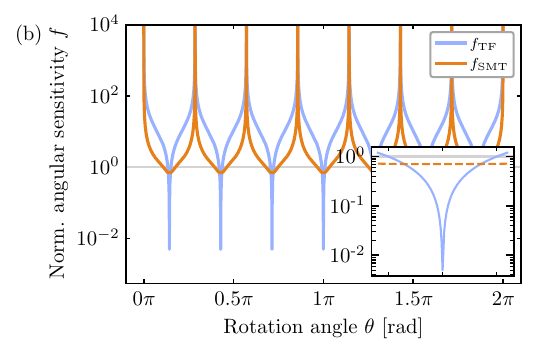}%
            }\\
            \subfloat{%
                \label{fig:f_hg_7_7}%
                \includegraphics[width=0.5\linewidth]{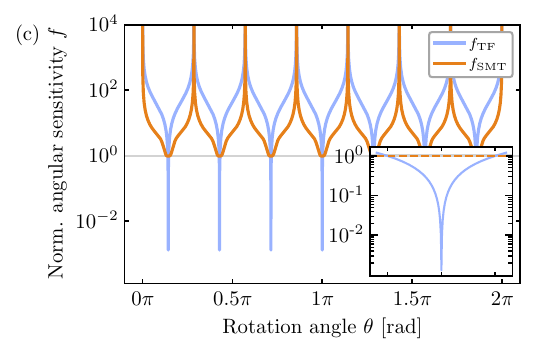}%
            }%
             \subfloat{%
                \label{fig:f_lg}%
                \includegraphics[width=0.5\linewidth]{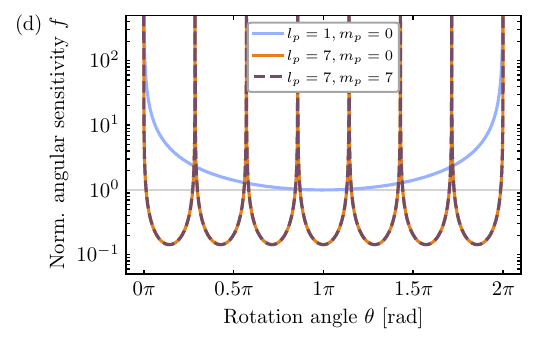}%
            }%
            \caption{%
                Normalized angular displacement
                sensitivities $f$ for the
                (a)-(c)~medium-gain regime
                ($G_{\mathrm{exp}}=2$) and
                (d)~low-gain regime
                ($G_{\mathrm{exp}}=0.01$). The pump
                parameters are (a)~$l_p=1$, $m_p=0$;
                (b)~$l_p=7$, $m_p=0$ and (c)~$l_p=7$, $m_p=7$.
                In (a)-(c), the insets show
                the supersensitivity regions around $\theta=\pi$. Here, 
                the outer x-axis ticks are located at the boundaries of
                the phase sensitivity region for the
                transfer-function approach.
                The widths
                (distances between the outer ticks)
                of said regions are
                (a)~$\SI{52.1}{\milli\rad}$,
                (b)~$\SI{53.3}{\milli\rad}$
                and (c)~$\SI{13.1}{\milli\rad}$.
                Note that the line for $f_{\mathrm{SMT}}$
                is dashed in the inset. 
                In~(d), for low-gain, $f$
                only depends on $l_p$ and $G$ and
                both $f_{\mathrm{TF}}$ and $f_{\mathrm{SMT}}$
                coincide, see \Cref{eq:delta_theta_lowgain}.
                For these figures, the average fitting
                constants $A$ were used, as described in the
                text. In all plots, the horizontal thin gray 
                line indicates the
                standard quantum limit $f=1$.
            }
            \label{fig:normd_angular_sens}
        \end{figure*}

 The results for~$f$ as a function
        of the rotation angle~$\theta$ for both approaches
        are presented
        in \Cref{fig:normd_angular_sens}.
        Note that, for simplicity, we average the values
        for the fitting constant $A$ for the low-
        and high-gain regime and use only the average value
        in this section. This is 
        justified by the experimental technique for
        measuring this constant, in which the relatively large
        measurement error exceeds the difference
        between the constants for the low
        and high-gain regime~\cite{Spasibko:12}.
        Furthermore, this approximation will greatly
        simplify the following steps, where we consider
        ranges of $G_{\mathrm{exp}}$ values.
        \par Here, \Cref{fig:f_lg} shows that, in the
        low-gain regime, the angular sensitivity does not depend
        on the radial pump number $m_p$, which is consistent with
        \Cref{eq:delta_theta_lowgain}. As the gain increases,
        according to the rigorous transfer functions approach,
        the angular sensitivity improves for all orbital numbers
        of the pump, see
        \Cref{fig:f_hg_1_0,fig:f_hg_7_0,fig:f_hg_7_7}.
        In all cases, $f$ is
        $2\pi/l_p$-periodic, which is expected
        from the analytic expressions of $\Delta\theta$
        [see \Cref{eq:delta_theta_tf,eq:delta_theta_tf_AA,eq:delta_theta_tf_BB}
        and also \Cref{eq:delta_theta_smt}].
        At the same time, the role of the radial pump number
        becomes visible: The angular sensitivity improves with
        increasing radial pump number when both the gain and the
        orbital pump number are fixed. However, in the
        high-gain regime, the angular
        sensitivity calculated using the Schmidt-mode approach
        (where the angular displacement is included directly
        in the TPA for the interferometer)
        differs from the prediction of the transfer
        functions approach due to the fact that the ordering
        in time in regard to the applied angular displacement phase
        is not accounted for, which restricts the applicability
        of the Schmidt-mode approach.
        \par
\begin{figure}[t!]%
            \includegraphics[width=\linewidth]{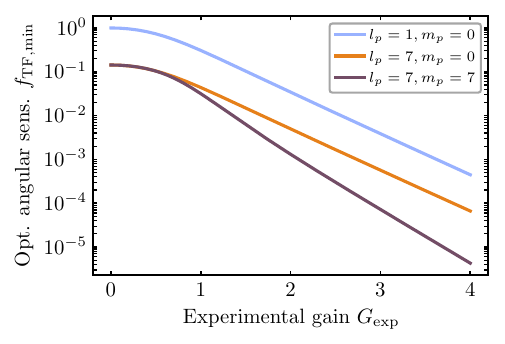}%
            \caption{%
                Optimal (minimal) SNL-normalized angular
                sensitivity $f_{\mathrm{TF,min}}$ over the
                experimental gain $G_{\mathrm{exp}}$
                as given by
                \Cref{eq:delta_theta_tf,eq:delta_theta_tf_AA,eq:delta_theta_tf_BB,eq:f_def}
                for selected values of $l_p$ and $m_p$.
                As $G_{\mathrm{exp}}\searrow 0$, $f_{\mathrm{TF,min}}$
                approaches $1/\left|l_p\right|$,
                see \Cref{eq:f_tf_G0}.
            }%
            \label{fig:f_min_G}
        \end{figure}

 Evidently, for $\theta\rightarrow 2\pi n/l_p$, $n\in\mathbb{Z}$, both
        $\Delta_{\mathrm{TF}}$ and $\Delta_{\mathrm{SMT}}$ diverge
        regardless of $G$,
        see \Cref{eq:delta_theta_tf} and \Cref{eq:delta_theta_smt}.
        Furthermore, $\Delta\theta_{\mathrm{TF}}$ in \Cref{eq:delta_theta_tf} 
        is minimized for any 
        $\theta=\pi\left(2n+1\right)\!/l_p$, $n\in\mathbb{Z}$,
        which is confirmed by the plots in \Cref{fig:normd_angular_sens}.
        Since the SNL does not depend on $\theta$, the
        optimal (minimal) SNL-normalized angular sensitivity $f$
        is also achieved at the aforementioned values for $\theta$
        and is given by
        \begin{align}
            f_{\mathrm{TF,min}} &= \frac{1}{\left|l_p\right|}
                \sqrt{\frac{\langle\hat{N}_{s,\mathrm{tot}}^{\left(1\right)}\rangle}{\mathcal{A}}}.
            \label{eq:f_TF_opt}
        \end{align}
        The optimal (minimal) SNL-normalized
        sensitivity for the transfer function approach
        $f_{\mathrm{TF,min}}$
        as a function of the experimental
        gain $G_{\mathrm{exp}}$ is shown
        in \Cref{fig:f_min_G}
        for selected values of $l_p$ and $m_p$.
        Clearly, the optimal SNL-normalized angular sensitivity improves strongly
        with increasing the gain. For low-gain, the quadratic contribution
        $\sum_{m,n}\left(\Lambda^{\left(1\right)}_{mn}\right)^2$ in $\mathcal{A}$
        [see \Cref{eq:delta_theta_tf_AA}] becomes negligible
        and $f_{\mathrm{TF,min}}$ behaves as:
        \begin{align}\label{eq:f_tf_G0}
            \lim_{G_{\mathrm{exp}}\searrow 0} f_{\mathrm{TF,min}} &= \frac{1}{\left|l_p\right|}.
        \end{align}       
        Indeed, in the low-gain regime, the shown
        normalized angular sensitivity does not
        depend on the pump radial number, see \Cref{fig:f_min_G}. 
        \par Furthermore, we may rewrite \Cref{eq:f_TF_opt} in terms of
        the number of effective modes~\cite{PRR5}:
        \begin{align}
            f_{\mathrm{TF},\mathrm{min}} &= \frac{1}{\left|l_p\right|}
                \frac{1}{\sqrt{1+\frac{\langle\hat{N}_{s,\mathrm{tot}}^{\left(1\right)}\rangle}{K^{\left(1\right)}}}},
        \end{align}
        with the Schmidt number 
        (effective mode number)
        $K^{\left(1\right)}$
        as defined in \Cref{eq:schmidt_number},
        which is here
        defined using the normalized first crystal eigenvalues
        ${\Lambda_{mn}^{\left(1\right)}}'$ [see
        \Cref{eq:renormal_eigenvalues}].
        \par The behavior of the optimal angular
        sensitivity for the transfer function
        approach in the multimode
        regime is similar to the behavior of the 
        optimal phase sensitivity discussed
        in \refcite{PRR5}: The angular
        sensitivity can be improved by increasing
        the intensity inside
        the interferometer and reducing the
        effective mode number.
        Interestingly, a reduction of both the number of
        OAM and radial modes leads to an improvement
        in the sensitivity. 
        \par
\begin{figure}[t!]%
            \includegraphics[width=\linewidth]{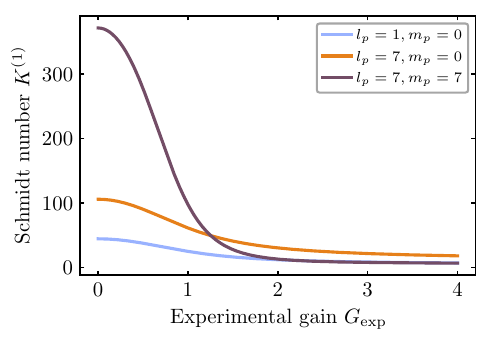}%
            \caption{%
                Schmidt number $K^{\left(1\right)}$
                as defined in \Cref{eq:schmidt_number}
                for the first crystal's high-gain
                eigenvalues $\Lambda_{mn}^{\left(1\right)}$.
                In the low-gain regime, the values for $l_p=7$
                coincide with the one presented in
                \Cref{fig:2D_eigen_Multi_7_0.0} (a small
                deviation is expected since we use averaged
                values for the fitting constant $A$ in this
                section, as discussed in the text).
                At increasing gain, highly populated
                modes are amplified more strongly,
                leading to a strong reduction in the effective
                number of modes~\cite{PRA91}. Therefore, for $l_p=m_p=7$,
                the number of effective modes may fall
                below $K^{\left(1\right)}$ for $l_p=7$, $m_p=0$
                at high gain, even though it
                is much larger for low gain.
            }%
            \label{fig:K_G}%
        \end{figure}

 Moreover, in the high-gain regime,
        the radial pump number strongly affects the angular
        sensitivity: The higher the radial pump number is, the
        better the angular sensitivity becomes at high
        gain for the
        selected pump OAM values, see \Cref{fig:f_min_G}.
        This might seem counterintuitive at first, since 
        increasing $m_p$ increases the Schmidt
        number $K^{\left(1\right)}$ as
        discussed in \Cref{sec:modes_struct_single_cryst}.
        However, with increasing gain, the number of
        effective modes is reduced since the most populated
        modes are amplified more strongly~\cite{PRA91}.
        This behavior is confirmed by the plots of
        the Schmidt numbers as a function of the experimental
        gain shown in \Cref{fig:K_G}.
        \par Similarly to \refcite{PhysRevLett.127.263601,
        PhysRevA.83.053829},
        we find a prefactor $1/\left|l_p\right|$ for the
        optimal sensitivity [\Cref{eq:f_TF_opt}].
        However, since the integral intensity
        $\langle\hat{N}_{s,\mathrm{tot}}^{\left(1\right)}\rangle$,
        the factor $\mathcal{A}$ and the fitting
        constant $A$ relating
        the experimental and theoretical gain
        also depend on $l_p$, it is not clear
        that the optimal sensitivity
        overall scales as $1/\left|l_p\right|$.
        Indeed, since we focus on multimode PDC radiation,
        the behavior of $f_{\mathrm{TF}}$ and
        $f_{\mathrm{TF},\mathrm{min}}$ as a
        function of
        $l_p$ is less obvious. Nevertheless,
        as shown above in \Cref{eq:f_tf_G0},
        this behavior is recovered exactly as
        $f_{\mathrm{TF,min}} = 1/\left|l_p\right|$
        in the low-gain limit and surpasses this value
        for high gain for the selected
        values of $l_p$ and $m_p$, see \Cref{fig:f_min_G}.
        \par
\begin{figure}[t!]%
            \includegraphics[width=\linewidth]{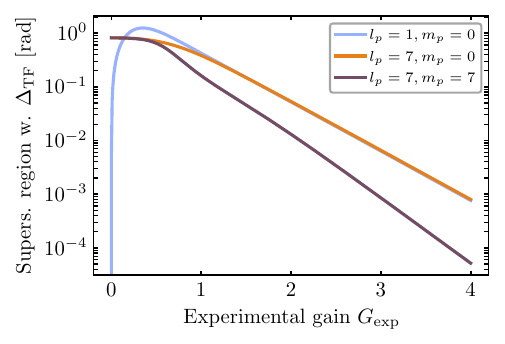}%
            \caption{%
                Supersensitivity region width $\Delta_{\mathrm{TF}}$ over
                the experimental gain
                $G_{\mathrm{exp}}$ for selected values of $l_p$ and $m_p$.
                Evidently, the width of the region decreases rapidly with
                increasing gain.
                As $G_{\mathrm{exp}}$ approaches $0$, the width only
                depends on $l_p$, see \Cref{eq:lim_Delta_tf_G0}.
                For any ${l_p\centernot\in \left\lbrace-1, 0, 1\right\rbrace}$,
                the width starts at a finite, nonzero value, while for
                $l_p=\pm 1$, it is
                exactly $0$, meaning the phase supersensitivity
                region quickly degenerates into
                a single point as the gain decreases.
            }%
            \label{fig:Delta_tf_G}
        \end{figure}

 For experimental applications, the width of the
        angular supersensitivity region~$\Delta$ is an
        important parameter since it must be large enough
        to allow for stable measurements.
        By setting $f_{\mathrm{TF}}\of{\theta}\eqset 1$, one can obtain
        the following expression for the width of said region:
        \begin{align}
            \Delta_{\mathrm{TF}} &= \frac{2\pi}{\left|l_p\right|} - 
            \frac{4}{\left|l_p\right|}\arctan\of{\sqrt{\frac{1+4 \frac{\mathcal{B}}{\mathcal{A}}}{2\mathcal{A}\,l_p^2 \left(\Delta\theta_{\mathrm{SNL}}\right)^2-1}}}.
        \end{align}
        Note that a supersensitivity region of nonzero width only exists for
        $2\mathcal{A}\,l_p^2 \left(\Delta\theta_{\mathrm{SNL}}\right)^2>1$,
        which is equivalent to
        $\left(\langle\hat{N}_{s,\mathrm{tot}}^{\left(1\right)}
        \rangle/K^{\left(1\right)}+1\right)l_p^2>1$ 
        and therefore always true for $l_p\neq 0$ and $G>0$.
        Figure~\ref{fig:Delta_tf_G} shows the width of
        the supersensitivity region for
        selected values of $l_p$
        and $m_p$ as a function of the
        experimental gain $G_{\mathrm{exp}}$.
        Clearly, the width of the supersensitivity
        region decreases strongly with
        increasing gain.
        \par
\begin{figure}[t!]%
            \includegraphics[width=\linewidth]{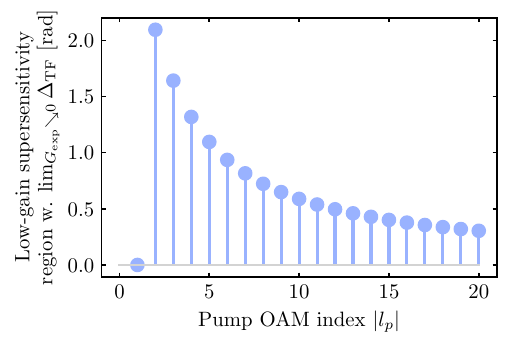}%
            \caption{%
                Low-gain limit of the supersensitivity
                region width
                $\lim_{G_{\mathrm{exp}}\searrow 0} \Delta_{\mathrm{TF}}$
                as given by \Cref{eq:lim_Delta_tf_G0}.
                For $l_p=0$, $\Delta\theta_{\mathrm{TF}}$
                and $\Delta_{\mathrm{TF}}$
                are undefined, as discussed in the text, while for
                $\left|l_p\right|=1$, the supersensitivity region
                is given by a single point (zero width)
                for $G_{\mathrm{exp}}\searrow 0$. Furthermore, for
                $\abs{l_p}=2$ the value is maximized with
                $\lim_{G_{\mathrm{exp}}\searrow 0} \Delta_{\mathrm{TF}}
                =\SI[fraction-function=\tfrac,parse-numbers=false]{%
                \frac{2\pi}{3}}{\rad}\approx \SI{2.09}{\rad}$.%
            }%
            \label{fig:lg_Delta}%
        \end{figure}

 For the low-gain limit, we find:
        \begin{align}\label{eq:lim_Delta_tf_G0}
            \lim_{G_{\mathrm{exp}}\searrow 0} \Delta_{\mathrm{TF}} &= \begin{cases}
                0 & l_p = \pm 1; \\
                \frac{2\pi}{\left|l_p\right|} - \frac{4}{\left|l_p\right|}\arctan\of{\frac{1}{\sqrt{l_p^2-1}}} & \left|l_p\right| > 1,
            \end{cases}
        \end{align}
        which explains the behavior of the curves shown in
        \Cref{fig:Delta_tf_G} near $G_{\mathrm{exp}}=0$.
        Note that the first case does not contradict
        the above statement regarding the existence of
        a supersensitivity region of nonzero width,
        since it only applies in the limit $G_{\mathrm{exp}}\searrow 0$.
        Figure~\ref{fig:lg_Delta} shows a plot of 
        \Cref{eq:lim_Delta_tf_G0} and indicates that in the
        low-gain regime the supersensitivity region width
        is maximized for $l_p=2$. Furthermore, even for 
        large values of $\left|l_p\right|$, this width still
        remains much larger compared to the observed widths
        in the high-gain regime, leading to an interesting
        interplay between the gain value and the pump OAM
        value to simultaneously achieve both high angular
        sensitivity and a wide
        supersensitivity region. Here,
        \Cref{fig:f_min_G,fig:Delta_tf_G} can serve as
        guidelines to determine
        optimal points.

        The behavior of $\Delta\theta_{\mathrm{SMT}}$
        [\Cref{eq:delta_theta_smt}]
        is less straightforward
        than $\Delta\theta_{\mathrm{TF}}$.
        It is not immediately obvious
        that the expression is minimized for
        the same rotation angles as $\Delta\theta_{\mathrm{TF}}$
        [$\theta=\pi\left(2n+1\right)\!/l_p$ for $n\in\mathbb{Z}$], although this is suggested by \Cref{fig:normd_angular_sens}.
        However, after normalization
        with respect to the SNL, one can show that
        \begin{align}
            \lim_{\theta\rightarrow \frac{\pi\left(2n+1\right)}{l_p}} \frac{\dd f_{\mathrm{SMT}}^2}{\dd\theta} = 0,
        \end{align}
        which implies that $f_{\mathrm{SMT}}$ has a local
        extremum as $\theta\rightarrow\pi\left(2n+1\right)\!/l_p$, since
        $\dd f_{\mathrm{SMT}}^2/\dd\theta=
        2 f_{\mathrm{SMT}}\,\dd f_{\mathrm{SMT}}/\dd\theta$
        and $f_{\mathrm{SMT}}\neq 0$ for any $\theta$.
        Moreover, at the mentioned extremal
        point\footnote{To evaluate
        this limit, one can apply L'Hospital's
        rule~\pagecite{56}{Bronshtein2015}
        twice to the square of the third fraction
        in \Cref{eq:delta_theta_smt}.},
        \begin{align}
            f_{\mathrm{SMT,min}}\leqdef\lim_{\theta\rightarrow \frac{\pi\left(2n+1\right)}{l_p}} f_{\mathrm{SMT}} &= 
            \frac{\sqrt{\langle\hat{N}^{(1)}_{s,\mathrm{tot}}\rangle}}{G\left|l_p\right|},
        \end{align}
        and as we show below, $f_{\mathrm{SMT}}>f_{\mathrm{SMT,min}}$
        for $\theta\neq \pi\left(2n + 1\right)\!/l_p$.
        \par Indeed, one can perform the
        estimation:
        \begin{widetext}
            \begin{subequations}
                \begin{align}
                    \sum_{mn}\sinh^2\ofb{4G\sqrt{\lambda_{mn}}\cos\of{\frac{l_p}{2}\theta}}
                    &\geq\left\lbrace\sum_{mn}
                    \sqrt{\lambda_{mn}}
                    \sinh\ofb{4G\sqrt{\lambda_{mn}}\left|\cos\of{\frac{l_p}{2}\theta}\right|}\right\rbrace^2 \\
                    &>\sin^2\of{\frac{l_p}{2}\theta}
                    \left\lbrace\sum_{mn}
                    \sqrt{\lambda_{mn}}
                    \sinh\ofb{4G\sqrt{\lambda_{mn}}\left|\cos\of{\frac{l_p}{2}\theta}\right|}\right\rbrace^2,
                \end{align}
            \end{subequations}
        \end{widetext}
        where in the first line the Cauchy-Schwarz
        inequality~\pagecite{31}{Bronshtein2015} has been
        applied and the normalization of the eigenvalues
        has been used ($\sum_{m,n}\lambda_{mn}=1$),
        while in
        the second line, it has been assumed that
        $\sin^2\of{\frac{l_p}{2}\theta}<1$
        meaning $\theta\neq \pi\left(2n + 1\right)\!/l_p$,
        where $n\in\mathbb{Z}$.
        According to \Cref{eq:delta_theta_smt}, 
        this proves that $f_{\mathrm{SMT}}>f_{\mathrm{SMT,min}}$
        and $f_{\mathrm{SMT,min}}$ is
        indeed the global minimum of $f_{\mathrm{SMT}}$.
        \par Finally, note that the predicted optimal
        phase sensitivity for the transfer function
        approach always (except at $G=0$)
        surpasses the phase sensitivity
        for the Schmidt mode theory:
        $f_{\mathrm{TF,min}}<f_{\mathrm{SMT,min}}$.
        To see this, one can first expand the
        $\Lambda_{mn}^{\left(1\right)}$ up to the second term and
        finds for the total intensity of the first crystal:
        \begin{align}
            \langle \hat{N}_{s,\mathrm{tot}}^{\left(1\right)}\rangle
            = \sum_{m,n}\Lambda_{mn}^{\left(1\right)}
            &= G^2 + \frac{G^4}{3 k} + \mathcal{R}\of{G},
        \end{align}
        where $k=\left(\sum_{m,n}\lambda_{mn}^2\right)^{-1}>0$
        is the low-gain Schmidt number (defined using the low-gain Schmidt
        eigenvalues $\lambda_{mn}$),
        analogously to the high-gain Schmidt number defined 
        in \Cref{eq:schmidt_number}, and $\mathcal{R}\of{G}>0$
        is the remainder term, which includes high-order $G$-terms.
        Then, for $G>0$,
        \begin{align}
            G < \sqrt{\langle \hat{N}_{s,\mathrm{tot}}\rangle} < \sqrt{\mathcal{A}},
        \end{align}
        which confirms the inequality $f_{\mathrm{TF,min}}<f_{\mathrm{SMT,min}}$.

    \section{CONCLUSION} \label{sec:conclusion}
        We have presented a theoretical framework to describe
        the parametric down-conversion process pumped by
        Laguerre-Gaussian beams. Our description is based on
        the Schmidt-mode approach and is valid for any radial
        and orbital pump number. From the perspective of
        Schmidt modes, we have analyzed the mode structure
        and eigenvalue distributions of the generated quantum
        light and found that increasing the orbital and radial
        pump numbers leads to a significant broadening of the
        eigenvalue distribution, which is centered around
        $l_p/2$.
        \par Furthermore, we have investigated an \textsuoo{}
        interferometer pumped by Laguerre-Gaussian beams and
        studied its properties depending on the distance
        between the two crystals. For any nonzero distance,
        we have surprisingly found a non-monotonic behavior
        in the eigenvalue distribution, which becomes more
        pronounced at the point of destructive interference
        (dark fringe). To explain this behavior, we have
        performed an analytical analysis based on the
        double-Gaussian approximation and found a
        complex interplay of Bessel functions of different
        orders that occurs for any nonzero distance and
        is responsible for this behavior.
        The observed non-monotonic behavior allows for configurations
        where PDC modes carrying OAM higher than the pump OAM
        $l_p$ have the largest population.
        These modes can then be 
        filtered by increasing the gain due to the repopulation 
        of the eigenvalues~\cite{PRA91}.
        In turn, non-trivial eigenvalue 
        distributions result in complex non-Gaussian shaped
        intensity 
        spectra for both the low- and high-gain regimes.

        \par Finally, we have analyzed the application of the
        \textsuoo{} interferometer for the detection of angular
        displacement. For this, we have considered a rotated
        Dove prism placed between the two crystals of the
        \textsuoo{} interferometer. To study the sensitivity of
        the angular displacement measurement, we have compared
        the Schmidt-mode and the transfer-function approaches.
        Both approaches are consistent for the low-gain regime,
        but result in different sensitivities in the case of high
        gain. This is a consequence of the way the angular
        displacement has been implemented: Inside the two-photon
        amplitude for the Schmidt-mode theory and into the
        transfer functions of the second crystal
        in the transfer-function
        approach, respectively. In the low-gain regime, the
        angular sensitivity depends only on the orbital pump
        number, however, at high-gain, the radial pump number
        also strongly affects the angular sensitivity
        due to the repopulation of the modes~\cite{PRA91}.
        The angular sensitivity improves as the gain increases
        and thus the effective number of modes (both radial and
        orbital) is reduced. At the same time, the width of the
        supersensitivity region decreases quickly with increasing
        gain. However, the interplay between the gain value and
        the pump OAM value suggests a promising method to 
        achieve both high angular sensitivity and a wide
        supersensitivity range.    

    \section*{ACKNOWLEDGEMENTS}
        We acknowledge financial support of the Deutsche
        Forschungsgemeinschaft (DFG) via Project
        SH 1228/3-1 and via the TRR 142/3 (Project No.
        231447078, Subproject No. C10). We also thank
        the PC2 (Paderborn Center for Parallel Computing)
        for providing computation time.

    \section*{AUTHOR DECLARATIONS}

        \subsection*{Conflict of Interest}
            \par The authors have no conflicts to disclose.

    \section*{DATA AVAILABILITY}
        \par The data that support the findings of
        this study are available from the 
        corresponding author upon reasonable request.

    \appendix

    \section{SUPPLEMENTARY FIGURES}

        \subsection{2D mode profiles}\label{sec:modeprof2d}
            \begin{figure*}[tbh!]%
                \centering%
                \includegraphics[width=\linewidth]{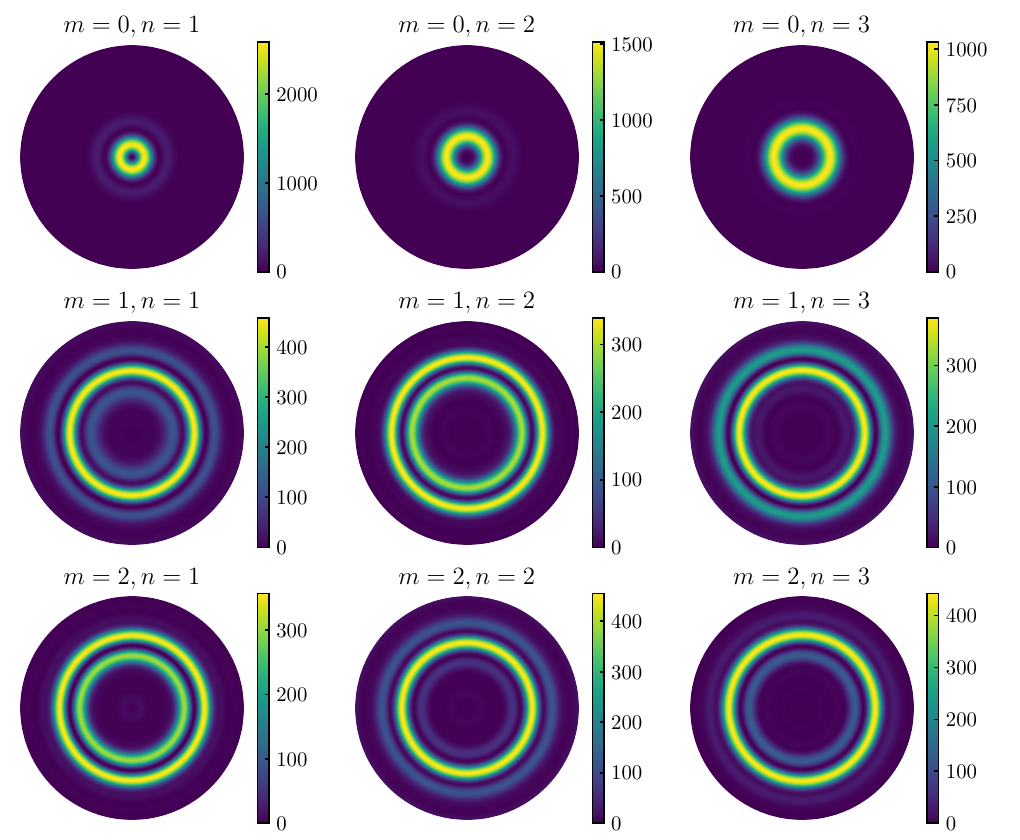}%
                \caption{%
                    Modulus squared $\left|u_{mn}\of{q}\right|^2\!/q$ (intensity profile) of the Schmidt modes  of the \textsuoo{}
                    interferometer
                    for $l_p=7$, $m_p=0$ and $d=\SI{0.42}{\centi\meter}$.
                    The radial mode index $m$  ranges from $0$ to $2$,
                    while the orbital index $n$ ranges from $1$ to $3$.
                    For an \textsuoo{} interferometer, the modes
                    are no longer similar to Laguerre-Gaussian modes.%
                }%
                \label{fig:multiplot_modes}%
            \end{figure*}
            Figure~\ref{fig:multiplot_modes} shows the modulus
            squared $\left|u_{mn}\of{q}\right|^2\!/q$ (intensity profile)
            of the Schmidt modes 
            of the \textsuoo{} interferometer
            normalized according to \Cref{eq:orthonorm_u}
            for $l_p=7$, $m_p=0$ and $d=\SI{0.42}{\centi\meter}$.
            The Schmidt modes are obtained by decomposing the
            expansion coefficients $\chi_n\of{q_s,q_i}$
            as defined in \Cref{eq:chi_n_sdecomp}.
            As one can see, the modes do not resemble
            Laguerre-Gaussian profiles. This is a consequence of
            the interference happening inside the \textsuoo{} 
            interferometer leading to a new set of basis modes.

        \subsection{Modal weight distributions}\label{sec:modal_weights_app}
            \begin{figure*}[tbh!]%
               \subfloat{%
                   \label{fig_Low_Lambda_cut_different_distances_18_0.5_a}%
                    \includegraphics[width=0.5\linewidth]{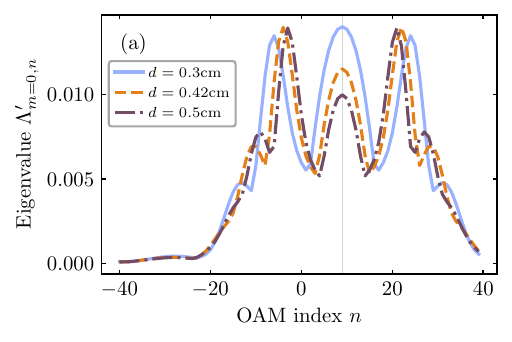}%
                }%
                \subfloat{%
                    \label{fig_Low_Lambda_cut_different_distances_18_0.5_b}%
                    \includegraphics[width=0.5\linewidth]{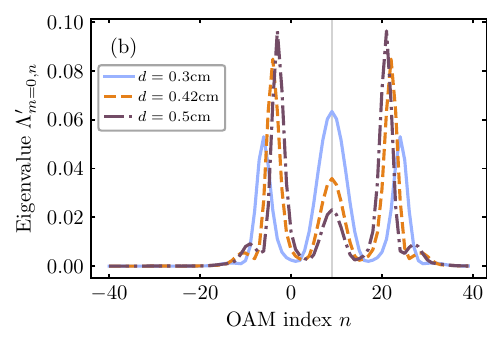}%
                }%
                \caption{%
                    Cuts of the normalized modal weights
                    $\Lambda'_{m=0,n}$ (${\sum_{m,n} \Lambda'_{mn}}=1$) for $l_p=18$, $m_p=0$ 
                    for the (a)~low-gain and (b)~high-gain regime
                    for different distances between the crystals around
                    the dark fringe point.
                    The thin gray lines indicate $n=l_p/2=9$.
                    For the distance $d=\SI{0.42}{\centi\meter}$ 
                    the most populated OAM mode
                    appears at $n = 22$, which is higher
                    than the pump OAM $l_p = 18$.
                    According to \Cref{eq:renormal_eigenvalues},
                    the most populated OAM modes
                    are strongly amplified in the high-gain
                    regime, leading to a reshaping 
                    of the eigenvalue spectrum~\cite{PRA91}.%
                }%
                \label{fig:Low_High_cut_different_distances_0.5_1}%
            \end{figure*}%
            Figure~\ref{fig:Low_High_cut_different_distances_0.5_1}
            shows the normalised (${\sum_{m,n} \Lambda'_{mn}}=1$)
            modal weights according to \Cref{eq:renormal_eigenvalues}
            for $l_p=18$, $m_p=0$ for the low-gain
            [\Cref{fig_Low_Lambda_cut_different_distances_18_0.5_a}]
            and high-gain [\Cref{fig_Low_Lambda_cut_different_distances_18_0.5_b}]
            regime for different distances between
            the crystals around the dark fringe point.
            In contrast to the case $l_p=7$ and $m_p=0$ 
            which was presented in \Cref{sec:two_crystal_setup} 
            in \Cref{fig:Low_High_cut_different_distances_0.5},
            where we have found the highest
            populated mode $n=11$ for the distances $d=\SI{0.42}{\centi\meter}$ 
            and $d=\SI{0.5}{\centi\meter}$, the highest populated modes
            we have found here
            for the case of $l_p=18$ and $m_p=0$
            is $n=22$  for the distance
            $d=\SI{0.42}{\centi\meter}$
             and $n=21$ for $d=\SI{0.5}{\centi\meter}$.
            Also, in this case, 
            the highest populated mode has a higher OAM
            than the pump OAM ($l_p=18$).

    \FloatBarrier

    \bibliography{bibliography}

\begin{thebibliography}{34}%
\makeatletter
\providecommand \@ifxundefined [1]{%
 \@ifx{#1\undefined}
}%
\providecommand \@ifnum [1]{%
 \ifnum #1\expandafter \@firstoftwo
 \else \expandafter \@secondoftwo
 \fi
}%
\providecommand \@ifx [1]{%
 \ifx #1\expandafter \@firstoftwo
 \else \expandafter \@secondoftwo
 \fi
}%
\providecommand \natexlab [1]{#1}%
\providecommand \enquote  [1]{``#1''}%
\providecommand \bibnamefont  [1]{#1}%
\providecommand \bibfnamefont [1]{#1}%
\providecommand \citenamefont [1]{#1}%
\providecommand \href@noop [0]{\@secondoftwo}%
\providecommand \href [0]{\begingroup \@sanitize@url \@href}%
\providecommand \@href[1]{\@@startlink{#1}\@@href}%
\providecommand \@@href[1]{\endgroup#1\@@endlink}%
\providecommand \@sanitize@url [0]{\catcode `\\12\catcode `\$12\catcode
  `\&12\catcode `\#12\catcode `\^12\catcode `\_12\catcode `\%12\relax}%
\providecommand \@@startlink[1]{}%
\providecommand \@@endlink[0]{}%
\providecommand \url  [0]{\begingroup\@sanitize@url \@url }%
\providecommand \@url [1]{\endgroup\@href {#1}{\urlprefix }}%
\providecommand \urlprefix  [0]{URL }%
\providecommand \Eprint [0]{\href }%
\providecommand \doibase [0]{https://doi.org/}%
\providecommand \selectlanguage [0]{\@gobble}%
\providecommand \bibinfo  [0]{\@secondoftwo}%
\providecommand \bibfield  [0]{\@secondoftwo}%
\providecommand \translation [1]{[#1]}%
\providecommand \BibitemOpen [0]{}%
\providecommand \bibitemStop [0]{}%
\providecommand \bibitemNoStop [0]{.\EOS\space}%
\providecommand \EOS [0]{\spacefactor3000\relax}%
\providecommand \BibitemShut  [1]{\csname bibitem#1\endcsname}%
\let\auto@bib@innerbib\@empty
\bibitem [{\citenamefont {Bliokh}\ \emph {et~al.}(2023)\citenamefont {Bliokh}
  \emph {et~al.}}]{Bliokh2023}%
  \BibitemOpen
  \bibfield  {author} {\bibinfo {author} {\bibfnamefont {K.~Y.}\ \bibnamefont
  {Bliokh}} \emph {et~al.},\ }\href {https://doi.org/10.1088/2040-8986/acea92}
  {\bibfield  {journal} {\bibinfo  {journal} {Journal of Optics}\ }\textbf
  {\bibinfo {volume} {25}},\ \bibinfo {pages} {103001} (\bibinfo {year}
  {2023})}\BibitemShut {NoStop}%
\bibitem [{\citenamefont {Forbes}(2022)}]{Forbes2022}%
  \BibitemOpen
  \bibfield  {author} {\bibinfo {author} {\bibfnamefont {A.}~\bibnamefont
  {Forbes}},\ }\href {https://doi.org/10.1088/2040-8986/aca109} {\bibfield
  {journal} {\bibinfo  {journal} {Journal of Optics}\ }\textbf {\bibinfo
  {volume} {24}},\ \bibinfo {pages} {124005} (\bibinfo {year}
  {2022})}\BibitemShut {NoStop}%
\bibitem [{\citenamefont {Allen}\ \emph {et~al.}(1992)\citenamefont {Allen},
  \citenamefont {Beijersbergen}, \citenamefont {Spreeuw},\ and\ \citenamefont
  {Woerdman}}]{PhysRevA.45.8185}%
  \BibitemOpen
  \bibfield  {author} {\bibinfo {author} {\bibfnamefont {L.}~\bibnamefont
  {Allen}}, \bibinfo {author} {\bibfnamefont {M.~W.}\ \bibnamefont
  {Beijersbergen}}, \bibinfo {author} {\bibfnamefont {R.~J.~C.}\ \bibnamefont
  {Spreeuw}},\ and\ \bibinfo {author} {\bibfnamefont {J.~P.}\ \bibnamefont
  {Woerdman}},\ }\href {https://doi.org/10.1103/PhysRevA.45.8185} {\bibfield
  {journal} {\bibinfo  {journal} {Phys. Rev. A}\ }\textbf {\bibinfo {volume}
  {45}},\ \bibinfo {pages} {8185} (\bibinfo {year} {1992})}\BibitemShut
  {NoStop}%
\bibitem [{\citenamefont {Wang}\ \emph {et~al.}(2021)\citenamefont {Wang},
  \citenamefont {Chen},\ and\ \citenamefont {Liu}}]{Wang2021}%
  \BibitemOpen
  \bibfield  {author} {\bibinfo {author} {\bibfnamefont {J.}~\bibnamefont
  {Wang}}, \bibinfo {author} {\bibfnamefont {S.}~\bibnamefont {Chen}},\ and\
  \bibinfo {author} {\bibfnamefont {J.}~\bibnamefont {Liu}},\ }\href
  {https://doi.org/10.1063/5.0049022} {\bibfield  {journal} {\bibinfo
  {journal} {APL Photonics}\ }\textbf {\bibinfo {volume} {6}},\ \bibinfo
  {pages} {060804} (\bibinfo {year} {2021})}\BibitemShut {NoStop}%
\bibitem [{\citenamefont {Mair}\ \emph {et~al.}(2001)\citenamefont {Mair},
  \citenamefont {Vaziri}, \citenamefont {Weihs},\ and\ \citenamefont
  {Zeilinger}}]{Mair2001}%
  \BibitemOpen
  \bibfield  {author} {\bibinfo {author} {\bibfnamefont {A.}~\bibnamefont
  {Mair}}, \bibinfo {author} {\bibfnamefont {A.}~\bibnamefont {Vaziri}},
  \bibinfo {author} {\bibfnamefont {G.}~\bibnamefont {Weihs}},\ and\ \bibinfo
  {author} {\bibfnamefont {A.}~\bibnamefont {Zeilinger}},\ }\href
  {https://doi.org/10.1038/35085529} {\bibfield  {journal} {\bibinfo  {journal}
  {Nature}\ }\textbf {\bibinfo {volume} {412}},\ \bibinfo {pages} {313}
  (\bibinfo {year} {2001})}\BibitemShut {NoStop}%
\bibitem [{\citenamefont {Shen}\ \emph {et~al.}(2019)\citenamefont {Shen},
  \citenamefont {Wang}, \citenamefont {Xie}, \citenamefont {Min}, \citenamefont
  {Fu}, \citenamefont {Liu}, \citenamefont {Gong},\ and\ \citenamefont
  {Yuan}}]{Shen2019}%
  \BibitemOpen
  \bibfield  {author} {\bibinfo {author} {\bibfnamefont {Y.}~\bibnamefont
  {Shen}}, \bibinfo {author} {\bibfnamefont {X.}~\bibnamefont {Wang}}, \bibinfo
  {author} {\bibfnamefont {Z.}~\bibnamefont {Xie}}, \bibinfo {author}
  {\bibfnamefont {C.}~\bibnamefont {Min}}, \bibinfo {author} {\bibfnamefont
  {X.}~\bibnamefont {Fu}}, \bibinfo {author} {\bibfnamefont {Q.}~\bibnamefont
  {Liu}}, \bibinfo {author} {\bibfnamefont {M.}~\bibnamefont {Gong}},\ and\
  \bibinfo {author} {\bibfnamefont {X.}~\bibnamefont {Yuan}},\ }\href
  {https://doi.org/10.1038/s41377-019-0194-2} {\bibfield  {journal} {\bibinfo
  {journal} {Light Sci Appl}\ }\textbf {\bibinfo {volume} {8}},\ \bibinfo
  {pages} {90} (\bibinfo {year} {2019})}\BibitemShut {NoStop}%
\bibitem [{\citenamefont {Sit}\ \emph {et~al.}(2017)\citenamefont {Sit},
  \citenamefont {Bouchard}, \citenamefont {Fickler}, \citenamefont
  {Gagnon-Bischoff}, \citenamefont {Larocque}, \citenamefont {Heshami},
  \citenamefont {Elser}, \citenamefont {Peuntinger}, \citenamefont
  {G\"{u}nthner}, \citenamefont {Heim}, \citenamefont {Marquardt},
  \citenamefont {Leuchs}, \citenamefont {Boyd},\ and\ \citenamefont
  {Karimi}}]{Sit:17}%
  \BibitemOpen
  \bibfield  {author} {\bibinfo {author} {\bibfnamefont {A.}~\bibnamefont
  {Sit}}, \bibinfo {author} {\bibfnamefont {F.}~\bibnamefont {Bouchard}},
  \bibinfo {author} {\bibfnamefont {R.}~\bibnamefont {Fickler}}, \bibinfo
  {author} {\bibfnamefont {J.}~\bibnamefont {Gagnon-Bischoff}}, \bibinfo
  {author} {\bibfnamefont {H.}~\bibnamefont {Larocque}}, \bibinfo {author}
  {\bibfnamefont {K.}~\bibnamefont {Heshami}}, \bibinfo {author} {\bibfnamefont
  {D.}~\bibnamefont {Elser}}, \bibinfo {author} {\bibfnamefont
  {C.}~\bibnamefont {Peuntinger}}, \bibinfo {author} {\bibfnamefont
  {K.}~\bibnamefont {G\"{u}nthner}}, \bibinfo {author} {\bibfnamefont
  {B.}~\bibnamefont {Heim}}, \bibinfo {author} {\bibfnamefont {C.}~\bibnamefont
  {Marquardt}}, \bibinfo {author} {\bibfnamefont {G.}~\bibnamefont {Leuchs}},
  \bibinfo {author} {\bibfnamefont {R.~W.}\ \bibnamefont {Boyd}},\ and\
  \bibinfo {author} {\bibfnamefont {E.}~\bibnamefont {Karimi}},\ }\href
  {https://doi.org/10.1364/OPTICA.4.001006} {\bibfield  {journal} {\bibinfo
  {journal} {Optica}\ }\textbf {\bibinfo {volume} {4}},\ \bibinfo {pages}
  {1006} (\bibinfo {year} {2017})}\BibitemShut {NoStop}%
\bibitem [{\citenamefont {Fickler}\ \emph {et~al.}(2012)\citenamefont
  {Fickler}, \citenamefont {Lapkiewicz}, \citenamefont {Plick}, \citenamefont
  {Krenn}, \citenamefont {Schaeff}, \citenamefont {Ramelow},\ and\
  \citenamefont {Zeilinger}}]{Fickler2012}%
  \BibitemOpen
  \bibfield  {author} {\bibinfo {author} {\bibfnamefont {R.}~\bibnamefont
  {Fickler}}, \bibinfo {author} {\bibfnamefont {R.}~\bibnamefont {Lapkiewicz}},
  \bibinfo {author} {\bibfnamefont {W.~N.}\ \bibnamefont {Plick}}, \bibinfo
  {author} {\bibfnamefont {M.}~\bibnamefont {Krenn}}, \bibinfo {author}
  {\bibfnamefont {C.}~\bibnamefont {Schaeff}}, \bibinfo {author} {\bibfnamefont
  {S.}~\bibnamefont {Ramelow}},\ and\ \bibinfo {author} {\bibfnamefont
  {A.}~\bibnamefont {Zeilinger}},\ }\href
  {https://doi.org/10.1126/science.1227193} {\bibfield  {journal} {\bibinfo
  {journal} {Science}\ }\textbf {\bibinfo {volume} {338}},\ \bibinfo {pages}
  {640} (\bibinfo {year} {2012})}\BibitemShut {NoStop}%
\bibitem [{\citenamefont {Karimi}\ \emph {et~al.}(2014)\citenamefont {Karimi},
  \citenamefont {Schulz}, \citenamefont {De~Leon}, \citenamefont {Qassim},
  \citenamefont {Upham},\ and\ \citenamefont {Boyd}}]{Karimi2014}%
  \BibitemOpen
  \bibfield  {author} {\bibinfo {author} {\bibfnamefont {E.}~\bibnamefont
  {Karimi}}, \bibinfo {author} {\bibfnamefont {S.~A.}\ \bibnamefont {Schulz}},
  \bibinfo {author} {\bibfnamefont {I.}~\bibnamefont {De~Leon}}, \bibinfo
  {author} {\bibfnamefont {H.}~\bibnamefont {Qassim}}, \bibinfo {author}
  {\bibfnamefont {J.}~\bibnamefont {Upham}},\ and\ \bibinfo {author}
  {\bibfnamefont {R.~W.}\ \bibnamefont {Boyd}},\ }\href
  {https://doi.org/10.1038/lsa.2014.48} {\bibfield  {journal} {\bibinfo
  {journal} {Light Sci. Appl.}\ }\textbf {\bibinfo {volume} {3}},\ \bibinfo
  {pages} {e167} (\bibinfo {year} {2014})}\BibitemShut {NoStop}%
\bibitem [{\citenamefont {{Miatto, F.M.}}\ \emph {et~al.}(2012)\citenamefont
  {{Miatto, F.M.}}, \citenamefont {{Brougham, T.}},\ and\ \citenamefont {{Yao,
  A.M.}}}]{Miatto2012}%
  \BibitemOpen
  \bibfield  {author} {\bibinfo {author} {\bibnamefont {{Miatto, F.M.}}},
  \bibinfo {author} {\bibnamefont {{Brougham, T.}}},\ and\ \bibinfo {author}
  {\bibnamefont {{Yao, A.M.}}},\ }\href
  {https://doi.org/10.1140/epjd/e2012-30063-y} {\bibfield  {journal} {\bibinfo
  {journal} {Eur. Phys. J. D}\ }\textbf {\bibinfo {volume} {66}},\ \bibinfo
  {pages} {183} (\bibinfo {year} {2012})}\BibitemShut {NoStop}%
\bibitem [{\citenamefont {Sharapova}\ \emph {et~al.}(2015)\citenamefont
  {Sharapova}, \citenamefont {P\'erez}, \citenamefont {Tikhonova},\ and\
  \citenamefont {Chekhova}}]{PRA91}%
  \BibitemOpen
  \bibfield  {author} {\bibinfo {author} {\bibfnamefont {P.}~\bibnamefont
  {Sharapova}}, \bibinfo {author} {\bibfnamefont {A.~M.}\ \bibnamefont
  {P\'erez}}, \bibinfo {author} {\bibfnamefont {O.~V.}\ \bibnamefont
  {Tikhonova}},\ and\ \bibinfo {author} {\bibfnamefont {M.~V.}\ \bibnamefont
  {Chekhova}},\ }\href {https://doi.org/10.1103/PhysRevA.91.043816} {\bibfield
  {journal} {\bibinfo  {journal} {Phys. Rev. A}\ }\textbf {\bibinfo {volume}
  {91}},\ \bibinfo {pages} {043816} (\bibinfo {year} {2015})}\BibitemShut
  {NoStop}%
\bibitem [{\citenamefont {Beltran}\ \emph {et~al.}(2017)\citenamefont
  {Beltran}, \citenamefont {Frascella}, \citenamefont {Perez}, \citenamefont
  {Fickler}, \citenamefont {Sharapova}, \citenamefont {Manceau}, \citenamefont
  {Tikhonova}, \citenamefont {Boyd}, \citenamefont {Leuchs},\ and\
  \citenamefont {Chekhova}}]{Beltran2017}%
  \BibitemOpen
  \bibfield  {author} {\bibinfo {author} {\bibfnamefont {L.}~\bibnamefont
  {Beltran}}, \bibinfo {author} {\bibfnamefont {G.}~\bibnamefont {Frascella}},
  \bibinfo {author} {\bibfnamefont {A.~M.}\ \bibnamefont {Perez}}, \bibinfo
  {author} {\bibfnamefont {R.}~\bibnamefont {Fickler}}, \bibinfo {author}
  {\bibfnamefont {P.~R.}\ \bibnamefont {Sharapova}}, \bibinfo {author}
  {\bibfnamefont {M.}~\bibnamefont {Manceau}}, \bibinfo {author} {\bibfnamefont
  {O.~V.}\ \bibnamefont {Tikhonova}}, \bibinfo {author} {\bibfnamefont {R.~W.}\
  \bibnamefont {Boyd}}, \bibinfo {author} {\bibfnamefont {G.}~\bibnamefont
  {Leuchs}},\ and\ \bibinfo {author} {\bibfnamefont {M.~V.}\ \bibnamefont
  {Chekhova}},\ }\href {https://doi.org/10.1088/2040-8986/aa600f} {\bibfield
  {journal} {\bibinfo  {journal} {Journal of Optics}\ }\textbf {\bibinfo
  {volume} {19}},\ \bibinfo {pages} {044005} (\bibinfo {year}
  {2017})}\BibitemShut {NoStop}%
\bibitem [{\citenamefont {Offer}\ \emph {et~al.}(2018)\citenamefont {Offer},
  \citenamefont {Stulga}, \citenamefont {Riis}, \citenamefont {Franke-Arnold},\
  and\ \citenamefont {Arnold}}]{Offer2018}%
  \BibitemOpen
  \bibfield  {author} {\bibinfo {author} {\bibfnamefont {R.~F.}\ \bibnamefont
  {Offer}}, \bibinfo {author} {\bibfnamefont {D.}~\bibnamefont {Stulga}},
  \bibinfo {author} {\bibfnamefont {E.}~\bibnamefont {Riis}}, \bibinfo {author}
  {\bibfnamefont {S.}~\bibnamefont {Franke-Arnold}},\ and\ \bibinfo {author}
  {\bibfnamefont {A.~S.}\ \bibnamefont {Arnold}},\ }\href
  {https://doi.org/10.1038/s42005-018-0077-5} {\bibfield  {journal} {\bibinfo
  {journal} {Commun Phys}\ }\textbf {\bibinfo {volume} {1}},\ \bibinfo {pages}
  {84} (\bibinfo {year} {2018})}\BibitemShut {NoStop}%
\bibitem [{\citenamefont {Cozzolino}\ \emph {et~al.}(2019)\citenamefont
  {Cozzolino}, \citenamefont {Bacco}, \citenamefont {Da~Lio}, \citenamefont
  {Ingerslev}, \citenamefont {Ding}, \citenamefont {Dalgaard}, \citenamefont
  {Kristensen}, \citenamefont {Galili}, \citenamefont {Rottwitt}, \citenamefont
  {Ramachandran},\ and\ \citenamefont
  {Oxenl\o{}we}}]{PhysRevApplied.11.064058}%
  \BibitemOpen
  \bibfield  {author} {\bibinfo {author} {\bibfnamefont {D.}~\bibnamefont
  {Cozzolino}}, \bibinfo {author} {\bibfnamefont {D.}~\bibnamefont {Bacco}},
  \bibinfo {author} {\bibfnamefont {B.}~\bibnamefont {Da~Lio}}, \bibinfo
  {author} {\bibfnamefont {K.}~\bibnamefont {Ingerslev}}, \bibinfo {author}
  {\bibfnamefont {Y.}~\bibnamefont {Ding}}, \bibinfo {author} {\bibfnamefont
  {K.}~\bibnamefont {Dalgaard}}, \bibinfo {author} {\bibfnamefont
  {P.}~\bibnamefont {Kristensen}}, \bibinfo {author} {\bibfnamefont
  {M.}~\bibnamefont {Galili}}, \bibinfo {author} {\bibfnamefont
  {K.}~\bibnamefont {Rottwitt}}, \bibinfo {author} {\bibfnamefont
  {S.}~\bibnamefont {Ramachandran}},\ and\ \bibinfo {author} {\bibfnamefont
  {L.~K.}\ \bibnamefont {Oxenl\o{}we}},\ }\href
  {https://doi.org/10.1103/PhysRevApplied.11.064058} {\bibfield  {journal}
  {\bibinfo  {journal} {Phys. Rev. Appl.}\ }\textbf {\bibinfo {volume} {11}},\
  \bibinfo {pages} {064058} (\bibinfo {year} {2019})}\BibitemShut {NoStop}%
\bibitem [{\citenamefont {Krenn}\ \emph {et~al.}(2016)\citenamefont {Krenn},
  \citenamefont {Handsteiner}, \citenamefont {Fink}, \citenamefont {Fickler},
  \citenamefont {Ursin}, \citenamefont {Malik},\ and\ \citenamefont
  {Zeilinger}}]{Krenn2016}%
  \BibitemOpen
  \bibfield  {author} {\bibinfo {author} {\bibfnamefont {M.}~\bibnamefont
  {Krenn}}, \bibinfo {author} {\bibfnamefont {J.}~\bibnamefont {Handsteiner}},
  \bibinfo {author} {\bibfnamefont {M.}~\bibnamefont {Fink}}, \bibinfo {author}
  {\bibfnamefont {R.}~\bibnamefont {Fickler}}, \bibinfo {author} {\bibfnamefont
  {R.}~\bibnamefont {Ursin}}, \bibinfo {author} {\bibfnamefont
  {M.}~\bibnamefont {Malik}},\ and\ \bibinfo {author} {\bibfnamefont
  {A.}~\bibnamefont {Zeilinger}},\ }\href
  {https://doi.org/10.1073/pnas.1612023113} {\bibfield  {journal} {\bibinfo
  {journal} {Proc. Natl. Acad. Sci. U.S.A.}\ }\textbf {\bibinfo {volume}
  {113}},\ \bibinfo {pages} {13648} (\bibinfo {year} {2016})}\BibitemShut
  {NoStop}%
\bibitem [{\citenamefont {Xie}\ \emph {et~al.}(2015)\citenamefont {Xie},
  \citenamefont {Li}, \citenamefont {Ren}, \citenamefont {Huang}, \citenamefont
  {Yan}, \citenamefont {Ahmed}, \citenamefont {Zhao}, \citenamefont {Lavery},
  \citenamefont {Ashrafi}, \citenamefont {Ashrafi}, \citenamefont {Bock},
  \citenamefont {Tur}, \citenamefont {Molisch},\ and\ \citenamefont
  {Willner}}]{Xie:15}%
  \BibitemOpen
  \bibfield  {author} {\bibinfo {author} {\bibfnamefont {G.}~\bibnamefont
  {Xie}}, \bibinfo {author} {\bibfnamefont {L.}~\bibnamefont {Li}}, \bibinfo
  {author} {\bibfnamefont {Y.}~\bibnamefont {Ren}}, \bibinfo {author}
  {\bibfnamefont {H.}~\bibnamefont {Huang}}, \bibinfo {author} {\bibfnamefont
  {Y.}~\bibnamefont {Yan}}, \bibinfo {author} {\bibfnamefont {N.}~\bibnamefont
  {Ahmed}}, \bibinfo {author} {\bibfnamefont {Z.}~\bibnamefont {Zhao}},
  \bibinfo {author} {\bibfnamefont {M.~P.~J.}\ \bibnamefont {Lavery}}, \bibinfo
  {author} {\bibfnamefont {N.}~\bibnamefont {Ashrafi}}, \bibinfo {author}
  {\bibfnamefont {S.}~\bibnamefont {Ashrafi}}, \bibinfo {author} {\bibfnamefont
  {R.}~\bibnamefont {Bock}}, \bibinfo {author} {\bibfnamefont {M.}~\bibnamefont
  {Tur}}, \bibinfo {author} {\bibfnamefont {A.~F.}\ \bibnamefont {Molisch}},\
  and\ \bibinfo {author} {\bibfnamefont {A.~E.}\ \bibnamefont {Willner}},\
  }\href {https://doi.org/10.1364/OPTICA.2.000357} {\bibfield  {journal}
  {\bibinfo  {journal} {Optica}\ }\textbf {\bibinfo {volume} {2}},\ \bibinfo
  {pages} {357} (\bibinfo {year} {2015})}\BibitemShut {NoStop}%
\bibitem [{\citenamefont {Hiekkam\"aki}\ \emph {et~al.}(2021)\citenamefont
  {Hiekkam\"aki}, \citenamefont {Bouchard},\ and\ \citenamefont
  {Fickler}}]{PhysRevLett.127.263601}%
  \BibitemOpen
  \bibfield  {author} {\bibinfo {author} {\bibfnamefont {M.}~\bibnamefont
  {Hiekkam\"aki}}, \bibinfo {author} {\bibfnamefont {F.}~\bibnamefont
  {Bouchard}},\ and\ \bibinfo {author} {\bibfnamefont {R.}~\bibnamefont
  {Fickler}},\ }\href {https://doi.org/10.1103/PhysRevLett.127.263601}
  {\bibfield  {journal} {\bibinfo  {journal} {Phys. Rev. Lett.}\ }\textbf
  {\bibinfo {volume} {127}},\ \bibinfo {pages} {263601} (\bibinfo {year}
  {2021})}\BibitemShut {NoStop}%
\bibitem [{\citenamefont {Jha}\ \emph {et~al.}(2011)\citenamefont {Jha},
  \citenamefont {Agarwal},\ and\ \citenamefont {Boyd}}]{PhysRevA.83.053829}%
  \BibitemOpen
  \bibfield  {author} {\bibinfo {author} {\bibfnamefont {A.~K.}\ \bibnamefont
  {Jha}}, \bibinfo {author} {\bibfnamefont {G.~S.}\ \bibnamefont {Agarwal}},\
  and\ \bibinfo {author} {\bibfnamefont {R.~W.}\ \bibnamefont {Boyd}},\ }\href
  {https://doi.org/10.1103/PhysRevA.83.053829} {\bibfield  {journal} {\bibinfo
  {journal} {Phys. Rev. A}\ }\textbf {\bibinfo {volume} {83}},\ \bibinfo
  {pages} {053829} (\bibinfo {year} {2011})}\BibitemShut {NoStop}%
\bibitem [{\citenamefont {P\'{e}rez}\ \emph {et~al.}(2014)\citenamefont
  {P\'{e}rez}, \citenamefont {Iskhakov}, \citenamefont {Sharapova},
  \citenamefont {Lemieux}, \citenamefont {Tikhonova}, \citenamefont
  {Chekhova},\ and\ \citenamefont {Leuchs}}]{Perez:14}%
  \BibitemOpen
  \bibfield  {author} {\bibinfo {author} {\bibfnamefont {A.~M.}\ \bibnamefont
  {P\'{e}rez}}, \bibinfo {author} {\bibfnamefont {T.~S.}\ \bibnamefont
  {Iskhakov}}, \bibinfo {author} {\bibfnamefont {P.}~\bibnamefont {Sharapova}},
  \bibinfo {author} {\bibfnamefont {S.}~\bibnamefont {Lemieux}}, \bibinfo
  {author} {\bibfnamefont {O.~V.}\ \bibnamefont {Tikhonova}}, \bibinfo {author}
  {\bibfnamefont {M.~V.}\ \bibnamefont {Chekhova}},\ and\ \bibinfo {author}
  {\bibfnamefont {G.}~\bibnamefont {Leuchs}},\ }\href
  {https://doi.org/10.1364/OL.39.002403} {\bibfield  {journal} {\bibinfo
  {journal} {Opt. Lett.}\ }\textbf {\bibinfo {volume} {39}},\ \bibinfo {pages}
  {2403} (\bibinfo {year} {2014})}\BibitemShut {NoStop}%
\bibitem [{\citenamefont {Manceau}\ \emph {et~al.}(2017)\citenamefont
  {Manceau}, \citenamefont {Khalili},\ and\ \citenamefont
  {Chekhova}}]{Manceau2017}%
  \BibitemOpen
  \bibfield  {author} {\bibinfo {author} {\bibfnamefont {M.}~\bibnamefont
  {Manceau}}, \bibinfo {author} {\bibfnamefont {F.}~\bibnamefont {Khalili}},\
  and\ \bibinfo {author} {\bibfnamefont {M.}~\bibnamefont {Chekhova}},\ }\href
  {https://doi.org/10.1088/1367-2630/aa53d1} {\bibfield  {journal} {\bibinfo
  {journal} {New Journal of Physics}\ }\textbf {\bibinfo {volume} {19}},\
  \bibinfo {pages} {013014} (\bibinfo {year} {2017})}\BibitemShut {NoStop}%
\bibitem [{\citenamefont {Chekhova}\ and\ \citenamefont
  {Ou}(2016)}]{Chekhova:16}%
  \BibitemOpen
  \bibfield  {author} {\bibinfo {author} {\bibfnamefont {M.~V.}\ \bibnamefont
  {Chekhova}}\ and\ \bibinfo {author} {\bibfnamefont {Z.~Y.}\ \bibnamefont
  {Ou}},\ }\href {https://doi.org/10.1364/AOP.8.000104} {\bibfield  {journal}
  {\bibinfo  {journal} {Adv. Opt. Photon.}\ }\textbf {\bibinfo {volume} {8}},\
  \bibinfo {pages} {104} (\bibinfo {year} {2016})}\BibitemShut {NoStop}%
\bibitem [{\citenamefont {Frascella}\ \emph {et~al.}(2019)\citenamefont
  {Frascella}, \citenamefont {Mikhailov}, \citenamefont {Takanashi},
  \citenamefont {Zakharov}, \citenamefont {Tikhonova},\ and\ \citenamefont
  {Chekhova}}]{Frascella:19}%
  \BibitemOpen
  \bibfield  {author} {\bibinfo {author} {\bibfnamefont {G.}~\bibnamefont
  {Frascella}}, \bibinfo {author} {\bibfnamefont {E.~E.}\ \bibnamefont
  {Mikhailov}}, \bibinfo {author} {\bibfnamefont {N.}~\bibnamefont
  {Takanashi}}, \bibinfo {author} {\bibfnamefont {R.~V.}\ \bibnamefont
  {Zakharov}}, \bibinfo {author} {\bibfnamefont {O.~V.}\ \bibnamefont
  {Tikhonova}},\ and\ \bibinfo {author} {\bibfnamefont {M.~V.}\ \bibnamefont
  {Chekhova}},\ }\href {https://doi.org/10.1364/OPTICA.6.001233} {\bibfield
  {journal} {\bibinfo  {journal} {Optica}\ }\textbf {\bibinfo {volume} {6}},\
  \bibinfo {pages} {1233} (\bibinfo {year} {2019})}\BibitemShut {NoStop}%
\bibitem [{\citenamefont {Sharapova}\ \emph {et~al.}(2020)\citenamefont
  {Sharapova}, \citenamefont {Frascella}, \citenamefont {Riabinin},
  \citenamefont {P\'erez}, \citenamefont {Tikhonova}, \citenamefont {Lemieux},
  \citenamefont {Boyd}, \citenamefont {Leuchs},\ and\ \citenamefont
  {Chekhova}}]{PRR2}%
  \BibitemOpen
  \bibfield  {author} {\bibinfo {author} {\bibfnamefont {P.~R.}\ \bibnamefont
  {Sharapova}}, \bibinfo {author} {\bibfnamefont {G.}~\bibnamefont
  {Frascella}}, \bibinfo {author} {\bibfnamefont {M.}~\bibnamefont {Riabinin}},
  \bibinfo {author} {\bibfnamefont {A.~M.}\ \bibnamefont {P\'erez}}, \bibinfo
  {author} {\bibfnamefont {O.~V.}\ \bibnamefont {Tikhonova}}, \bibinfo {author}
  {\bibfnamefont {S.}~\bibnamefont {Lemieux}}, \bibinfo {author} {\bibfnamefont
  {R.~W.}\ \bibnamefont {Boyd}}, \bibinfo {author} {\bibfnamefont
  {G.}~\bibnamefont {Leuchs}},\ and\ \bibinfo {author} {\bibfnamefont {M.~V.}\
  \bibnamefont {Chekhova}},\ }\href
  {https://doi.org/10.1103/PhysRevResearch.2.013371} {\bibfield  {journal}
  {\bibinfo  {journal} {Phys. Rev. Res.}\ }\textbf {\bibinfo {volume} {2}},\
  \bibinfo {pages} {013371} (\bibinfo {year} {2020})}\BibitemShut {NoStop}%
\bibitem [{\citenamefont {Baghdasaryan}\ \emph {et~al.}(2022)\citenamefont
  {Baghdasaryan}, \citenamefont {Sevilla-Guti\'errez}, \citenamefont
  {Steinlechner},\ and\ \citenamefont {Fritzsche}}]{PhysRevA.106.063711}%
  \BibitemOpen
  \bibfield  {author} {\bibinfo {author} {\bibfnamefont {B.}~\bibnamefont
  {Baghdasaryan}}, \bibinfo {author} {\bibfnamefont {C.}~\bibnamefont
  {Sevilla-Guti\'errez}}, \bibinfo {author} {\bibfnamefont {F.}~\bibnamefont
  {Steinlechner}},\ and\ \bibinfo {author} {\bibfnamefont {S.}~\bibnamefont
  {Fritzsche}},\ }\href {https://doi.org/10.1103/PhysRevA.106.063711}
  {\bibfield  {journal} {\bibinfo  {journal} {Phys. Rev. A}\ }\textbf {\bibinfo
  {volume} {106}},\ \bibinfo {pages} {063711} (\bibinfo {year}
  {2022})}\BibitemShut {NoStop}%
\bibitem [{\citenamefont {Karan}\ \emph {et~al.}(2020)\citenamefont {Karan},
  \citenamefont {Aarav}, \citenamefont {Bharadhwaj}, \citenamefont {Taneja},
  \citenamefont {De}, \citenamefont {Kulkarni}, \citenamefont {Meher},\ and\
  \citenamefont {Jha}}]{Karan2020}%
  \BibitemOpen
  \bibfield  {author} {\bibinfo {author} {\bibfnamefont {S.}~\bibnamefont
  {Karan}}, \bibinfo {author} {\bibfnamefont {S.}~\bibnamefont {Aarav}},
  \bibinfo {author} {\bibfnamefont {H.}~\bibnamefont {Bharadhwaj}}, \bibinfo
  {author} {\bibfnamefont {L.}~\bibnamefont {Taneja}}, \bibinfo {author}
  {\bibfnamefont {A.}~\bibnamefont {De}}, \bibinfo {author} {\bibfnamefont
  {G.}~\bibnamefont {Kulkarni}}, \bibinfo {author} {\bibfnamefont
  {N.}~\bibnamefont {Meher}},\ and\ \bibinfo {author} {\bibfnamefont {A.~K.}\
  \bibnamefont {Jha}},\ }\href {https://doi.org/10.1088/2040-8986/ab89e4}
  {\bibfield  {journal} {\bibinfo  {journal} {Journal of Optics}\ }\textbf
  {\bibinfo {volume} {22}},\ \bibinfo {pages} {083501} (\bibinfo {year}
  {2020})}\BibitemShut {NoStop}%
\bibitem [{\citenamefont {Klyshko}(1988)}]{klyshko1988photons}%
  \BibitemOpen
  \bibfield  {author} {\bibinfo {author} {\bibfnamefont {D.}~\bibnamefont
  {Klyshko}},\ }\href {https://doi.org/10.1201/9780203743508} {\emph {\bibinfo
  {title} {Photons and {N}onlinear {O}ptics}}}\ (\bibinfo  {publisher} {Taylor
  \& Francis},\ \bibinfo {address} {New York},\ \bibinfo {year}
  {1988})\BibitemShut {NoStop}%
\bibitem [{\citenamefont {Pampaloni}\ and\ \citenamefont
  {Enderlein}(2004)}]{Pampaloni2004}%
  \BibitemOpen
  \bibfield  {author} {\bibinfo {author} {\bibfnamefont {F.}~\bibnamefont
  {Pampaloni}}\ and\ \bibinfo {author} {\bibfnamefont {J.}~\bibnamefont
  {Enderlein}},\ }\href@noop {} {\bibinfo {title} {{G}aussian,
  {H}ermite-{G}aussian, and {L}aguerre-{G}aussian beams: {A} primer}} (\bibinfo
  {year} {2004}),\ \Eprint {https://arxiv.org/abs/physics/0410021}
  {arXiv:physics/0410021} \BibitemShut {NoStop}%
\bibitem [{\citenamefont {Bronshtein}\ \emph {et~al.}(2015)\citenamefont
  {Bronshtein}, \citenamefont {Semendyayev}, \citenamefont {Musiol},\ and\
  \citenamefont {M\"{u}hlig}}]{Bronshtein2015}%
  \BibitemOpen
  \bibfield  {author} {\bibinfo {author} {\bibfnamefont {I.}~\bibnamefont
  {Bronshtein}}, \bibinfo {author} {\bibfnamefont {K.}~\bibnamefont
  {Semendyayev}}, \bibinfo {author} {\bibfnamefont {G.}~\bibnamefont
  {Musiol}},\ and\ \bibinfo {author} {\bibfnamefont {H.}~\bibnamefont
  {M\"{u}hlig}},\ }\href {https://doi.org/10.1007/978-3-662-46221-8} {\emph
  {\bibinfo {title} {Handbook of Mathematics}}}\ (\bibinfo  {publisher}
  {Springer Berlin Heidelberg},\ \bibinfo {year} {2015})\BibitemShut {NoStop}%
\bibitem [{\citenamefont {Watson}(1995)}]{watson}%
  \BibitemOpen
  \bibfield  {author} {\bibinfo {author} {\bibfnamefont {G.~N.}\ \bibnamefont
  {Watson}},\ }\href@noop {} {\emph {\bibinfo {title} {A Treatise on the Theory
  of Bessel Functions}}}\ (\bibinfo  {publisher} {Cambridge University Press},\
  \bibinfo {year} {1995})\BibitemShut {NoStop}%
\bibitem [{\citenamefont {{Bateman Manuscript Project}}\ \emph
  {et~al.}(1954)\citenamefont {{Bateman Manuscript Project}}, \citenamefont
  {Bateman},\ and\ \citenamefont {Erd{\'e}lyi}}]{bateman}%
  \BibitemOpen
  \bibfield  {author} {\bibinfo {author} {\bibnamefont {{Bateman Manuscript
  Project}}}, \bibinfo {author} {\bibfnamefont {H.}~\bibnamefont {Bateman}},\
  and\ \bibinfo {author} {\bibfnamefont {A.}~\bibnamefont {Erd{\'e}lyi}},\
  }\href@noop {} {\emph {\bibinfo {title} {Tables of Integral Transforms}}},\
  Vol.~\bibinfo {volume} {2}\ (\bibinfo  {publisher} {McGraw-Hill},\ \bibinfo
  {year} {1954})\BibitemShut {NoStop}%
\bibitem [{\citenamefont {Fontaine}\ \emph {et~al.}(2019)\citenamefont
  {Fontaine}, \citenamefont {Ryf}, \citenamefont {Chen}, \citenamefont
  {Neilson}, \citenamefont {Kim},\ and\ \citenamefont
  {Carpenter}}]{Fontaine2019}%
  \BibitemOpen
  \bibfield  {author} {\bibinfo {author} {\bibfnamefont {N.~K.}\ \bibnamefont
  {Fontaine}}, \bibinfo {author} {\bibfnamefont {R.}~\bibnamefont {Ryf}},
  \bibinfo {author} {\bibfnamefont {H.}~\bibnamefont {Chen}}, \bibinfo {author}
  {\bibfnamefont {D.~T.}\ \bibnamefont {Neilson}}, \bibinfo {author}
  {\bibfnamefont {K.}~\bibnamefont {Kim}},\ and\ \bibinfo {author}
  {\bibfnamefont {J.}~\bibnamefont {Carpenter}},\ }\href
  {https://doi.org/10.1038/s41467-019-09840-4} {\bibfield  {journal} {\bibinfo
  {journal} {Nat. Commun.}\ }\textbf {\bibinfo {volume} {10}},\ \bibinfo
  {pages} {1865} (\bibinfo {year} {2019})}\BibitemShut {NoStop}%
\bibitem [{\citenamefont {Scharwald}\ \emph {et~al.}(2023)\citenamefont
  {Scharwald}, \citenamefont {Meier},\ and\ \citenamefont {Sharapova}}]{PRR5}%
  \BibitemOpen
  \bibfield  {author} {\bibinfo {author} {\bibfnamefont {D.}~\bibnamefont
  {Scharwald}}, \bibinfo {author} {\bibfnamefont {T.}~\bibnamefont {Meier}},\
  and\ \bibinfo {author} {\bibfnamefont {P.~R.}\ \bibnamefont {Sharapova}},\
  }\href {https://doi.org/10.1103/PhysRevResearch.5.043158} {\bibfield
  {journal} {\bibinfo  {journal} {Phys. Rev. Res.}\ }\textbf {\bibinfo {volume}
  {5}},\ \bibinfo {pages} {043158} (\bibinfo {year} {2023})}\BibitemShut
  {NoStop}%
\bibitem [{\citenamefont {Spasibko}\ \emph {et~al.}(2012)\citenamefont
  {Spasibko}, \citenamefont {Iskhakov},\ and\ \citenamefont
  {Chekhova}}]{Spasibko:12}%
  \BibitemOpen
  \bibfield  {author} {\bibinfo {author} {\bibfnamefont {K.~Y.}\ \bibnamefont
  {Spasibko}}, \bibinfo {author} {\bibfnamefont {T.~S.}\ \bibnamefont
  {Iskhakov}},\ and\ \bibinfo {author} {\bibfnamefont {M.~V.}\ \bibnamefont
  {Chekhova}},\ }\href {https://doi.org/10.1364/OE.20.007507} {\bibfield
  {journal} {\bibinfo  {journal} {Opt. Express}\ }\textbf {\bibinfo {volume}
  {20}},\ \bibinfo {pages} {7507} (\bibinfo {year} {2012})}\BibitemShut
  {NoStop}%
\bibitem [{\citenamefont {Gonz\'{a}lez}\ \emph {et~al.}(2006)\citenamefont
  {Gonz\'{a}lez}, \citenamefont {Molina-Terriza},\ and\ \citenamefont
  {Torres}}]{Gonzalez:06}%
  \BibitemOpen
  \bibfield  {author} {\bibinfo {author} {\bibfnamefont {N.}~\bibnamefont
  {Gonz\'{a}lez}}, \bibinfo {author} {\bibfnamefont {G.}~\bibnamefont
  {Molina-Terriza}},\ and\ \bibinfo {author} {\bibfnamefont {J.~P.}\
  \bibnamefont {Torres}},\ }\href {https://doi.org/10.1364/OE.14.009093}
  {\bibfield  {journal} {\bibinfo  {journal} {Opt. Express}\ }\textbf {\bibinfo
  {volume} {14}},\ \bibinfo {pages} {9093} (\bibinfo {year}
  {2006})}\BibitemShut {NoStop}%
\end{thebibliography}%

\end{document}